\documentclass[a4paper,11pt]{article} 
\usepackage[margin=0.75in]{geometry} 
\usepackage{amssymb}
\usepackage{amsmath,epsfig, epsf, epic, graphicx} 
\usepackage{fancybox, lscape} 
\usepackage{bm}
\usepackage{natbib}
\usepackage{graphicx}
\DeclareGraphicsExtensions{.eps,.pdf,.jpeg,.png}
\usepackage{caption}
\usepackage[margin=2cm]{caption}
\usepackage{setspace}
\usepackage{cleveref}
\usepackage{xcolor}
\usepackage{array}
\usepackage{wrapfig}
\usepackage{multirow}
\usepackage{url}
\usepackage{tabularx}
\allowdisplaybreaks

\usepackage{tikz} 
\usetikzlibrary{positioning, calc}
\usepackage{floatrow}
\usepackage{subfig}
\usepackage[normalem]{ulem}

\newtheorem{theorem}{Theorem}

\newtheorem{assumption}{Assumption}

\newcommand\independent{\protect\mathpalette{\protect\independenT}{\perp}}
\def\independenT#1#2{\mathrel{\rlap{$#1#2$}\mkern2mu{#1#2}}}

\DeclareMathOperator*{\argmax}{arg\,max}

\title{Principal stratification with continuous treatments and continuous post-treatment variables}
\author{Joseph Antonelli\thanks{Department of Statistics, University of Florida, Gainesville, FL, USA},  Minxuan Wu\thanks{Department of Statistics, University of Florida, Gainesville, FL, USA}, Fabrizia Mealli\thanks{Department of Economics, European University Institute, Florence, Italy}, Brenden Beck\thanks{School of Criminal Justice, Rutgers University, Newark, NJ, USA}, and Alessandra Mattei\thanks{Department of Statistics and Data Science, University of Florence, Florence, Italy}}
\date{}

\begin{document}

\maketitle

\begin{abstract}
    
Principal stratification (PS) is a commonly used approach for understanding the mechanisms through which a treatment affects an outcome. The goal of this work is to extend the PS framework to studies with continuous treatments, which introduces a number of both challenges and opportunities in terms of defining causal effects and performing inference. This manuscript provides multiple key methodological contributions: 1) we introduce principal causal estimands for continuous treatments that provide insights into different causal mechanisms, 2) we show that nonparametric identification is possible under a principal ignorability assumption, but only under a restrictive assumption on the joint distribution of potential mediators, which can be dropped under certain parametric assumptions, 3)  we utilize nonparametric Bayesian models for the joint distribution of the potential mediating variables to mitigate issues from model misspecification, and 4) we provide theoretical justification for utilizing an outcome model to identify the joint distribution of the potential mediating variables, and show that this is only possible if a principal ignorability assumption is violated. Lastly, we illustrate our methodology by applying it to a novel study of the relationship between the economy and arrest rates, and how this is potentially mediated by police capacity. 

\end{abstract}

\section{Introduction and background} 

In the potential outcome approach to causal inference \cite[][]{rubin1974estimating}, principal stratification \cite[PS,][]{frangakis2002principal} is a useful approach to conceptualize the mediating role of intermediate variables in the treatment-outcome relationship \cite[e.g.,][]{ten2012review, baccini2017bayesian,  kim2019bayesian}. Principal stratification partitions units into groups, called principal strata, based on the joint potential values of a mediator under different treatment levels. Because principal stratum membership is unaffected by treatment, it can be viewed as an intrinsic latent characteristic of each unit. Principal causal effects (PCEs), defined as comparisons of potential outcomes within a principal stratum (or union of principal strata), characterize treatment effect heterogeneity with respect to the mediator and can provide insights into underlying causal mechanisms. \cite[][]{mealli2012refreshing}. As an example, one can look at treatment effects within principal strata for which the treatment either does or does not causally affect the mediator, sometimes referred to as associative or dissociative effects. Dissociative PCEs can be interpreted as the principal strata \emph{direct} effects of the treatment on the outcome not channeled through the mediator, whereas associative PCEs combine both direct and indirect effects of the treatment through the mediating variable \cite[][]{vanderweele2008simple}. See \cite{mealli2003assumptions, rubin2004direct,mattei2011augmented, baccini2017bayesian} for a discussion. Comparing the magnitude of these two effects can provide insights into whether the treatment effects are mediated or not \cite[e.g.,][]{zigler2012estimating, forastiere2016identification}. 


Principal stratification is well studied for binary treatments and has been applied to a wide range of post-treatment variable types spanning binary, multivariate, continuous and time-dependent mediators \cite[][]{imbens1997bayesian, hirano2000assessing, mealli2003assumptions, jin2008principal, 
sjolander2009sensitivity, frumento2012evaluating, zigler2012bayesian,mattei2020assessing, bia2022assessing}. Assumptions on the treatment assignment mechanism alone are insufficient to identify PCEs, motivating additional assumptions whose plausibility depends on the empirical setting. Arguably the most common of these are monotonicity and exclusion restriction assumptions \cite[][]{angrist1996identification, hirano2000assessing, bia2022assessing}, which eliminate certain principal strata and enforce that some PCEs are zero. Alternative assumptions include principal ignorability  \cite[][]{jo2011use, jiang2021identification, mattei2023assessing, lu2023principal},  or conditional independence between an auxiliary variable and the outcome given principal stratum membership and covariates \cite[e.g.,][]{jiang2021identification}. If explicit identification is not possible, some authors have derived partial identification results for PCEs \cite[e.g.,][]{zhang2003estimation,  cheng2006bounds, imai2008sharp, lee2009training, mattei2011augmented, mealli2016identification, yang2016using}. When structural assumptions alone are insufficient for nonparametric identification, they can be combined with parametric assumptions \cite[e.g.,][]{hirano2000assessing, mattei2007application, jin2008principal, ma2011causal, schwartz2011bayesian, bartolucci2011modeling, frumento2012evaluating,  zigler2012bayesian, kim2019bayesian}. For a comprehensive overview of identification results for PCEs with binary treatments, we point readers to \cite{jiang2021identification}. In addition to the binary treatment setting, extensions exist for both longitudinal treatments \cite[][]{frangakis2004methodology, ricciardi2020bayesian} and multi-valued treatments \cite[][]{cheng2006bounds, feller2016compared}. 
 
Despite this extensive literature, principal stratification with continuous treatments remains largely unexplored. This seemingly straightforward extension introduces substantial challenges for both identification and inference. Addressing these challenges is important because continuous treatments are common in many applied fields, including economics. 
The combination of a continuous treatment and a continuous mediator raises challenges because there are infinitely many principal strata defined by the treatment-mediator trajectories. Moreover, principal causal estimands become treatment-response functions within strata defined by treatment-mediator trajectories, complicating the definition of associative, dissociative, and other scientifically meaningful effects. Even when meaningful estimands can be defined, identification remains challenging, as assumptions such as monotonicity and exclusion restrictions do not yield nonparametric identification of PCEs.

We develop a framework for principal stratification with continuous treatments, introducing estimands that capture associative, dissociative, and other principal causal effects to investigate the mediating role of a continuous mediator. While we focus on the specific scenario of continuous treatments and continuous mediators, these ideas apply to any situation with a continuous treatment, regardless of the type of mediator. We establish nonparametric identification under principal ignorability, which requires knowledge of the typically unknown joint distribution of potential mediators, and show that identification without principal ignorability is possible under certain parametric constraints. Importantly, however, we show that in this situation, one can also learn the joint distribution of the potential mediators by incorporating information from an outcome model. We provide theoretical results around identification, and the ability to estimate association parameters between potential mediators at different treatment levels. 
We also develop a Bayesian approach to estimation, which is particularly well-suited to the substantial missing-data problem arising with continuous treatments and mediators \citep{li2023bayesian}. This framework enables imputation of unobserved trajectories while jointly accounting for uncertainty in latent principal strata, model parameters, and causal estimands, without relying on large-sample approximations. We employ flexible Bayesian nonparametric models where possible to mitigate concerns about model misspecification. Lastly, we study an important problem in criminology, which is how the economic prosperity of a region affects arrest rates in that area, and whether this effect is channeled through changes in police capacity.
\section{Principal stratification framework with continuous treatments and intermediate variables}\label{sec:ps}

Throughout, we observe $n$ independent observations $\boldsymbol{D}_i=(Y_i,S_i,T_i,\boldsymbol{X}_i)$, $i=1,\dots,n$, where $Y_i$ is the outcome, $S_i$ the mediator, $T_i$ the treatment, and $\boldsymbol{X}_i$ a $p$-dimensional vector of pre-treatment covariates. We focus on continuous treatments and mediators with supports $\mathcal{T}$ and $\mathcal{S}$, respectively. Let $S_i(t)$ and $Y_i(t)$ denote the potential mediator and outcome under treatment level $t\in\mathcal{T}$. We assume the Stable Unit Treatment Value Assumption \cite[SUTVA,][]{rubin1980randomization}, which rules out interference between units and multiple versions of treatment, implying $S_i=S_i(T_i)$ and $Y_i=Y_i(T_i)$. We define $\boldsymbol{S}_i=\{S_i(t) \}_{t \in \mathcal{T}}$ as the entire treatment-mediator trajectory for unit $i$, distinguishing it from the single, observed mediator $S_i$. We omit the unit $i$ subscript when unnecessary.

\subsection{Principal Stratification and Causal Estimands}

Information on the overall effect of treatment $T$ on outcome $Y$ can be obtained from the average dose-response function, $\mu(t)=E[Y(t)]$, for $t\in\mathcal{T}$. Our interest, however, lies in understanding how treatment effects vary with the potential mediator values, for which we use principal stratification \cite[][]{frangakis2002principal}. Principal stratification partitions units into groups, called principal strata, defined by their joint potential mediator values. With continuous treatments, a principal stratum consists of units sharing the same treatment-mediator trajectory, ${i:\boldsymbol{S}_i=\boldsymbol{s}}$, for some $\boldsymbol{s}$. These trajectories describe how the mediator responds to changes in treatment, allowing us to investigate treatment effects among units with different mediator responses. The causal estimands of interest are principal causal effects (PCEs), which compare potential outcomes within a particular principal stratum or union of strata. PCEs therefore characterize treatment effect heterogeneity across subpopulations defined by how their mediators respond to treatment. In our setting, a natural estimand is the principal stratum-specific exposure-response curve,
$$
\mu(t,\boldsymbol{s})=E[Y(t)\mid\boldsymbol{S}=\boldsymbol{s}], \qquad t\in\mathcal{T}.
$$

With continuous treatments and mediators, there are infinitely many principal strata, making it useful to consider unions of strata defined by features of the mediator trajectories. Of particular interest are dissociative effects, corresponding to approximately constant mediator trajectories, and associative effects, corresponding to trajectories that vary with treatment \cite[e.g.,][]{kim2019bayesian}. We therefore consider estimands of the form
\begin{align}
E[Y(t) \mid a < g(\boldsymbol{S}) < b], \label{eqn:EstimandDef}
\end{align}
as many principal strata of interest can be expressed as the set of individuals for whom $a < g(\boldsymbol{S}) < b$, for suitable functions $g(\cdot)$. Many features of the potential intermediate variables, such as the shape, slope, range, or more complex estimands can be written in this format, and the choice of function depends on the empirical application and what scientific questions are being addressed. If interest lies in associative and dissociative effects, then one could look at principal strata defined by $g(\boldsymbol{S}) \equiv \text{range}(\boldsymbol{S}) = \max_{t \in \mathcal{T}}S(t) - \min_{t \in \mathcal{T}} S(t)$. One could focus on effects of the form $E[Y(t) \mid \text{range}(\boldsymbol{S}) \leq \epsilon] \ \forall \ t,$  which looks at the exposure-response curve among units for which there is little to no effect of $T$ on $\boldsymbol{S}$.  This would be analogous to a direct effect in this subpopulation whose value of $\boldsymbol{S}$ is unaffected by treatment (dissociative PCE).  Alternatively, one can examine $E[Y(t) \mid \text{range}(\boldsymbol{S}) \geq \delta]  \ \forall \ t$, to see the exposure-response curve among units for which there is a large effect of $T$ on $\boldsymbol{S}$ (associative PCE).

\subsection{Assumptions and identification}
Inference on causal effects typically first requires assumptions on the treatment assignment mechanism. We first assume \textit{strong unconfoundedness}:
\begin{assumption}\label{ass:s_unconfoundedness} (Strong unconfoundedness). The assignment mechanism is strongly unconfounded, given pre-treatment variables  $\boldsymbol{X}$, if
       $$\{S(t), Y(t)\}_{t \in \mathcal{T}} \independent T \mid \boldsymbol{X}.$$
\end{assumption}
We also require that the treatment assignment mechanism is probabilistic:
\begin{assumption}\label{ass:overlap} (Overlap). For all possible values of the pre-treatment variables $\boldsymbol{X}$, the conditional probability density function of receiving any possible
treatment level  $t \in \mathcal{T}$ given $\boldsymbol{X}$ is positive.
\end{assumption}

Unlike other identifying assumptions, potential violations of overlap can be investigated empirically using diagnostics based on the generalized propensity score \citep{hirano2004propensity}, which we recommend before proceeding with analysis. When overlap is limited, trimming observations in regions of poor support can improve estimation by reducing model extrapolation, although this changes the target estimand \citep{crump2009dealing,branson2023causal}. We describe overlap diagnostics and trimming in Appendix E. Importantly, our assumption requires positivity of the treatment assignment mechanism but not of the mediator, unlike other approaches to mediation analysis \citep{nguyen2022clarifying}. Strong unconfoundedness holds by design in randomized experiments but requires the absence of unmeasured treatment-mediator and treatment-outcome confounding in observational studies. Its plausibility therefore depends on the observed covariates and subject-matter knowledge.

Strong unconfoundedness implies that the distribution of principal strata trajectories, $\boldsymbol{S}$, is independent of treatment conditional on $\boldsymbol{X}$. Moreover, it implies that $\{Y(t)\}_{t \in \mathcal{T}} \independent T \mid \{S(t)\}_{t \in \mathcal{T}}, \boldsymbol{X}$, allowing comparisons of potential outcomes across treatment levels among units in the same principal stratum with the same covariates. However, strong unconfoundedness and overlap alone are insufficient for nonparametric identification of PCEs. We show in Appendix A that under these assumptions, $E[Y(t)\mid a<g(\boldsymbol{S})<b]$ can be expressed as
{\small
\begin{align}
\int_{\boldsymbol{x}} \int_{\boldsymbol{s}} E[Y   \mid   T=t, \boldsymbol{S}=\boldsymbol{s}, \boldsymbol{X} = \boldsymbol{x}]  \dfrac{f(\boldsymbol{s}\mid \boldsymbol{X} = \boldsymbol{x}) 1(a < g(\boldsymbol{s}) <b)}{P(a < g(\boldsymbol{S}) <b\mid \boldsymbol{X} = \boldsymbol{x})} \,d \boldsymbol{s}\, dF_{\boldsymbol{X}\mid a < g(\boldsymbol{S}) < b}(\boldsymbol{x}), \label{eq:estimand}
\end{align}}where $f(\boldsymbol{s}\mid \boldsymbol{X} = \boldsymbol{x})$ is the conditional density function of the potential mediators. If we could observe the principal stratum $\boldsymbol{S}$ for each unit, we could identify both $ E[Y \mid T=t, \boldsymbol{S} = \boldsymbol{s}, \boldsymbol{X} = \boldsymbol{x}]$ and $f(\boldsymbol{s}\mid \boldsymbol{X} = \boldsymbol{x})$, which would imply nonparametric identification of the principal causal effect. However, this quantity is only partially observed as we only observe one $S$ associated with the observed value of the treatment $T$, and therefore we need to impose further assumptions to draw inference on principal strata exposure-response curves, as is frequently done in the principal stratification literature. In the standard setting with binary treatments and binary intermediates, common assumptions such as monotonicity or exclusion restriction assumptions are made to obtain nonparametric identification \cite[e.g.,][] 
{imbens1994identification,angrist1996identification,imbens1997bayesian}. 
In the more complex setting of binary treatments with continuous mediators, these are not sufficient for identification as there are infinitely many principal strata. Moreover, exclusion restriction assumptions may be difficult to justify when the focus is on understanding the mechanisms through which a treatment affects an outcome because they rule out a priori effects of interest, such as direct effects \cite[e.g.][]{mealli2012refreshing}.
Papers focusing on nonparametric identification in this setting typically make two assumptions: 1) a type of principal ignorability that requires conditional (mean) independence between (some) principal strata and (some) potential outcomes for the primary outcome given covariates, and 2) an assumption on the joint distribution of the two potential intermediates \citep{lu2023principal, zhang2024semiparametric}. The second of these two assumptions commonly assumes that the joint distribution of the two potential intermediates is governed by a \textit{known} copula with a \textit{known} correlation parameter. This second assumption is a very strong parametric assumption, and therefore inference is usually made under a range of correlation parameters to see if results are sensitive to this choice. 

Extensions of these two assumptions can be used to obtain identification in our setting, which we detail in the following section. We will see, however, that there are crucial differences when working with continuous treatments. In particular, if principal ignorability does not hold, then there are additional avenues to obtain identification. If one is willing to make certain parametric assumptions on the conditional mean of the observed outcome, then we are able to identify the correlation between different potential intermediate values, though only when principal ignorability is violated. Related strategies using an outcome model to identify the correlation between potential intermediates have been used heuristically in settings with binary treatments \citep{bartolucci2011modeling,schwartz2011bayesian}. Recent theoretical results, however, have shown this strategy only works in the binary treatment setting when very restrictive parametric assumptions are made \citep{wu2024partial}. We provide theoretical support for this strategy in our setting in Section \ref{sec:OutcomeCorrelation} and show that identification relies on weaker parametric assumptions when using continuous treatments.

\section{Inference under principal ignorability}

In this section, we explore identification of the estimand defined in \eqref{eqn:EstimandDef} under a principal ignorability assumption, which is defined in the following:
\begin{assumption}\label{ass:PIcontinuous} (Principal ignorability). For all values $t \in \mathcal{T}$ , we have that
$$E(Y(t) \mid \boldsymbol{S} = \boldsymbol{s}, \boldsymbol{X} = \boldsymbol{x}) = E(Y(t) \mid S(t) = s, \boldsymbol{X} = \boldsymbol{x})$$
\end{assumption}

This assumption states that the mean of the potential outcome at treatment level $t$ depends on the potential intermediate at the same treatment level, $S(t)$, rather than on the entire intermediate curve given by $\boldsymbol{S}$. The distinction is important for identification. For units receiving treatment level $t$, $S(t)$ is observed, whereas the remaining potential intermediate values $S(t')$, $t'\neq t$, are unobserved. Under strong unconfoundedness, principal ignorability therefore implies that
\[
E(Y \mid T=t,\boldsymbol{S}=\boldsymbol{s},X=x) = E(Y \mid T=t,S(t)=s,X=x),
\]
so the relevant outcome regression depends only on observed quantities and can be estimated flexibly, including nonparametrically. In contrast, if the outcome model depends on the entire intermediate curve $\boldsymbol{S}$, including unobserved potential intermediate values, it cannot, in general, be identified nonparametrically\footnote{It is useful to contrast principal ignorability with the sequential ignorability assumptions commonly invoked in modern mediation analysis (e.g., Imai et al., 2010). Modern mediation analysis typically defines causal effects in terms of nested potential outcomes, such as $Y(t,s)$ and $Y(t,S(t'))$, corresponding to hypothetical interventions on the mediator. Sequential ignorability provides conditions under which causal effects defined in terms of these nested potential outcomes can be identified from the observed-data distribution. Principal stratification takes a different approach: it does not conceptualize the mediator as an intervention and does not invoke nested potential outcomes. Instead, principal ignorability restricts the association between Y(t) and the latent principal-stratum trajectory by assuming that, conditional on $S(t)$ and observed covariates, the mean of $Y(t)$ does not depend further on the remaining potential mediator values, $S(t')$, $t' \neq t'$.}.
This assumption rules out unmeasured factors that jointly influence the potential outcome and the potential intermediate trajectory, after accounting for the observed covariates. Therefore, principal ignorability is violated whenever such unmeasured common causes exist. We provide a simple, illustrative example of such a setting that leads to failure of principal ignorability in Appendix F.
Principal Ignorability alone does not provide identification of \eqref{eqn:EstimandDef}, however, and we must make an assumption on the joint distribution of the potential intermediates. 
\begin{assumption}\label{ass:IntermediateDist} The joint distribution of the potential intermediates is given by
$$\boldsymbol{S} \mid \boldsymbol{X} = \boldsymbol{x} \sim \mathcal{GP}(m(t, \boldsymbol{x}), K(t, t')).$$
with known kernel function $K(\cdot, \cdot)$.
\end{assumption}
Arguably, the strongest part of this assumption is that the kernel function, which includes its functional form and any correlation parameters dictating the joint distribution between potential intermediates, is known. This result establishes that, under principal ignorability, the dependence structure among the potential mediator values must be treated as a sensitivity parameter to achieve nonparametric identification of principal causal effects when both the treatment and the mediator are continuous. Analogous results were obtained by \cite{lu2023principal} for continuous intermediates and binary treatments. Our work differs from this existing work by examining continuous treatments and the corresponding estimands defined in Section \ref{sec:ps}. Additionally, we show in subsequent sections that the continuous treatment setting allows for other possibilities for identification that relax this assumption. 

In practice, when doing inference we discretize $\mathcal{T}$ to a grid of values so that this assumption assumes that the joint distribution of the potential intermediates is a multivariate normal distribution with known correlation parameters. We discuss sensitivity to the size of the grid in Appendix H and find that results are robust to this choice as long as the grid is dense enough to cover all reasonable values of the treatment. We can estimate the marginal variances of the potential intermediates under Assumption \ref{ass:s_unconfoundedness}, but the correlation parameters are not identified and must be assumed known for now. Knowledge of this joint distribution is not plausible in most settings. At best, one might be willing to assume that the correlation between $S(t)$ and $S(t')$ is positive and is a decreasing function of $|t - t'|$, but knowing the degree of this correlation or how it decays with distance is not likely. For this reason, if one invokes Assumption \ref{ass:IntermediateDist}, it should be coupled with sensitivity analyses assessing the robustness of results to different functional forms or magnitudes of correlation. We perform these sensitivity analyses for our policing analysis of Section \ref{sec:application} in Appendix H and find that results are relatively robust across a range of reasonable specifications. As an alternative, we show in subsequent sections that these crucial correlation parameters can actually be estimated from the data, but only if principal ignorability fails to hold and certain parametric assumptions are made. One might also be inclined to consider alternative, weaker assumptions about the distribution of $\boldsymbol{S} \mid \boldsymbol{X} = \boldsymbol{x}$, such as monotonicity with respect to $t$ or other shape constraints. These do not lead to nonparametric identifiction in this setting, however, so we do not expand upon them here. Such shape constraints can still be useful from a finite-sample estimation perspective, and we detail one such approach in Appendix D. Now we present our formal identification result under the aforementioned assumptions. 
\begin{theorem} \label{thm:PIidentificationFormula}
    Under Assumptions \ref{ass:s_unconfoundedness} - \ref{ass:IntermediateDist}, the estimand $E[Y(t) \mid a < g(\boldsymbol{S}) < b]$ is identified as
{\small
\begin{align}
\int_{\boldsymbol{x}} \int_{\boldsymbol{s}} E[Y \mid T=t, S(t)=s, \boldsymbol{X} = \boldsymbol{x}]  \dfrac{f(\boldsymbol{s}\mid \boldsymbol{X} = \boldsymbol{x}) 1(a < g(\boldsymbol{s}) <b)}{P(a < g(\boldsymbol{S}) <b\mid \boldsymbol{X} = \boldsymbol{x})} \,d \boldsymbol{s}\, dF_{\boldsymbol{X}\mid a < g(\boldsymbol{S}) < b}(\boldsymbol{x}).
\label{eqn:EstimandPI}
\end{align}}
\end{theorem}
A proof of this result can be found in Appendix A. The quantity $E[Y \mid T=t, S(t)=s, \boldsymbol{X} = \boldsymbol{x}]$ is identifiable as it depends only on observable variables. The marginal mean and variance of the intermediates given covariates are identified by Assumption \ref{ass:s_unconfoundedness}, and Assumption \ref{ass:IntermediateDist} ensures that the full joint distribution $f(\boldsymbol{s}\mid \boldsymbol{X} = \boldsymbol{x})$ is identifiable. Inference can proceed by fitting models for both the outcome and intermediate distributions above. We describe this strategy and our choice of models in Section \ref{sec:Bayesian}, though we first discuss identification when principal ignorability does not hold as this provides novel insights on identification of principal causal effects.

\section{Identification when principal ignorability is violated}
\label{sec:IdentificationMain}

While Assumption \ref{ass:PIcontinuous} is plausible in certain settings, it is still an unverifiable assumption that can be violated. For instance, if there are unmeasured confounders of the $S \rightarrow Y$ relationship, this assumption would be violated, and any inferences assuming principal ignorability would be invalid. Here we show that this assumption can indeed be dropped, but generally this only leads to partial identification of causal effects. However, under certain parametric assumptions on the outcome model, not only can we drop Assumption \ref{ass:PIcontinuous} and obtain point identification, but additionally we are able to weaken Assumption \ref{ass:IntermediateDist} as the correlation parameters become identifiable in this setting. It is important to emphasize, however, that while we are able to remove or weaken Assumptions \ref{ass:PIcontinuous} and \ref{ass:IntermediateDist}, this comes at the expense of additional parametric assumptions on the outcome model, which we detail in the following sections. Given that Assumption \ref{ass:IntermediateDist} is itself a strong parametric assumption on the intermediate variable, one can view this alternative strategy as instead placing assumptions on the outcome process. In practice, one should reason through each  of these assumptions and the trade-offs they provide for identification before proceeding with analysis. One can also proceed under both strategies, and if there are conflicting results between the two approaches, that would provide some evidence that at least one identifying or parametric assumption is invalid. Throughout this section, we assume the potential outcomes follow 
$$Y(t) = f(t, \boldsymbol{s}, \boldsymbol{x}) + \epsilon(t),$$
where $E[Y(t) \mid \boldsymbol{S}=\boldsymbol{s}, \boldsymbol{X}=\boldsymbol{x}] = E[Y \mid T=t, \boldsymbol{S} = \boldsymbol{s}, \boldsymbol{X} = \boldsymbol{x}] = f(t, \boldsymbol{s}, \boldsymbol{x})$, and $\epsilon(t)$ denotes the random error of $Y(t)$. When principal ignorability is not assumed, then inference on $E[Y(t) \mid a < g(\boldsymbol{S}) < b]$ crucially depends on being able to identify the full outcome model given by $f(t, \boldsymbol{s}, \boldsymbol{x})$, which we discuss in subsequent sections. 

\subsection{Partial identification with a nonparameteric potential outcome model}
\label{sec:partial_identificationa_nonpara}

In this section, we show that nonparametric identification of $E[Y(t) \mid \boldsymbol{S}=\boldsymbol{s}, \boldsymbol{X}=\boldsymbol{x}]$ is in general not possible due to the incomplete information on $\boldsymbol{S}$. Partial identification, meaning that a set of values of the parameters (and therefore causal effects) are equally supported by the data, is possible in this setting. We formalize this with Theorem \ref{thm:partial_identification}, which we prove in Appendix A. 
\begin{theorem} \label{thm:partial_identification}
    If there exists an invertible function $M_{t_0, \boldsymbol{x}}$ such that $M_{t_0, \boldsymbol{x}}(\boldsymbol{S})\mid S(t_0) = s, \boldsymbol{X} = \boldsymbol{x}$ and $\boldsymbol{S}\mid S(t_0) = s, \boldsymbol{X} = \boldsymbol{x}$ have the same distribution for some $t_0 \in \mathcal{T}$, then for $Y'(t_0) = f(t_0, M_{t_0, \boldsymbol{x}}^{-1}(\boldsymbol{s}), \boldsymbol{x}) + \epsilon(t_0)$, marginally, $(Y'(t_0), S(t_0))|\boldsymbol{X} = \boldsymbol{x}$ and $(Y(t_0), S(t_0))\mid \boldsymbol{X} = \boldsymbol{x}$ follow the same distribution.
\end{theorem}

\begin{figure}[htbp]
    \centering
    \includegraphics[width=0.85\linewidth]{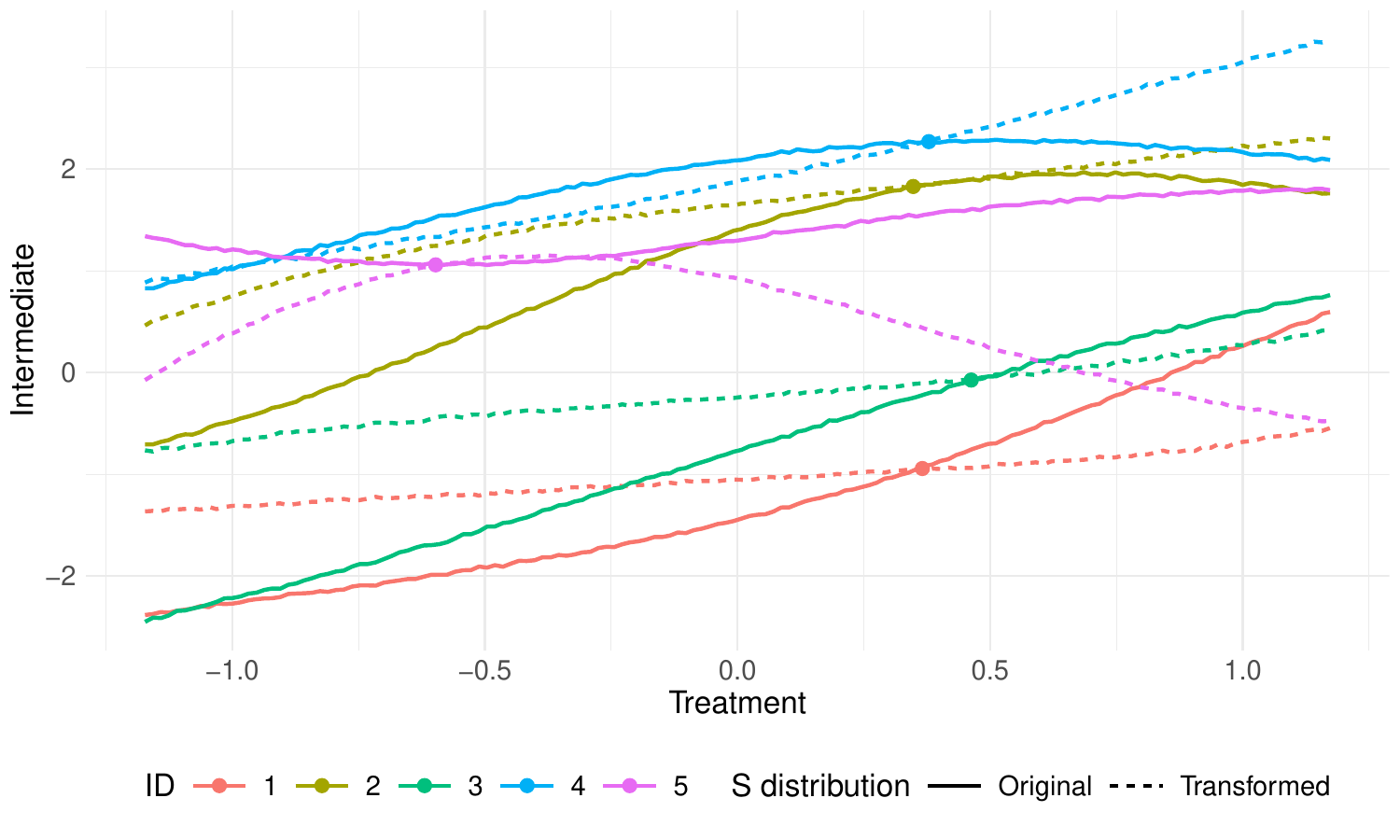}
    \caption{Values of $\boldsymbol{S}$ and $M_{t_0, \boldsymbol{x}}(\boldsymbol{S})$ for five observations from a simulated data set. The solid lines correspond to the original trajectories, while the dashed lines correspond to the transformed trajectories. }
    \label{fig:AlternateTrajectories}
\end{figure}

Theorem \ref{thm:partial_identification} shows that if there exists a function $M_{t_0, \boldsymbol{x}}$ such that $M_{t_0, \boldsymbol{x}}(\boldsymbol{S})$ and $\boldsymbol{S}$ have the same conditional distribution, these can lead to different potential outcomes, but the observed data distribution will be the same. Because they lead to the same observed data distribution, yet lead to different causal effects, the causal effects are partially identified in this setting. As an illustration of this, assume $\boldsymbol{S}\mid \boldsymbol{X}$ follows a Gaussian process. Let $m_{t_0, \boldsymbol{x}}(t, \boldsymbol{x})$ denote the conditional mean function of $S(t)|S(t_0), \boldsymbol{X}$. Then, an invertible function can be defined as $M_{t_0, \boldsymbol{x}}: S(t) \to 2m_{t_0, \boldsymbol{x}}(t, \boldsymbol{x}) - S(t)$ for any $t_0 \in \mathcal{T}$. It is straightforward to verify in this setting that both $M_{t_0, \boldsymbol{x}}(\boldsymbol{S})\mid S(t_0), \boldsymbol{X}$ and $\boldsymbol{S}\mid S(t_0), \boldsymbol{X}$ have the same distribution. They will lead to very different causal effects, however, as the sign of $\boldsymbol{S}$ has been flipped. One can see an illustration of this using simulated data in Figure \ref{fig:AlternateTrajectories} where we show the true intermediate trajectory $\boldsymbol{S}$ as well as $M_{t_0, \boldsymbol{x}}(\boldsymbol{S})$ for five randomly chosen observations.
One can see that the two trajectories for the same observation can differ substantially, and would lead to very different causal effects if principal ignorability does not hold. Note that we have constructed and illustrated one such invertible function here, but there are infinitely many such functions, and these together form the partial identification region for the causal effect. Generally, however, it is very difficult to derive this partial identification region as many of the functions $M_{t_0, \boldsymbol{x}}$ are overly complex, or even discontinuous, functions of $\boldsymbol{S}$, which lead to very complex potential outcome models for $Y'(t)$. One could attempt to rectify the situation using shape constraints, such as smoothness or monotonicity, on the treatment-intermediate trajectories. One can see in Figure \ref{fig:AlternateTrajectories}, however, that such restrictions would generally not lead to point identification as there are situations where both $\boldsymbol{S}$ and $M_{t_0, \boldsymbol{x}}(\boldsymbol{S})$ are smooth and monotonic. Due to the aforementioned issues with identification, we now focus on parametric settings, which preclude such overly complex models, and can lead to identification of principal causal effects. 

\subsection{Identification with a parametric potential outcome model}
\label{sec:identification_para}

In this section, we detail how imposing certain parametric constraints, even relatively flexible ones, can lead to identification of all model parameters and therefore the causal effect of interest. As we will see in subsequent sections, this has an additional benefit beyond identification of the outcome model. In these scenarios, it can additionally lead to identification of the correlation parameter governing the kernel function of the Gaussian process for $\boldsymbol{S} \mid \boldsymbol{X}$. Let us consider a linear functional for $\boldsymbol{s}$ in $f(t,\boldsymbol{s},\boldsymbol{x})$ as follows:
\begin{equation*}
f(t,\boldsymbol{s},\boldsymbol{x})=\lambda(t,\boldsymbol{x})+\int_{t'\in\mathcal{T}} \widetilde{\beta}(t,t')s(t')dt'.
\end{equation*}
In practice, we discretize $\boldsymbol{s}$ to a grid containing a finite set of values and obtain
\begin{equation*}
    f(t,\boldsymbol{s},\boldsymbol{x}) = \lambda(t,\boldsymbol{x}) + \sum_{t'\in\mathcal{T}_d} \widetilde{\beta}(t,t')s(t'),
\end{equation*}
where $\mathcal{T}_d$ denotes the collection of discretization points. For the sake of simplicity in notation, we continue to use $\widetilde{\beta}(t,t')$, but it differs from the one with a functional $\boldsymbol{s}$. Without loss of generality, we can re-write this function in the following manner: 
\begin{equation}
    f(t,\boldsymbol{s},\boldsymbol{x}) = \lambda(t,\boldsymbol{x}) + \psi(t)s(t)+\sum_{t'\in\mathcal{T}_d} \beta(t,t')s(t'), \label{eqn:ParametricModelSpec}
\end{equation}
where for identifiability, we enforce that $\beta(t, t) = 0$ for any $t \in \mathcal{T}$. This allows for a more explicit characterization of the principal ignorability assumption within the model, as principal ignorability would impose that $\beta(t, t') = 0$ for any $t,t' \in \mathcal{T}$. Lastly, we assume that $\lambda(t,\boldsymbol{x})$, $\psi(t)$ and $\beta(t,t')$ belong to some finite-dimensional function spaces spanned by basis functions $\boldsymbol{b}_{\lambda}$, $\boldsymbol{b}_{\psi}$, and $\boldsymbol{b}_{\beta}$, respectively. Let $\boldsymbol{\zeta}_{\lambda}$, $\boldsymbol{\zeta}_{\psi}$, and $\boldsymbol{\zeta}_{\beta}$ denote the coordinates of $\lambda(t,\boldsymbol{x})$, $\psi(t)$, and $\beta(t,t')$ with respect to $\boldsymbol{b}_{\lambda}$, $\boldsymbol{b}_{\psi}$, and $\boldsymbol{b}_{\beta}$, respectively. In Appendix A, we prove the following result regarding identifiability.
\begin{theorem}
(Identification of a parametric outcome model)\label{thm:identification_para}
    Suppose that Assumptions \ref{ass:s_unconfoundedness}, \ref{ass:overlap}, and \ref{ass:IntermediateDist} hold. Then, under some mild regularity conditions described in Appendix A, we have that $\boldsymbol{\zeta}_{\lambda}$, $\boldsymbol{\zeta}_{\psi}$, and $\boldsymbol{\zeta}_{\beta}$ are identifiable.
\end{theorem} 

Theorem \ref{thm:identification_para} shows that if we consider a flexible parametric outcome model instead of a nonparametric one, then $f(t, \boldsymbol{s}, \boldsymbol{x})$ is indeed identifiable. Note that while Theorem \ref{thm:identification_para} only shows identification for the parameters of the outcome model, this in turn leads to identification of $E[Y(t) \mid a < g(\boldsymbol{S}) < b]$, though we require the model in \eqref{eqn:ParametricModelSpec} to be correctly specified in order to obtain valid inference on principal causal effects. 
Additionally, the proof of Theorem \ref{thm:identification_para} reveals the possibility of identifying the covariance function of the Gaussian process. One can show that if $K(\cdot, \cdot)$ has a parametric form and certain mild conditions are assumed, then identification for $f(t, \boldsymbol{s}, \boldsymbol{x})$ still holds, and identification for the parameters in $K(\cdot, \cdot)$ also holds. In other words, one can replace the condition that $K(\cdot, \cdot)$ is known with the condition that $K(\cdot, \cdot)$ has a parametric form. This is a hugely important result, given that most analyses rely on the overly strong assumption that this joint distribution is known, but under certain parametric assumptions, we are able to estimate this crucial quantity. Note also that this differs substantially from the binary treatment setting, where neither the outcome model parameters or the association parameters of the post-treatment variable are identifiable under the same parametric assumptions \citep{wu2024partial}. 

The parametric model in \eqref{eqn:ParametricModelSpec} posits an additive structure that assumes both 1) the effect of $s(t)$ on the outcome does not depend on the treatment or covariates, and 2) the effect of the full trajectory $\boldsymbol{S}$ on the outcome is linear and does not depend on the other observed variables. The first of these two assumptions is made for ease of exposition, and it is straightforward to include $s(t)$ inside of the $\lambda(\cdot)$ function to allow interactions between $s(t)$, $t$, and $\boldsymbol{x}$ as long as they can be modeled with a finite-dimensional basis expansion. The second assumption is a more subtle issue that requires additional discussion. Theoretically, the result in Theorem \ref{thm:identification_para} would continue to hold even if the effects of $\boldsymbol{S}$ were nonlinear or interacted with other variables, as long as they could be modeled with a finite-dimensional basis expansion. In practice, however, there are other issues to consider which lead us to recommend the additive structure above. First, we argue that such a linear and additive structure may be a reasonable approximation to the truth in many settings, and we detail an example in Appendix F showing how this structure arises from an unmeasured confounder violating principal ignorability under a commonly assumed unmeasured confounding model. Another key practical issue is that Markov chain Monte Carlo (MCMC) would become significantly more challenging if the linearity assumption were dropped, as it leads to closed-form Gibbs sampling updates for imputing the missing $\boldsymbol{S}_i$ for all $i$. More importantly, however, we have found there to be a trade-off between identification and robustness to model misspecification when deciding how to model the effect of $\boldsymbol{S}$. Increasing the number of parameters and model complexity corresponding to the effect of $\boldsymbol{S}$ would mitigate issues from model misspecification, but we have found these parameters to be only weakly identified as the number of parameters grows. Allowing for interactions or nonlinearities for the effect of $\boldsymbol{S}$ would only exacerbate such issues, and would make inference challenging.  We expand on the issue of weak identification in the next section, which also has connections to the results of the previous section on partial identification, as the cause of partial identification can also lead to weak identification in parametric settings. Lastly, we provide additional simulation results in misspecified settings where the linearity assumption no longer holds in Appendix G to assess the robustness of causal conclusions to this assumption.

\subsection{Weak identification and multi-modality}
\label{ssec:WeakIdentification}

While the previous result suggests that the outcome model becomes identifiable once a finite-dimensional parametric model is utilized for estimation, one still may face issues in small sample sizes as very different outcome models can lead to similar observed data distributions. This is effectively an issue of multi-modality of the likelihood function, as very different model parameters lead to similar likelihoods, and hence posterior distributions when doing Bayesian inference. For the remainder of the paper, we focus specifically on Bayesian inference for these parameters, as this is often adopted in PS analysis where inference involves techniques for incomplete data 
\cite[e.g.][]{jin2008principal, schwartz2011bayesian, kim2019bayesian, mattei2020assessing}.

To illustrate the issue of multi-modality, we utilize the same example as in Section \ref{sec:partial_identificationa_nonpara}, which led to the sign of $\boldsymbol{S}$ flipping.  While in principle there are many other possibilities that lead to weak identifiability or multi-modality, this is the one we have most often seen empirically in simulations, and therefore we focus attention to it. This also highlights how the issues that led to a lack of identification in Section \ref{sec:partial_identificationa_nonpara} also can lead to weak identification in parametric settings. For simplicity of illustration, assume the true outcome model is given by an additive function of the form
\begin{align*}
    f(t,\boldsymbol{s}, \boldsymbol{x}) &= \lambda(t) + \boldsymbol{x} \boldsymbol{\gamma} + \psi(t)s(t) +\sum_{t'\in \mathcal{T}_{d}} \beta(t,t')s(t') = \lambda(t) + \boldsymbol{x} \boldsymbol{\gamma} + \psi(t)s(t) + \boldsymbol{s} \boldsymbol{\beta}^\ast
\end{align*}
We again can define an invertible function as $M_{t_0, \boldsymbol{x}}: S(t) \to 2m_{t_0, \boldsymbol{x}}(t, \boldsymbol{x}) - S(t)$. Plugging $M_{t_0, \boldsymbol{x}}(\boldsymbol{s})$ into this equation in place of $\boldsymbol{s}$, we obtain
\begin{align*}
    \lambda(t) + \boldsymbol{x} \boldsymbol{\gamma} + \psi(t)s(t) + (2 \boldsymbol{m}_{t_0, \boldsymbol{x}}(\boldsymbol{t}, \boldsymbol{x}) -\boldsymbol{s}) \boldsymbol{\beta}^\ast = \lambda'(t, \boldsymbol{x}) + \psi'(t) s(t) - \boldsymbol{s} \boldsymbol{\beta}^{\ast},
\end{align*}
where $\lambda'(t, \boldsymbol{x}) + \psi'(t) s(t)=
\lambda(t) + \boldsymbol{x} \boldsymbol{\gamma}+2 \boldsymbol{m}_{t_0, \boldsymbol{x}}(\boldsymbol{t}, \boldsymbol{x})\boldsymbol{\beta}^\ast + \psi(t)s(t)$.
Note that the coefficient for $\boldsymbol{s}$ in this model is now $-\boldsymbol{\beta}^{\ast}$ instead of $\boldsymbol{\beta}^{\ast}$, showing that the effect of $\boldsymbol{s}$ has flipped. If one were to specify an additive model, the data would correctly favor the true model which has a positive sign for the coefficient, but in small sample sizes the approximation using the flipped sign may be good enough to produce likelihoods similar to the true parameters. In large samples, the likelihood should be much higher at the true values of the parameter, but issues can still occur due to mixing issues with MCMC algorithms getting stuck in suboptimal local modes. 

To address this issue, we propose to estimate the model on a large number of MCMC chains that have different starting values. We then aggregate results across all MCMC chains, but instead of the traditional approach of assigning each MCMC chain equal weight, we assign each MCMC chain a weight that is proportional to its marginal likelihood. Specifically, if we have $M$ MCMC chains, then our final posterior distribution for the entire vector of parameters $\boldsymbol{\theta}$ is given by
\begin{align*}
    P(\boldsymbol{\theta} \vert \mathcal{D}) = \sum_{m=1}^M \pi_m P_{m}(\boldsymbol{\theta} \vert \mathcal{D}),
\end{align*}
where $\pi_m$ is the weight assigned to each MCMC chain, and $P_{m}(\boldsymbol{\theta} \vert \mathcal{D})$ is the posterior distribution from each MCMC chain. To ensure that we assign weight to each chain that is proportional to how much they are favored by the data, we set
$$\pi_m = \frac{L(\mathcal{D} \vert E_m(\boldsymbol{\theta} \vert \mathcal{D}))}{\sum_{k=1}^M L(\mathcal{D} \vert E_k(\boldsymbol{\theta} \vert \mathcal{D}))},$$
where $L(\mathcal{D} \vert \boldsymbol{\theta})$ is the marginal likelihood of the data evaluated at $\boldsymbol{\theta}$ and $E_k$ represents the posterior mean of the $k$th chain $(k=1,2,…,M)$. Assuming that there is a nonzero probability of our MCMC chains falling into the mode that corresponds to the true parameter vector, this construction of the posterior should converge to the true posterior, as $\pi_m \rightarrow 0$ for any chains $m$ that fall in suboptimal modes. We explore this approach in simulation in Section \ref{sec:SimulationStudies} and find that it leads to improved performance relative to standard strategies that assign equal weight to each MCMC chain.

\section{Nonparametric Bayesian modeling framework}\label{sec:Bayesian}

As seen in Equation \eqref{eq:estimand}, under Assumptions 1 and 2, our estimand of interest can be written as a function of two quantities: $E[Y \mid T=t, \boldsymbol{S}=\boldsymbol{s}, \boldsymbol{X} = \boldsymbol{x}]$ and $f(\boldsymbol{s}\mid \boldsymbol{X} = \boldsymbol{x})$. In this section we describe a joint Bayesian model for these two quantities, and derive the posterior distributions of the causal estimand of interest using an MCMC algorithm with data augmentation. We first describe modeling of $f(\boldsymbol{s}\mid \boldsymbol{X} = \boldsymbol{x})$, which is the same regardless as to whether principal ignorability (Assumption \ref{ass:PIcontinuous}) holds. We then detail modeling of $E[Y \mid T=t, \boldsymbol{S}=\boldsymbol{s}, \boldsymbol{X} = \boldsymbol{x}]$ separately based on whether principal ignorability holds or not. Estimation is much simpler under principal ignorability where nonparametric Bayesian methods can be used and point identification can still be obtained. Without principal ignorability, we propose flexible parametric specifications that require data augmentation to impute the missing values of the potential intermediate variables at each step of the MCMC algorithm, and all parameters are updated conditionally on these values. Critically, these missing values are updated using information from the outcome model as well, which we expand upon in Section \ref{sec:OutcomeCorrelation}. 

\subsection{Nonparametric Bayesian modeling for potential intermediate}
\label{sec:Smodel}

We first need to posit a model for the joint distribution of the potential intermediate values, conditional on the observed covariates, i.e. $\boldsymbol{S} \mid \boldsymbol{X}$. 
We want to estimate this model as flexibly as possible so that we do not misspecify the distribution of the potential intermediate values, which would lead to incorrect imputations of the missing potential intermediate values and estimates of principal causal effects. 
As in Assumption \ref{ass:IntermediateDist}, we assume that the set of potential intermediate values follows a Gaussian process defined by
$$\boldsymbol{S} \mid \boldsymbol{X} = \boldsymbol{x} \sim \mathcal{GP}(m(t, \boldsymbol{x}), K(t, t')).$$
The mean function $m(t, \boldsymbol{x})$ allows the mean of $S(t)$ to vary by both $t$ and $\boldsymbol{x}$. We can use a standard kernel function, such as 
\begin{align*}
  K(t, t') = \sigma_S^2 \ e^{-\frac{(t - t')^2}{\rho}},  
\end{align*}
which enforces that two values of the potential intermediate at treatment values $t$ and $t'$ will be more correlated when the difference $t-t'$ is small in magnitude. An important parameter in this specification is $\rho$, which dictates the magnitude of this correlation, and we discuss it further in subsequent sections. In practice, we discretize the set of potential intermediate values to look at and only consider $\boldsymbol{t} = [t_1, t_2, \dots, t_M]$. The Gaussian process specification above implies that  
$$\Big(S(t_1), \dots, S(t_M) \mid \boldsymbol{X} = \boldsymbol{x} \Big) \sim \mathcal{N} \Big( (m(t_1, \boldsymbol{x}), \dots, m(t_M, \boldsymbol{x})), \boldsymbol{\Sigma}_S \Big),$$
where the $(i,j)$ element of $\boldsymbol{\Sigma}_S$ is given by $K(t_i, t_j)$. The main choice to be made is how to model the mean function of the Gaussian process, denoted by $m(t, \boldsymbol{x})$. We take a nonparametric Bayesian approach, and place a soft Bayesian additive regression tree (SoftBART, \cite{linero2018bayesian}) prior distribution on $m(\cdot, \cdot)$. SoftBART priors are an extension of the popular BART prior \citep{chipman2010bart}, which has been shown to work well for principal stratification \citep{kim2024bayesian}, where the decision trees follow probabilistic, instead of deterministic, paths. This has been shown to lead to improved empirical performance, particularly for estimating smooth functions, and they have improved theoretical properties in terms of posterior contraction rates. 

Note that monotonicity, which in this case would imply that $S_i(t) \geq S_i(t')$ when $t \geq t'$, is commonly assumed in principal stratification contexts. While it does not lead to nonparametric identification in this setting, it could be incorporated to improve efficiency in the estimation of this model or the imputation of the missing $S(t)$ values. This is difficult to embed within the proposed SoftBART framework above, but we explore one approach to incorporating monotonicity within a semiparametric model in Appendix D. 

\subsection{Specification of the outcome model}
\label{sec:Ymodel}

If principal ignorability is assumed, then identification of causal effects relies on $E[Y \mid T=t, S(t)=s, \boldsymbol{X} = \boldsymbol{x}]$, which is fairly straightforward to estimate given that it is a function of observable quantities. Under this assumption, we therefore model this conditional expectation using the aforementioned SoftBART prior that was used in estimation of the model for the potential intermediate. We focus attention for the rest of the manuscript on the more difficult setting, where principal ignorability is violated, and identification relies on $E[Y \mid T=t, \boldsymbol{S} = \boldsymbol{s}, \boldsymbol{X} = \boldsymbol{x}]$, where now $\boldsymbol{S}$ represents the entire vector of potential intermediates, most of which are unobserved. This will require fitting this model jointly with the model for the potential intermediates, along with a data augmentation step that imputes the missing potential intermediates. Interestingly, the outcome model not only helps impute the missing potential intermediates, but also can inform the correlation parameter $\rho$ dictating the correlation among the potential intermediates.  We provide more details on the nature of this in Section \ref{sec:OutcomeCorrelation}.

As discussed in the identifiability results in Section \ref{sec:IdentificationMain}, we can not fit an arbitrarily flexible model for $E[Y \mid T=t, \boldsymbol{S} = \boldsymbol{s}, \boldsymbol{X} = \boldsymbol{x}]$ as it will not be identified. For this reason, we focus on the following additive decomposition of the model for the outcome:
$$E[Y \mid T=t, \boldsymbol{S} = \boldsymbol{s}, \boldsymbol{X} = \boldsymbol{x}] = \lambda(t,\boldsymbol{x}) + \psi(t)s(t)+\sum_{t'\in\mathcal{T}_d} \beta(t,t')s(t').$$
We do not discuss estimation of $\lambda(t,\boldsymbol{x})$ or $\psi(t)$ in great detail as a variety of standard approaches could apply here. Nonparametric Bayesian priors could be used for $\lambda(t, \boldsymbol{x})$, and standard one-dimensional function estimates such as smoothing splines could be used for $\psi(t)$. For the analyses in Sections \ref{sec:SimulationStudies} and \ref{sec:application}, we include covariates linearly, and use basis function expansions for any smooth functions of $t$ based on natural cubic splines. Of most interest is how we specify the crucial $\beta(t,t')$ function, which allows for deviations from principal ignorability. For model identifiability, we require that $\beta(t,t) = 0$ for all $t \in \mathcal{T}$. One way to enforce this constraint is to assume a basis function expansion, defined by
$$\beta(t,t') = \sum_{j=1}^J \zeta_j z_j(t - t'), $$
for basis functions $z_j(\cdot)$ defined such that $z_j(0) = 0$ for all $j$. Possible functions include $(t - t')^d$ for some power $d$, or $e^{d(t - t')^2} - 1$ for positive constant $d \geq 0$. This allows for deviations from principal ignorability, ensures that the identifiabililty constraint is maintained, and ensures that $\beta(t, t')$ gets farther from zero only when $|t - t'|$ gets larger. 

\subsection{Posterior inference}

We estimate the aforementioned models within the Bayesian paradigm and therefore inference on all treatment effects of interest will be based on the posterior distribution of the unknown parameters. As a reminder, our estimand $E[Y(t) \mid a < g(\boldsymbol{S}) < b]$ can be written as in Equation \eqref{eq:estimand}. Note that each of the conditional expectations and densities in Equation \eqref{eq:estimand} is a function of the unknown parameters for which we have a posterior distribution. The outer expectation taken over the distribution of $\boldsymbol{X}$ given $a < g(\boldsymbol{S}) < b$ can be approximated by using the empirical distribution of the covariates among the units in the sample for which $a < g(\boldsymbol{S}) < b$. Specifically, if we let the superscript $(b)$ denote the $b^{th}$ posterior sample of an unknown quantity, then we can calculate
\begin{align*}
    & \Bigg( \frac{1}{\sum_{j=1}^n \omega_j^{(b)} 1(a < g(\boldsymbol{S}_j^{(b)}) < b)} \Bigg) \Bigg\{ \sum_{i=1}^n \omega_i^{(b)} 1(a < g(\boldsymbol{S}_i^{(b)}) < b) \\
    \times & \int_{\boldsymbol{s}} E^{(b)}(Y \mid T=t, \boldsymbol{X}=\boldsymbol{X}_i, \boldsymbol{S}=\boldsymbol{s}) \frac{f^{(b)}(\boldsymbol{s} \mid \boldsymbol{X}=\boldsymbol{X}_i) 1(a < g(\boldsymbol{s}) < b)}{P^{(b)}(a < g(\boldsymbol{S}) < b \vert \boldsymbol{X}=\boldsymbol{X}_i)} d \boldsymbol{s} \Bigg\}.
\end{align*}
Here we have introduced weights for each data point $(\omega_1^{(b)}, \omega_2^{(b)},\dots \omega_n^{(b)})$, which follow a $\text{Dirichlet}(1, 1, \dots, 1)$ distribution to account for uncertainty stemming from the observed sample being used to approximate the expectation over the covariate distribution as in the Bayesian bootstrap \citep{rubin1981bayesian}. This is a standard approach to accounting for covariate uncertainty for population estimands in Bayesian causal inference \citep{linero2023and}. In practice, the integral with respect to $\boldsymbol{s}$ is done using a Monte Carlo approximation, which we outline in Appendix A. We obtain this quantity for all posterior samples $b = 1, \dots, B$, and inference can proceed using traditional Bayesian approaches once a posterior distribution is obtained. 

We do not discuss technical details of Markov chain Monte Carlo (MCMC) sampling or other computational concerns here, and these can be found in Appendix C. All parameters have closed form full conditional distributions, which facilitates sampling through a Gibbs sampler. Updating the correlation parameter for the intermediates, denoted by $\rho$, is greatly improved by utilizing a conditional distribution that marginalizes over the potential intermediates $\boldsymbol{S}$. We discuss this in more detail in the following section. Additionally, prior distributions are assumed to be relatively flat, non-informative prior distributions unless specified otherwise above. Computation time for MCMC algorithms can also be found in Appendix I, which shows how computation time scales with sample size and the size of the discretization grid of $\boldsymbol{S}$.

\section{Using the outcome model to identify correlation between principal strata}
\label{sec:OutcomeCorrelation}

One difficulty in the estimation of principal causal effects in this setting is that we do not know the underlying correlation between $S_i(t)$ and $S_i(t')$, and we only get to observe one value of the intermediate for each unit in the sample. In prior work on principle stratification, this parameter is frequently treated as a sensitivity parameter and the analysis is run for different values of the sensitivity parameter to see how results vary \cite[e.g.,][]{jiang2021identification}. In other work, it is assumed that the correlation can be learned by incorporating information from the outcome model, but only heuristic arguments for this strategy have been made \cite[e.g.,][]{bartolucci2011modeling, schwartz2011bayesian}. Throughout the manuscript, we take the approach of learning the correlation through information from the outcome model, and in this section, we provide justification for this procedure. 
For simplicity, in this section, we focus on $P(\boldsymbol{S}, \rho \vert \mathcal{D})$ the posterior distribution of $(\boldsymbol{S}, \rho)$ given the observed data, and treat the other model parameters as fixed at their true values. The same ideas would hold if the other model parameters were being estimated simultaneously, but our goal here is to show the interplay between $\rho$ and the outcome model, and uncover when and how the outcome model informs this parameter. To study what happens to the correlation parameter, we can examine $P(\rho \vert \mathcal{D})$, the marginal posterior of the correlation parameter. We show in Appendix B that this marginal posterior is given by

{\footnotesize
\begin{align*}
P(\rho \vert \mathcal{D}) &= P(\rho) \left\{ \prod_{i=1}^n \bigg( 1 + \frac{\boldsymbol{\beta}_i^T \widetilde{\boldsymbol{\Sigma}}_{si} \boldsymbol{\beta}_i}{\sigma^2} \bigg)^{-1/2} \right\} \\
&\times \ \text{exp} \Bigg\{ -\frac{1}{2} \Bigg[ \widetilde{\boldsymbol{\mu}}_{si}^T \widetilde{\boldsymbol{\Sigma}}_{si} \widetilde{\boldsymbol{\mu}}_{si} - \bigg(\frac{\widetilde{Y}_i \boldsymbol{\beta}_i}{\sigma^2} + \widetilde{\boldsymbol{\Sigma}}_{si}^{-1} \widetilde{\boldsymbol{\mu}}_{si} \bigg)^T \bigg(\frac{\boldsymbol{\beta}_i \boldsymbol{\beta}_i^T}{\sigma^2} + \widetilde{\boldsymbol{\Sigma}}_{si}^{-1} \bigg)^{-1} \bigg(\frac{\widetilde{Y}_i \boldsymbol{\beta}_i}{\sigma^2} + \widetilde{\boldsymbol{\Sigma}}_{si}^{-1} \widetilde{\boldsymbol{\mu}}_{si} \bigg) \Bigg] \Bigg\}
\end{align*}
}

\noindent Here, we have used that $\widetilde{Y}_i = Y_i - \lambda(T_i, \boldsymbol{X}_i) - \psi(T_i) S_i$. We have also adopted the notation that $\boldsymbol{\beta}_i = (\beta(T_i, t_1), \dots, \beta(T_i, t_M))$. We also defined the mean and variance of the conditional distributions of the potential intermediates $\boldsymbol{S}_i$ given $S_i$, the observed intermediate, as
\begin{align*}
\widetilde{\boldsymbol{\mu}}_{si} &=  m(\boldsymbol{t}, \boldsymbol{X}_i) + \boldsymbol{\Sigma}_{i.} (S_i - m(T_i, \boldsymbol{X}_i)) / \sigma_S^2 \\
\widetilde{\boldsymbol{\Sigma}}_{si} &= \boldsymbol{\Sigma}_S - \boldsymbol{\Sigma}_{i.} \boldsymbol{\Sigma}_{i.}^T / \sigma_S^2
\end{align*}
where here $\boldsymbol{\Sigma}_{i.}$ is a vector of values given by $K(T_i, \boldsymbol{t})$. Note that while not explicitly stated, both $\widetilde{\boldsymbol{\mu}}_{si}$ and $\widetilde{\boldsymbol{\Sigma}}_{si}$ are functions of $\rho$.

This result has a number of important implications. For one, it greatly facilitates sampling as we have found that MCMC sampling algorithms have greatly improved performance and require far fewer iterations when iteratively sampling between $P(\rho \vert \mathcal{D})$ and $P(\boldsymbol{S} \vert \rho, \mathcal{D})$ instead of a standard Gibbs sampling algorithm that iterates between $P(\rho \vert \boldsymbol{S}, \mathcal{D})$ and $P(\boldsymbol{S} \vert \rho, \mathcal{D})$. Of more relevance to this section is that it also has consequences for understanding identification of $\rho$. One can see from the expression that if $\boldsymbol{\beta}_i = \boldsymbol{0}$ for all $i$, the posterior distribution reduces to $P(\rho)$, the prior distribution. This shows that we have no information in the data to learn about $\rho$ when principal ignorability holds. If, however, principal ignorability is violated, then there exists information in the data to estimate this parameter. 

It is then of interest to study whether the information from the outcome model is correct in the sense that our posterior will concentrate around the true value of $\rho$ when principal ignorability is violated. This result is seen in Theorem \ref{thm:CorrelationResults}, where we denote the true value of the correlation parameter by $\rho^*$. 
\begin{theorem} \label{thm:CorrelationResults}
    (Identification of correlation parameters)
    Assume the outcome model in \eqref{eqn:ParametricModelSpec} is correctly specified and we use an improper flat prior on $\rho$. Additionally, assume there exists $t,t'$ such that $\beta(t,t')\neq0$. Under mild regularity conditions,
$$ \rho^* = \argmax_{\rho} E \Big[ \log P(\rho \vert \mathcal{D}) \Big]. $$
\end{theorem}
A proof of this can be found in Appendix B. This result shows that the posterior distribution is maximized at the true correlation parameter, and that we obtain pointwise identification of this parameter. Under mild conditions on the prior distribution for $\rho$, this result also shows that we obtain consistent estimates of $\rho$ asymptotically. Note that while we have simplified the correlation structure to depend on a single parameter, $\rho$, the same result would hold regardless of the kernel function of the Gaussian process and the corresponding structure of $\boldsymbol{\Sigma}_S$.  This result, along with the fact that it requires principal ignorability to be violated, may seem counterintuitive at first. To gain intuition for them, it is important to understand that the only information we have on $\rho$ comes indirectly from the residuals in our outcome model. If principal ignorability holds, then $\boldsymbol{S}$ does not enter the outcome model, and the residuals should contain only random variation. If, however, principal ignorability is violated, then the full vector $\boldsymbol{S}$ is included in the outcome model residuals, and the variance of these residuals therefore depends on the covariance matrix of $\boldsymbol{S}$, which provides information on $\rho$. The result therefore depends on correctly specifying the outcome model. If the outcome model is misspecified, the limiting value of the posterior of $\rho$ would depend on both the true correlation structure of $\boldsymbol{S}$ and the variability of the approximation error present due to model misspecification.

\begin{figure}[htbp]
    \centering
    \includegraphics[width=0.9\linewidth]{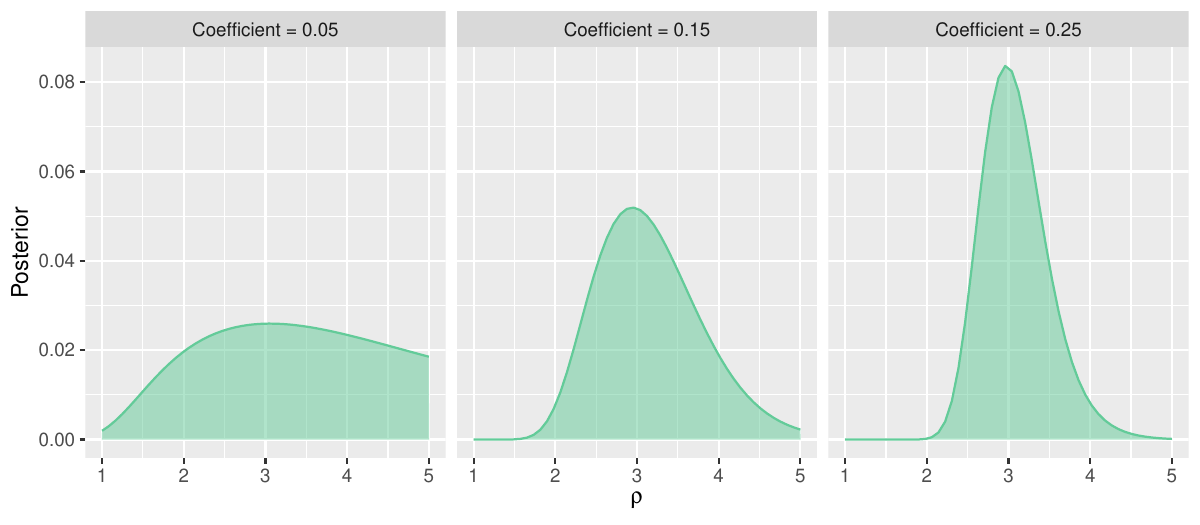}
    \caption{Average marginal posterior distribution across 100 simulated data sets when $\rho^* = 3$. Each panel corresponds to a different magnitude of association between the intermediate and the outcome.}
    \label{fig:PosteriorRho}
\end{figure}

One final point to make is that the shape of the posterior distribution depends on the magnitude of $\boldsymbol{\beta}_i$ for $i=1, \dots, n$. Figure \ref{fig:PosteriorRho} shows $P(\rho \vert \mathcal{D})$ averaged over 100 data sets of size $n=500$, under three scenarios for the true effect of the intermediate on the outcome. We set $\boldsymbol{\beta}_i = c \boldsymbol{1}_M$ and varied $c \in \{0.05, 0.15, 0.25\}$. We see that on average, the posterior distribution is maximized at the true value of 3, but its shape is greatly affected by the magnitude of $\boldsymbol{\beta}_i$. As the degree of principal ignorability violations becomes larger, there is more information to update $\rho$ and the posterior becomes more concentrated. While we assumed all other parameters were fixed and known here, we also see this empirically in Section \ref{sec:SimulationStudies} when all parameters are estimated simultaneously.

\section{Simulation studies}
\label{sec:SimulationStudies}

Here we investigate the performance of the proposed approach to target principal causal effects in a range of simulated scenarios. We first explore scenarios where principal ignorability holds, highlighting the ability of the proposed nonparametric Bayesian models to estimate causal effects in the presence of complex data generating processes. Next, we explore simulation scenarios where principal ignorability is violated. We illustrate the identification results derived in Section \ref{sec:identification_para}, and show how weak identifiability can manifest in MCMC algorithms getting stuck in local modes, but that this can be addressed through the weighting strategy proposed in Section \ref{ssec:WeakIdentification}. Additionally, in Appendix G we provide simulations when principal ignorability is violated but the model is misspecified. 

\subsection{Inference under principal ignorability}

Throughout, we set $n=500$ and $p=5$ and generate covariates from independent standard normal distributions. The treatment is generated from a normal distribution with mean $0.3 X_2$ and variance 1. We discretize the set of $t$ values considered to 10 locations on an equally spaced grid between -1.35 and 1.5, and generate the potential intermediate $\boldsymbol{S} = (S(t_1), \dots, S(t_{10}))$ from a multivariate normal distribution with mean and covariance defined in Section \ref{sec:Smodel} with $\rho = 4$. The mean of $S(t)$ is given by $m(t, \boldsymbol{x}) = 0.3 t + 0.4 t^2 + 0.4 x_4 + 0.2 t x_2 - 0.3 t x_4.$ The conditional mean of the outcome is given by $E(Y \mid T=t, \boldsymbol{S}=\boldsymbol{s}, \boldsymbol{X} = \boldsymbol{x}) = t + 0.4e^{0.75t} - 0.3 s_2 - 0.2 s_3 + 0.3 s_5 + t x_4 + t s$. Importantly, in both of these models, the effect of $t$ is nonlinear and heterogeneous with respect to the observed covariates. The residual variance in the outcome model is given by $\sigma^2 = 0.5$. Throughout, we assess performance for estimating two different principal causal effects: 1) $E(Y(t) \mid \text{range}(\boldsymbol{S}) < 1.5)$, and 2) $E(Y(t) \mid \text{range}(\boldsymbol{S}) > 2.5)$. The two principal causal effects correspond to effects in subgroups of the population that have small and large effects of the treatment on the intermediate, respectively. For simplicity we discretize these exposure-response curves to estimating the effect at five exposure levels between -1 and 1. 

\begin{figure}[ht!]
    \centering
    \includegraphics[width=0.95\linewidth]{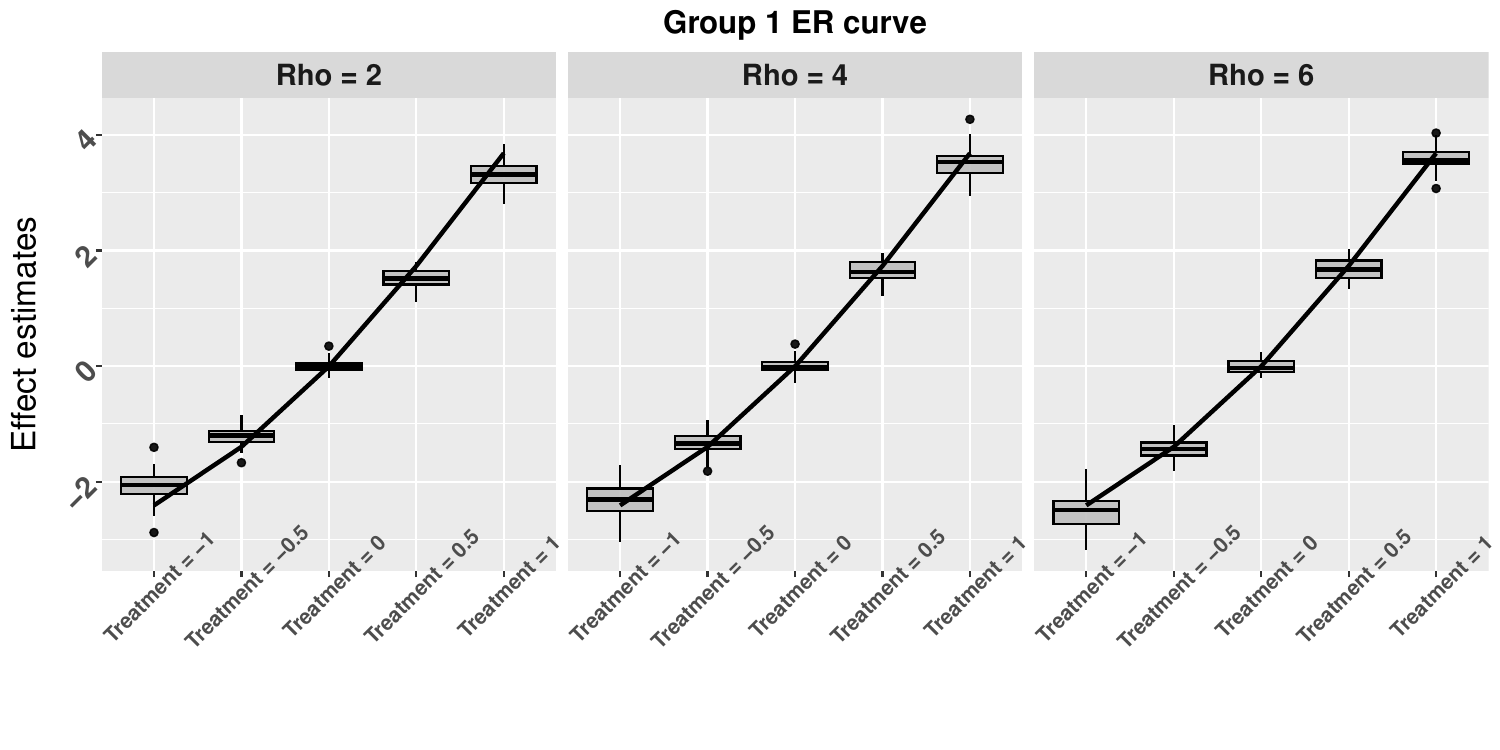} \\
    \includegraphics[width=0.95\linewidth]{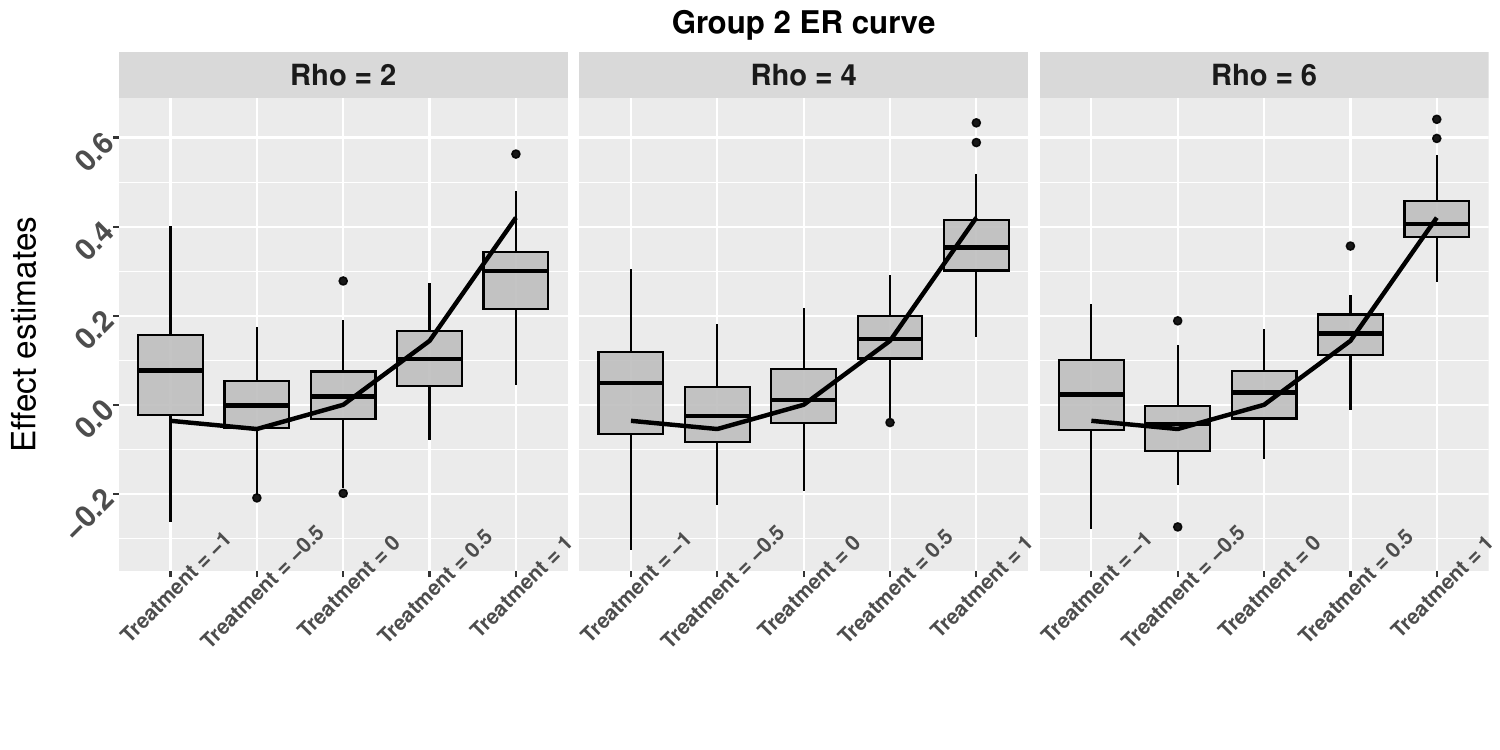}
    \caption{Boxplots of posterior mean estimates of the exposure-response curve across simulated datasets, evaluated at five treatment levels under principal ignorability. The true value at each treatment level is indicated by the black line.}
    \label{fig:PIsimGroup1}
\end{figure}

The results from the simulation study across 250 simulated data sets is given in Figure \ref{fig:PIsimGroup1}. We see that the proposed approach is able to capture the nonlinear and heterogeneous effects present in the data generating mechanisms and provide nearly unbiased results across all treatment levels and principal strata considered. Importantly, results do not appear to be sensitive to the choice of $\rho$, as we obtain reasonable results even under misspecified values where $\rho \in \{2,6\}$. 

\subsection{Violations of principal ignorability}
Now we consider the situation where principal ignorability is violated. We generate covariates and treatment in the same manner as in the previous section. We set $\rho = 3$ and the mean function to be $m(t,\boldsymbol{x})=0.3t+0.4tx_4+0.7tx_2$. Finally, we generate the outcome from the model in Section \ref{sec:Ymodel}, where we set $\lambda(t,\boldsymbol{x}) = t-0.3x_2-0.2x_3+0.3x_5$, $\psi(t)=0.5t$, and $\beta(t,t')=0.05(t-t')+0.025(t-t')^2$. The residual variance in the outcome model is 1.

\begin{figure}[ht!]
    \centering  \includegraphics[width=0.95\linewidth]{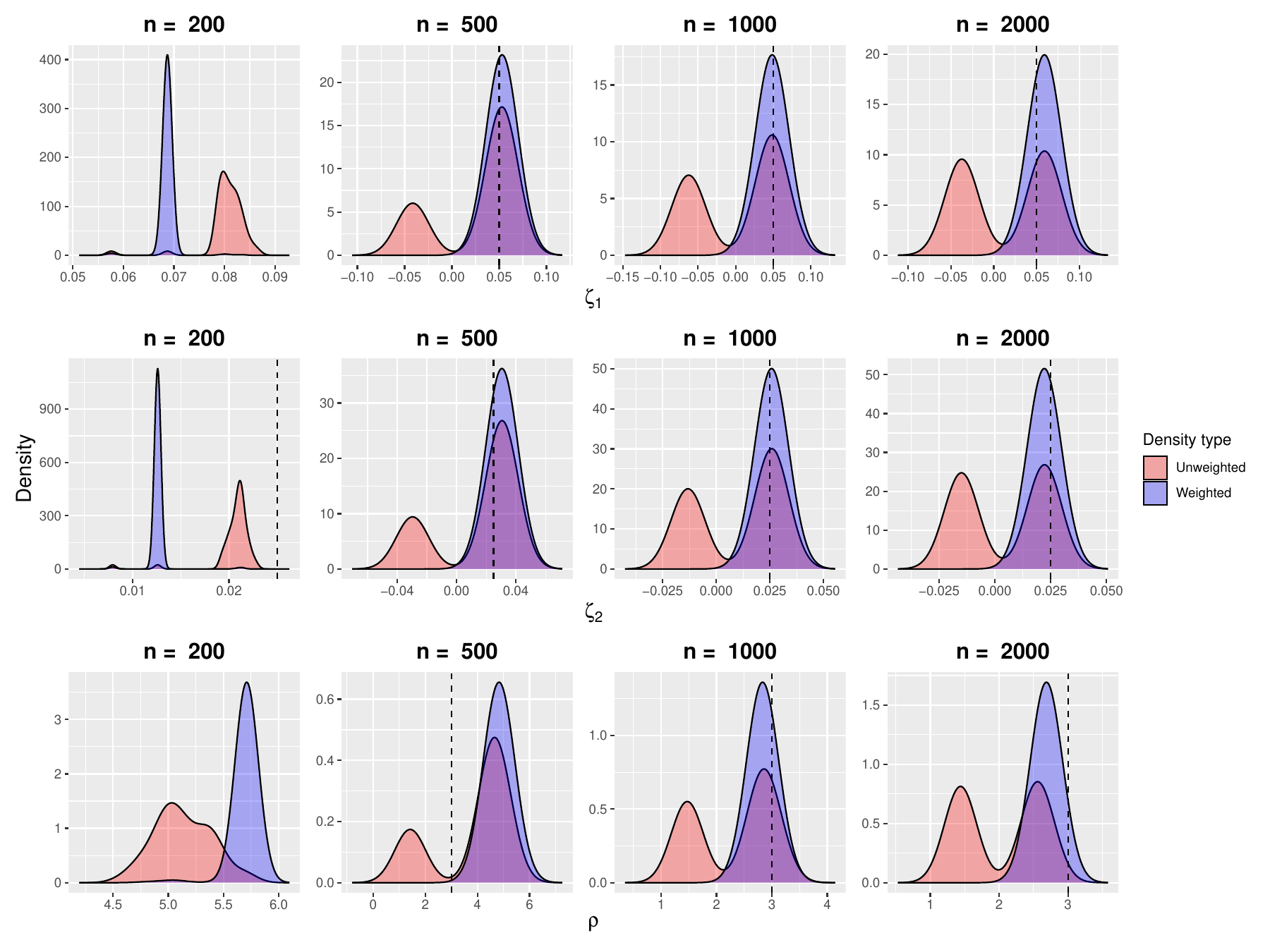} 
    \caption{Posterior distributions of $(\zeta_1,\zeta_2,\rho)$ for a single data set across sample sizes. The two histograms represent the posterior distribution both with and without the weighting scheme of Section \ref{ssec:WeakIdentification}. }
    \label{fig:simWithoutPI1}
\end{figure}

\begin{figure}[ht!]
    \centering
    \includegraphics[width=0.95\linewidth]{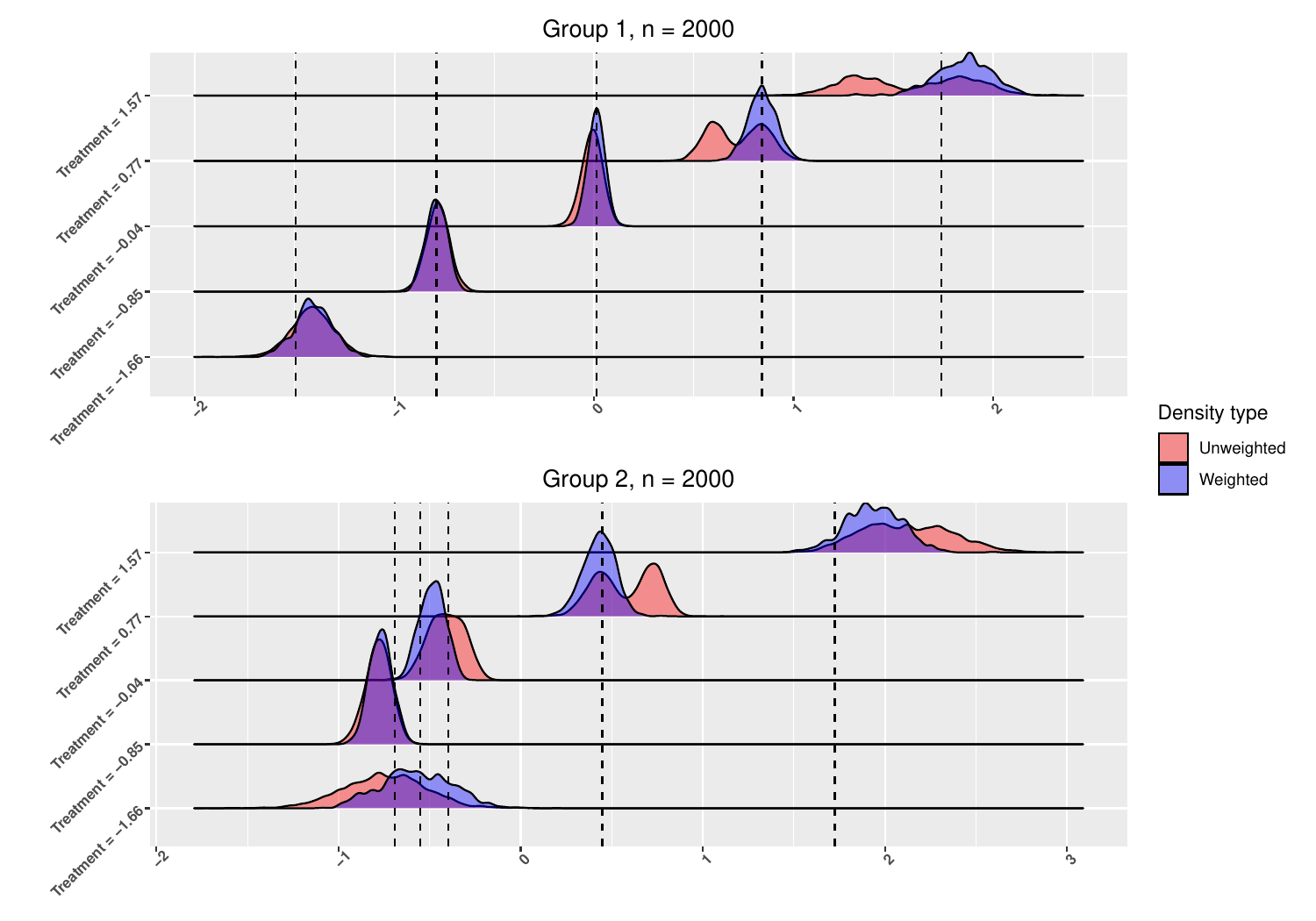}
    \caption{Posterior distributions of  $E(Y(t)\mid \text{range}(\boldsymbol{S})<1.5)$ and $E(Y(t)\mid\text{range}(\boldsymbol{S})>2.5)$, referred to as group 1 and 2, respectively, across five exposure values when $n=2000$. The dashed black vertical lines represent the true values of the estimands. The two histograms represent the posterior distribution both with and without the weighting scheme of Section \ref{ssec:WeakIdentification}. }
    \label{fig:simWithoutPI2}
\end{figure}

To investigate how the aforementioned weak identification influences posterior sampling across different sample sizes, we vary $n$ from 200 to 2000, generate one dataset for each sample size, and run 50 chains for each dataset. For all parameters, we explore two different posterior distributions: 1) a weighted posterior distribution described in Section \ref{ssec:WeakIdentification} that assigns weights $\pi_m$ proportional to the marginal likelihood of chain $m$, and 2) an unweighted posterior distribution that assigns equal weight to each MCMC chain, i.e. $\pi_m = 1/M$. The causal estimands of interest are 1) $E(Y(t) \mid \text{range}(\boldsymbol{S}) < 1.5)$, and 2) $E(Y(t) \mid \text{range}(\boldsymbol{S}) >2.5)$, for $t\in\{-1.66,-0.85,-0.04,0.77,1.57\}$. Figure \ref{fig:simWithoutPI1} shows the posterior distribution of $(\zeta_1, \zeta_2, \rho)$ and we see that there is a clear issue with MCMC chains getting stuck in local modes. We see that for both $\zeta_1$ and $\zeta_2$, the MCMC chains get stuck in the local mode described in Section \ref{ssec:WeakIdentification} where the values are the opposite sign of the true values. This also leads to a bimodal posterior distribution for $\rho$, which is affected by the values of $\boldsymbol{\zeta}$. The weighted posterior distribution addresses this issue, however, as it assigns much more weight to the correct mode, leading to improved inferences on the parameters of interest. As a reminder, these results are for a single simulated data set and therefore it is not expected for any one posterior distribution to be perfectly centered around the true value, but we see as the sample size grows, it is concentrating around the true values of the parameters. Similar findings are found for the causal estimands of interest in the $n=2000$ setting, as illustrated in Figure \ref{fig:simWithoutPI2}. We see that the unweighted posterior distributions are bimodel due to chains getting stuck in local, suboptimal modes, but the proposed weighted posterior distributions are closer to the true values generally. Overall, these results highlight the identifiability of these crucial parameters of interest, and the ability to infer the unknown dependence parameter $\rho$, which is typically left as a sensitivity parameter. 

\section{The role of the economy on arrest rates}\label{sec:application}

Understanding how the economy shapes the criminal justice system is an important question that has been studied since at least Rusche and Kirchheimer’s 1939 book ``Punishment and Social Structure'' \citep{rusche1939punishment}, which posited that recessions increase coercive control because the state needs to maintain order in the face of economic crises. Empirical examinations, however, have produced mixed results, perhaps because they have not accounted for the multiple causal pathways through which the economy can impact criminal justice \citep{sutton2004political, carmichael2014persistent}. The present study examines police capacity as a potential mechanism through which economic conditions could impact arrest rates. Results from this study can help policymakers and criminologists situate police in their macro-economic context. 

Our data consist of all cities in the United States, though we restrict attention to those with at least 50,000 people in the year 2010, since data are more reliable in larger cities. After removing observations with missing data, we have $n=596$ cities. The treatment variable is log gross domestic product (GDP) in 2011, which is measured in inflation-adjusted thousands of dollars, and comes from the Bureau of Economic Analysis. The intermediate variable is the number of police officers per 1000 individuals in 2012 and it comes from the Law Enforcement Officers Killed and Assaulted database. The outcome is the arrest rate per 1000 people in 2013 and comes from the FBI’s Uniform Crime Reports. Pre-treatment covariates consist of percent Latino, percent Black, percent with a bachelor's degree, percent men aged 15-34, percent of vacant housing, percent foreign born, population size, and violent crime rates, all taken from the year 2010. Demographic variables come from the Census Bureau’s decennial census and American Community Survey (ACS), while violent crime rates also come from the FBI's Uniform Crime Reports. We measure a wide range of important features of cities in order to capture confounders of the GDP, police capacity, and arrest rate relationships.  Nonetheless, causal conclusions from this study should be interpreted with caution as one cannot rule out the presence of unmeasured confounders affecting results. Regarding other assumptions, positivity is assessed in Appendix E, and we believe that SUTVA is plausible in this study as we focused on large cities that are sufficiently separated to avoid issues with interference.  

We target estimands with two different $g(\cdot)$ functions to focus on different PCEs. We examine the average potential intermediate given by $g(\boldsymbol{S}) =  \overline{\boldsymbol{S}}=\frac{1}{M} \sum_{m=1}^M S(t_m)$, as well as $g(\boldsymbol{S}) = \text{range}(\boldsymbol{S})$. We look at principal strata defined by units with either low or high values of these two functions. Specifically, if we define $c_m$ and $c_s$ to be the sample mean and standard deviation of $S_i$, we look at principal strata exposure-response curves defined by $E(Y(t) \mid \overline{\boldsymbol{S}} < c_m)$ and $E(Y(t) \mid \overline{\boldsymbol{S}} \geq c_m)$, as well as $E(Y(t) \mid \text{range}(\boldsymbol{S}) < c_s)$ and $E(Y(t) \mid \text{range}(\boldsymbol{S}) \geq c_s)$. The first set of principal strata looks at groups of cities having high or low average values of police capacity, regardless of the shape of $\boldsymbol{S}$ and whether GDP alters police capacity. The range estimand instead looks at cities with relatively flat or steep curves for $S(t)$ as a function of $t$. These can be viewed as associative and dissociative principal strata, e.g.,  cities that have steep treatment-intermediate trajectories are those cities that alter their police capacity in response to economic conditions. Differences in the exposure-response curves in the associative and dissociative principal strata would be suggestive of the fact that police capacity mediates the impact of GDP on arrest rates.

Throughout, we use our proposed model with prior distributions described above. Assuming principal ignorability, we run two chains with 10,000 iterations, discarding the first 4,000 as burn-in and thinning every 20th sample. When violations are allowed, we run 20 chains with 10,000 iterations, discarding the first 2,000 and thinning every 20th sample.

\subsection{Results}

Here we estimate the causal estimands of interest both with and without the assumption of principal ignorability. When principal ignorability is assumed, we use the nonparametric Bayesian SoftBART approach for estimation of all necessary conditional expectations. When principal ignorability is assumed,  we fix it at $\rho=4$. We explored other values of $\rho$ in Appendix H and found similar results across all choices.  When principal ignorability is allowed to be violated, we continue to use SoftBART for the mean of the potential intermediates, given by $m(t, \boldsymbol{x})$, but we now posit a parametric model for the conditional expectation of the outcome. We assume this model is additive in the covariates, treatment, and intermediate and allow the effects of the covariates to be linear, while we allow for nonlinear effects of treatment and the intermediate through basis function expansions. As in the simulations, we allow for violations of principal ignorability by setting $\beta(t,t') = \sum_{j=1}^J \zeta_j (t - t')^j$ for $J=2$.  

\begin{figure}[ht!]
    \centering

    \includegraphics[width=0.95\linewidth]{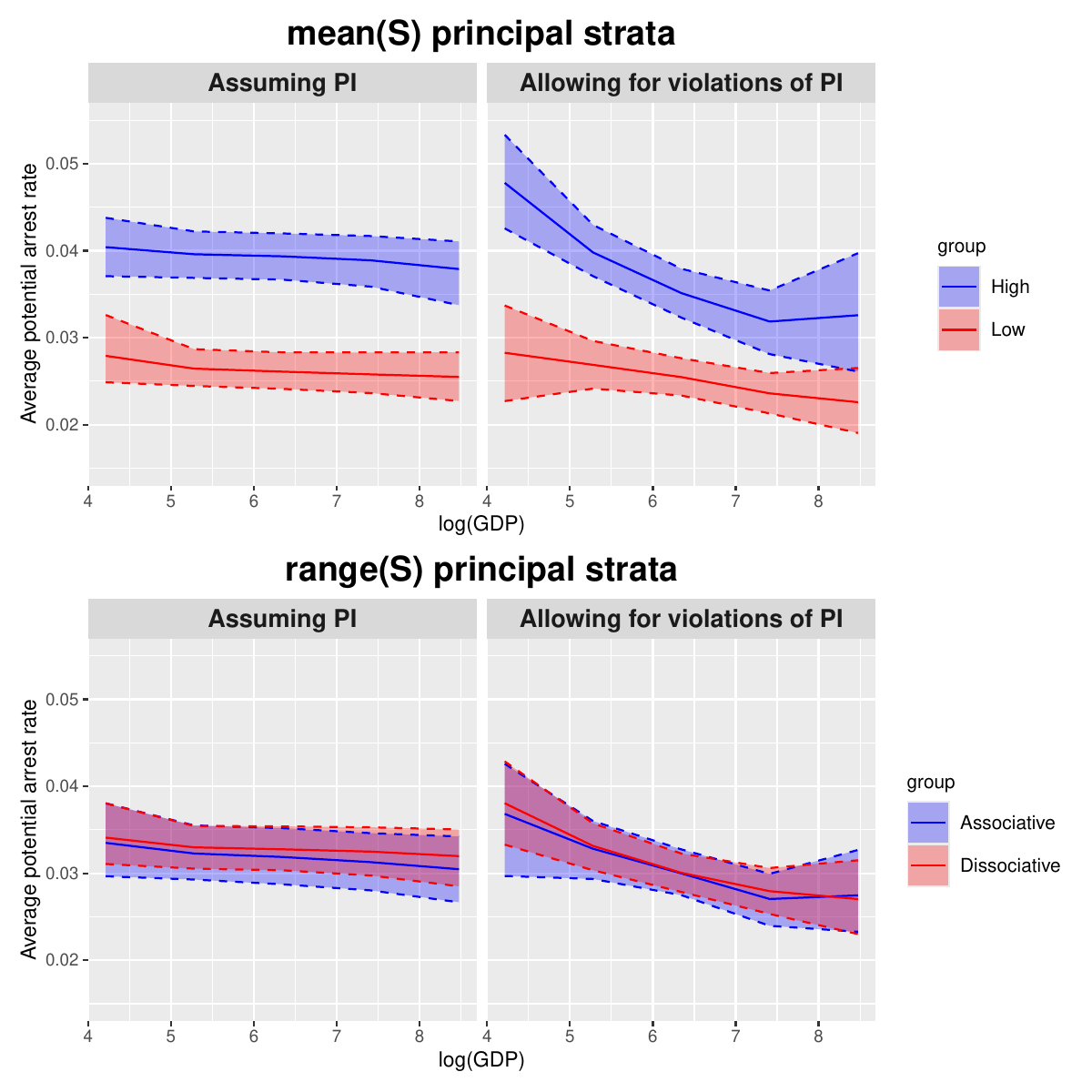}
    \caption{Exposure-response curves across principal strata. Solid lines correspond to posterior means and dashed lines represent 95\% credible intervals. The left column corresponds to results that assume principal ignorability, while the right column are results when it is allowed to be violated. The first row looks at principal strata defined by mean$(\boldsymbol{S})$, and the second row looks at principal strata defined by range$(\boldsymbol{S})$.}
    \label{fig:AnalysisResultsAll}
\end{figure}

All results for causal estimands of interest can be found in Figure \ref{fig:AnalysisResultsAll}. For our model that allows for violations of principal ignorability, we obtain 95\% credible intervals of (-0.068, -0.011) for $\zeta_1$ and (0.013, 0.034) for $\zeta_2$. This suggests that principal ignorability is indeed violated as both of these intervals do not contain zero, which is the value corresponding to principal ignorability. We also obtain a 95\% credible interval of (0.25, 4.65) for the dependence parameter $\rho$ suggesting that a wide range of potential dependencies are supported by the observed data. In terms of causal effects, the top panel of Figure \ref{fig:AnalysisResultsAll} shows that cities with higher police capacity (high values of mean$(\boldsymbol{S})$) tend to have higher arrest rates than those with low police capacity, which is to be expected. Interestingly, when principal ignorability is assumed, we estimate very flat exposure-response curves in both principal strata indicating that there is little effect of the economy on arrest rates. Once principal ignorability is allowed to be violated, however, we estimate that the exposure-response curve has a negative slope, particularly in the cities with higher police capacity. This suggests that an improved economy can reduce arrest rates, but that this effect is more pronounced in cities with higher police capacity. This is an interesting result, that is also fairly intuitive, as it is unlikely for the economy to affect arrest rates in cities with low police capacity who already have lower arrest rates. Looking at the range estimand, we see that regardless of the assumption of principal ignorability, cities in either of the associative or dissociative strata have similar arrest rates and similar exposure-response curves. Overall, these results paint an interesting picture about the effects of both the economy and policy capacity on arrest rates, and the potential pathways by which these variables can affect arrest rates. Both models show that increased police capacity is associated with higher arrest rates. Additionally, under our models that do not assume principal ignorability, we see that an improved economy tends to reduce arrest rates. Given that the effects are similar in both the associative and dissociative strata, it does not appear that this relationship is mediated by police capacity, however, and this is likely because GDP's effects on police capacity are small (for this see Appendix H). 

\section{Discussion} \label{sec:discussion}

This manuscript developed methodology to estimate principal causal effects when both the treatment and intermediate variables are continuous. We formalized assumptions and estimands in this setting, and provided a flexible, nonparametric Bayesian framework to estimate all estimands of interest. One key aspect of our results regards identification when the commonly used principal ignorability assumption is violated. While this would seem to be problematic at first, we show it provides possibilities to identify the dependence parameters of the missing potential intermediates, along with identification of parametric outcome models. This is in sharp contrast from standard, binary treatment settings, where these parameters are not identifiable even when the most basic parametric models are assumed \citep{wu2024partial}. Lastly, we analyzed the relationships between GDP, police capacity, and arrest rates to provide an improved understanding of the relationship between the economy and arrest rates. 

While we showed that principal ignorability need not be assumed, this only works when certain parametric models hold for the outcome model. While these can be made to be semiparametric to some degree due to the inclusion of nonlinear functions within the regression model, this approach is reliant on a well-specified model. Future work could study the interplay between the flexibility of the outcome model, and the ability to identify all model parameters in order to reduce sensitivity to model misspecification while still identifying (either fully or partially) all parameters of interest. 
In addition to model specification, future work could develop further sensitivity analyses to assess the robustness of findings to unmeasured confounding \citep{franks2020flexible}. Another potential area of future research would be to extend the ideas presented here to panel data settings where each data point is observed repeatedly over time as this could potentially weaken assumptions about unmeasured confounding or principal ignorability.

\appendix

\section{Identification of causal estimands}
\label{sec:IdentificationEstimands}

Throughout this section we focus on estimands of the form $E[Y(t) \mid a < g(\boldsymbol{S}) < b]$ for some function $g(\cdot)$, though analogous calculations would apply for other estimands. First we can see using iterated expectations that
\begin{align*}
    E[Y(t) \mid a < g(\boldsymbol{S}) < b] &= E_{\boldsymbol{X} \vert a < g(\boldsymbol{S}) < b} E[Y(t) \mid a < g(\boldsymbol{S}) < b, \boldsymbol{X}]
\end{align*}
The outer expectation is relatively easy to work with, so we focus on the inner expectation now, which is slightly more complex since we are conditioning on $\boldsymbol{S}$ satisfying a constraint, rather than being set to a fixed value that we can plug into the outcome regression equation. This inner expectation can be decomposed as
\begin{align*}
    E[Y(t) \mid a < g(\boldsymbol{S}) < b, \boldsymbol{X}] &= E_{S \mid \boldsymbol{X}, a < g(\boldsymbol{S}) < b} E[Y(t) \mid a < g(\boldsymbol{S}) < b, \boldsymbol{X}, \boldsymbol{S}=\boldsymbol{s}] \\
    &= E_{S \mid \boldsymbol{X}, a < g(\boldsymbol{S}) < b} E[Y(t) \mid \boldsymbol{X}, \boldsymbol{S}=\boldsymbol{s}] \\
    &= E_{S \mid \boldsymbol{X}, a < g(\boldsymbol{S}) < b} E[Y(t) \mid T=t, \boldsymbol{X}, \boldsymbol{S}=\boldsymbol{s}] \\
    &= E_{S \mid \boldsymbol{X}, a < g(\boldsymbol{S}) < b} E[Y \mid T=t, \boldsymbol{X}, \boldsymbol{S}=\boldsymbol{s}],
\end{align*}
where the third equality relied on the strong unconfoundedness assumption, and the fourth equality relied on SUTVA.  The inner of these two expectations is straightforward to calculate since we have an outcome model for $E[Y \mid T=t, \boldsymbol{X}, \boldsymbol{S}=\boldsymbol{s}]$. We simply need to be able to average this outcome model with respect to the distribution of $\boldsymbol{S} \mid \boldsymbol{X}=\boldsymbol{x}, a < g(\boldsymbol{S}) < b$. This density can be written as
\begin{align*}
    f(\boldsymbol{s} \mid \boldsymbol{X}=\boldsymbol{x}, a < g(\boldsymbol{S}) < b) &= \frac{f(\boldsymbol{s} \mid  \boldsymbol{X}=\boldsymbol{x}) P(a < g(\boldsymbol{S}) < b \vert  \boldsymbol{X}=\boldsymbol{x}, \boldsymbol{S} = \boldsymbol{s})}{P(a < g(\boldsymbol{S}) < b \vert  \boldsymbol{X}=\boldsymbol{x})} \\
    &=\frac{f(\boldsymbol{s} \mid \boldsymbol{X}=\boldsymbol{x}) 1(a < g(\boldsymbol{s}) < b)}{P(a < g(\boldsymbol{S}) < b \vert \boldsymbol{X}=\boldsymbol{x})}
\end{align*}
Both the numerator and the denominator can be calculated from the model for $\boldsymbol{S} \mid X$. The denominator can be calculated empirically by sampling a large number of times from $\boldsymbol{S} \mid X$ and calculating the proportion of draws for which the inequality is true. In practice, we will calculate $E[Y \mid T=t, a < g(\boldsymbol{S}) < b, \boldsymbol{X}]$ by sampling draws of $\boldsymbol{S}$ from the distribution of $\boldsymbol{S} | X$, calculating $E[Y \mid T=t, \boldsymbol{S} = \boldsymbol{s}, X]$ for each draw, and taking a weighted average of these values with weights given by $\frac{1(a < g(\boldsymbol{s}) < b)}{P(a < g(\boldsymbol{S}) < b \vert \boldsymbol{X}=\boldsymbol{x})}.$ Combining all of these steps, we can see that
\begin{align*}
    & E(Y(t) \mid a < g(\boldsymbol{S}) < b) \\
    =& E_{\boldsymbol{X} \vert a < g(\boldsymbol{S}) < b} \bigg\{ E_{S \mid \boldsymbol{X}, a < g(\boldsymbol{S}) < b} \Big[ E(Y \mid T=t, \boldsymbol{X}, \boldsymbol{S}=\boldsymbol{s}) \Big] \bigg\} \\
    =& E_{\boldsymbol{X} \vert a < g(\boldsymbol{S}) < b} \bigg\{ \int_{\boldsymbol{s}} E(Y \mid T=t, \boldsymbol{X}, \boldsymbol{S}=\boldsymbol{s}) f(\boldsymbol{s} \mid \boldsymbol{X}=\boldsymbol{x}, a < g(\boldsymbol{S}) < b) d \boldsymbol{s}  \bigg\} \\
    =& E_{\boldsymbol{X} \vert a < g(\boldsymbol{S}) < b} \bigg\{ \int_{\boldsymbol{s}} E(Y \mid T=t, \boldsymbol{X}, \boldsymbol{S}=\boldsymbol{s}) \frac{f(\boldsymbol{s} \mid \boldsymbol{X}=\boldsymbol{x}) 1(a < g(\boldsymbol{s}) < b)}{P(a < g(\boldsymbol{S}) < b \vert \boldsymbol{X}=\boldsymbol{x})} d \boldsymbol{s}  \bigg\} \\
    =& \int_{\boldsymbol{x}} \int_{\boldsymbol{s}} E(Y \mid T=t, \boldsymbol{X}, \boldsymbol{S}=\boldsymbol{s}) \frac{f(\boldsymbol{s} \mid \boldsymbol{X}=\boldsymbol{x}) 1(a < g(\boldsymbol{s}) < b)}{P(a < g(\boldsymbol{S}) < b \vert \boldsymbol{X}=\boldsymbol{x})} d \boldsymbol{s} \  dF_{\boldsymbol{X} \mid a < g(\boldsymbol{S}) < b}(\boldsymbol{x}),
\end{align*}
and we get the final expression seen in the manuscript. Obtaining the identification result seen in Theorem 3.1 relies on the principal ignorability assumption, which allows us to replace 
$E(Y \mid T=t, \boldsymbol{X}, \boldsymbol{S}=\boldsymbol{s})$ with $E(Y \mid T=t, \boldsymbol{X}, S(t)=s)$, leading to a fully identifiable quantity when Assumptions 1-4 are satisfied.  
\subsection{Proof of Theorem 4.1}

    The fact that $M_{t_0, \boldsymbol{x}}(\boldsymbol{S})\mid S(t_0),\boldsymbol{X}$ and $\boldsymbol{S}\mid S(t_0),\boldsymbol{X}$ follow the same distribution implies $M_{t_0, \boldsymbol{x}}(\boldsymbol{S})(t_0) = S(t_0)$. Let $\boldsymbol{S}'$ denote $M_{t_0, \boldsymbol{x}}(\boldsymbol{S})$ and let $p(Y(t)=y \mid S(t_0), \boldsymbol{X})$ denote the conditional density of $Y(t)$ evaluated at $y$. Additionally, let $p_{\epsilon(t_0)}$ represent the density of the random error term. 
\begin{equation*}
    \begin{split}
        p(Y'(t_0)=y_{t_0} \mid S(t_0),\boldsymbol{X})&=\int p(Y'(t_0)=y_{t_0}\mid \boldsymbol{S}=\boldsymbol{s},\boldsymbol{X})dP_{\boldsymbol{S}\mid S(t_0),\boldsymbol{X}}(\boldsymbol{s})\\
        &=\int p_{\epsilon(t_0)}(y_{t_0}-f(t_0,M_{t_0, \boldsymbol{x}}^{-1}(\boldsymbol{s}),\boldsymbol{X}))dP_{\boldsymbol{S}\mid S(t_0),\boldsymbol{X}}(\boldsymbol{s})\\
        &=\int p_{\epsilon(t_0)}(y_{t_0}-f(t_0,M_{t_0, \boldsymbol{x}}^{-1}(\boldsymbol{s}'),\boldsymbol{X}))dP_{\boldsymbol{S'}\mid S(t_0),\boldsymbol{X}}(\boldsymbol{s}')\\
        &=\int p_{\epsilon(t_0)}(y_{t_0}-f(t_0,\boldsymbol{s},\boldsymbol{X}))dP_{\boldsymbol{S}\mid S(t_0),\boldsymbol{X}}(\boldsymbol{s})\\
        &=P(Y(t_0)=y_{t_0}\mid S(t_0),\boldsymbol{X}).
    \end{split}
\end{equation*}

The third equality always holds since $\boldsymbol{S}'\mid S(t_0),\boldsymbol{X}$ and $\boldsymbol{S}\mid S(t_0),\boldsymbol{X}$ have the same distribution (note that we also change the notation from $\boldsymbol{s}$ to $\boldsymbol{s}'$). The fourth equality holds since we apply the change of variable. Note that the subscript with $\boldsymbol{x}$ emphasizes that it relates to $\boldsymbol{x}$, and the subscript with $t_0$ emphasizes that it relates to $t_0$ and $S_{t_0}$.

\subsection{Proof of theorem 4.2}
Before we start the proof, we first introduce one required mild condition.

\renewcommand{\theassumption}{A\arabic{assumption}}
\begin{assumption}\label{ass:mildConditionForOutcome}
Assume the collection of basis functions of $\boldsymbol{b}_\lambda$, $ s(t)\boldsymbol{b}_\psi$, and \\ $
\sum_{t'\in\mathcal{T}_d}\boldsymbol{b}_{\beta}(m(t',\boldsymbol{x})+K(t,t')(s(t)-m(t,\boldsymbol{x}))$, as functions of $(t,s(t),\boldsymbol{x})$, are linearly independent.
\end{assumption}

This assumption ensures that the design matrix used in the outcome regression model is full rank. Without such an assumption one would be able to write one covariate as a linear combination of the others, which would lead to a lack of identification of the corresponding regression parameters. One can also show that in most realistic settings, this is expected to hold. The linear independence of functions of $\boldsymbol{b}_\lambda$ and $ s(t)\boldsymbol{b}_\psi$ is provided by $ s(t)\boldsymbol{b}_\psi$ being a function of $s(t)$. The covariance function $K$, for example, can be an exponential function of $t$, which could also provide linear independence of the third set of basis functions and the previous two sets of basis functions, when considering spline basis functions. Thus, Assumption \ref{ass:mildConditionForOutcome} is mild and holds in most common situations.

    Let $t_i$ denote the observed values for $T_i$. By strong unconfoundedness, $E[Y(t_i)\mid S(t_i), \boldsymbol{X}, T=t_i]=E[Y(t_i)\mid S(t_i), \boldsymbol{X}]$, thus $E[Y(t_i)\mid S(t_i), \boldsymbol{X}]$ is identifiable for any $t_i\in\mathcal{T}$. Given the parametric outcome model and the Gaussian process for the potential intermediates, the conditional mean of $Y(t_i)$ given $S(t_i)$ and $\boldsymbol{X}$ is 
    {\small
\begin{equation*}
    \begin{split}
        E[Y(t_i)\mid S(t_i), \boldsymbol{X}]&=\lambda(t_i,\boldsymbol{X})+
        \psi(t_i)S(t_i)+
        \sum_{t'\in\mathcal{T}_d}\beta(t_i,t')E[S(t')\mid S(t_i),\boldsymbol{X}]\\
        &=\lambda(t_i,\boldsymbol{X})+
        \psi(t_i)S(t_i)+
        \sum_{t'\in\mathcal{T}_d}\beta(t_i,t')(m(t',\boldsymbol{X}) + K(t_i,t')(S(t_i)-m(t_i,\boldsymbol{X})))\\
        &=\boldsymbol{b}_\lambda^T\boldsymbol{\eta} + S(t_i)\boldsymbol{b}_\psi^T\boldsymbol{\zeta}_\psi+
        \sum_{t'\in\mathcal{T}_d}\boldsymbol{b}_{\beta}^T\boldsymbol{\zeta}_{\beta}(m(t',\boldsymbol{X})+K(t_i,t')(S(t_i)-m(t_i,\boldsymbol{X})).
    \end{split}
\end{equation*}
}

Since there are infinitely many $t_i \in \mathcal{T}$ and $E[Y(t_i)\mid S(t_i), \boldsymbol{X}]$ can be expressed as a linear function in terms of $(\boldsymbol{\eta}, \boldsymbol{\zeta}_\psi,\boldsymbol{\zeta}_{\beta})$, as long as the coefficient matrix is full rank, then $(\boldsymbol{\eta}, \boldsymbol{\zeta}_\psi,\boldsymbol{\zeta}_{\beta})$ can be expressed in terms of $E[Y(t_i)\mid S(t_i), \boldsymbol{X}]$ and are thus identifiable. A full-rank coefficient matrix is equivalent to the corresponding functions being linearly independent, which is exactly Assumption \ref{ass:mildConditionForOutcome}. 


\section{Proof of results on correlation parameter}
\label{sec:AppendixCorrProofs}

In this section, we provide technical details for the derivation of the marginal posterior of $\rho$, as well as a proof of the fact that this posterior is maximized at the true correlation parameter.

\subsection{Deriving marginal posterior of $\rho$}

Utilizing the notation that the potential intermediates are given by $\boldsymbol{S}_i(t) = \boldsymbol{S}_i$, and the observed intermediate is $S_i$, the marginal posterior distribution can be written as

\begin{align*}
    P(\rho \vert \mathcal{D}) &\propto P(\rho) \prod_{i=1}^n \Bigg\{ P \Big(Y_i \vert T_i, \boldsymbol{X}_i, \boldsymbol{S}_i, S_i, \rho \Big) P \Big(\boldsymbol{S}_i \vert T_i, S_i, \boldsymbol{X}_i, \rho \Big) \Bigg\} \\
    & \propto P(\rho) \Bigg\{ \prod_{i=1}^n |\widetilde{\boldsymbol{\Sigma}}_{si}|^{-1/2} \Bigg\} \ \text{exp} \Bigg\{- \frac{\sum_{i=1}^n (\widetilde{Y}_i - \boldsymbol{S}_i^T \boldsymbol{\beta}_i)^2}{2 \sigma^2} \Bigg\} \\
    & \times \ \text{exp} \Bigg\{- \frac{1}{2} \sum_{i=1}^n (\boldsymbol{S}_i - \widetilde{\boldsymbol{\mu}}_{si})^T \widetilde{\boldsymbol{\Sigma}}_{si}^{-1} (\boldsymbol{S}_i - \widetilde{\boldsymbol{\mu}}_{si}) \Bigg\} \\
    & \propto P(\rho) \Bigg\{ \prod_{i=1}^n |\widetilde{\boldsymbol{\Sigma}}_{si}|^{-1/2} \Bigg\} \\
    & \times \text{exp} \Bigg\{- \frac{1}{2} \sum_{i=1}^n -2 \boldsymbol{S}_i^T \bigg(\frac{\widetilde{Y}_i \boldsymbol{\beta}_i}{\sigma^2} + \widetilde{\boldsymbol{\Sigma}}_{si}^{-1} \widetilde{\boldsymbol{\mu}}_{si} \bigg) + \boldsymbol{S}_i^T \bigg(\frac{\boldsymbol{\beta}_i \boldsymbol{\beta}_i^T}{\sigma^2} + \widetilde{\boldsymbol{\Sigma}}_{si}^{-1} \bigg) \boldsymbol{S}_i + \widetilde{\boldsymbol{\mu}}_{si}^T \widetilde{\boldsymbol{\Sigma}}_{si} \widetilde{\boldsymbol{\mu}}_{si} \Bigg\}
\end{align*}

The left two terms in the exponential represent the kernel of a multivariate normal distribution for $\boldsymbol{S}_i$ with mean given by 
$$\boldsymbol{M} = \bigg(\frac{\boldsymbol{\beta}_i \boldsymbol{\beta}_i^T}{\sigma^2} + \widetilde{\boldsymbol{\Sigma}}_{si}^{-1} \bigg)^{-1} 
 \bigg(\frac{\widetilde{Y}_i \boldsymbol{\beta}_i}{\sigma^2} + \widetilde{\boldsymbol{\Sigma}}_{si}^{-1} \widetilde{\boldsymbol{\mu}}_{si} \bigg)$$ 
 and variance given by
 $$\boldsymbol{V} = \bigg(\frac{\boldsymbol{\beta}_i \boldsymbol{\beta}_i^T}{\sigma^2} + \widetilde{\boldsymbol{\Sigma}}_{si}^{-1} \bigg)^{-1}.$$
 With that in mind, we can multiply and divide this expression by the determinant term of the corresponding multivariate normal distribution, as well as adding and subtracting the relevant constants into the exponential term to get the full multivariate normal density with this mean and variance. The determinant term that we will multiply and divide by is given by
 \begin{align*}
     \Bigg\{ \prod_{i=1}^n \bigg|\bigg(\frac{\boldsymbol{\beta}_i \boldsymbol{\beta}_i^T}{\sigma^2} + \widetilde{\boldsymbol{\Sigma}}_{si}^{-1} \bigg)^{-1} \bigg|^{-1/2} \Bigg\} &= \Bigg\{ \prod_{i=1}^n \bigg|\bigg(\frac{\boldsymbol{\beta}_i \boldsymbol{\beta}_i^T}{\sigma^2} + \widetilde{\boldsymbol{\Sigma}}_{si}^{-1} \bigg) \bigg|^{1/2} \Bigg\}.
 \end{align*}
 The constant term that we will add and subtract to the exponential term of this posterior distribution is given by
 $$\bigg(\frac{\widetilde{Y}_i \boldsymbol{\beta}_i}{\sigma^2} + \widetilde{\boldsymbol{\Sigma}}_{si}^{-1} \widetilde{\boldsymbol{\mu}}_{si} \bigg)^T \bigg(\frac{\boldsymbol{\beta}_i \boldsymbol{\beta}_i^T}{\sigma^2} + \widetilde{\boldsymbol{\Sigma}}_{si}^{-1} \bigg)^{-1} \bigg(\frac{\widetilde{Y}_i \boldsymbol{\beta}_i}{\sigma^2} + \widetilde{\boldsymbol{\Sigma}}_{si}^{-1} \widetilde{\boldsymbol{\mu}}_{si} \bigg).$$
After the inclusion of these terms, we are left with
\begin{align*}
& P(\rho) \left\{ \prod_{i=1}^n \frac{|\widetilde{\boldsymbol{\Sigma}}_{si}^{-1}|^{1/2}}{\Bigg|\bigg(\frac{\boldsymbol{\beta}_i \boldsymbol{\beta}_i^T}{\sigma^2} + \widetilde{\boldsymbol{\Sigma}}_{si}^{-1} \bigg) \Bigg|^{1/2}} \right\} \\
&\times \ \text{exp} \Bigg\{ -\frac{1}{2} \Bigg[ \widetilde{\boldsymbol{\mu}}_{si}^T \widetilde{\boldsymbol{\Sigma}}_{si} \widetilde{\boldsymbol{\mu}}_{si} - \bigg(\frac{\widetilde{Y}_i \boldsymbol{\beta}_i}{\sigma^2} + \widetilde{\boldsymbol{\Sigma}}_{si}^{-1} \widetilde{\boldsymbol{\mu}}_{si} \bigg)^T \bigg(\frac{\boldsymbol{\beta}_i \boldsymbol{\beta}_i^T}{\sigma^2} + \widetilde{\boldsymbol{\Sigma}}_{si}^{-1} \bigg)^{-1} \bigg(\frac{\widetilde{Y}_i \boldsymbol{\beta}_i}{\sigma^2} + \widetilde{\boldsymbol{\Sigma}}_{si}^{-1} \widetilde{\boldsymbol{\mu}}_{si} \bigg) \Bigg] \Bigg\} \\
&\times \ \prod_{i=1}^n  f(\boldsymbol{S}_i ; \boldsymbol{M}, \boldsymbol{V})
\end{align*}

where $f(\boldsymbol{S}_i ; \boldsymbol{M}, \boldsymbol{V})$ represents the density of a normal distribution with mean $\boldsymbol{M}$ and variance $\boldsymbol{V}$ evaluated at $\boldsymbol{S}_i$. Only the final term contains $\boldsymbol{S}_i$ and it is a density function, so if we integrate this expression with respect to $\boldsymbol{S}_i$, the final term integrates to 1 and we are left with the remaining terms. To obtain the final desired result, we can use a result on determinants of matrices that are the sum of a positive definite and a rank 1 matrix to see that
\begin{align*}
    \Bigg| \bigg(\frac{\boldsymbol{\beta}_i \boldsymbol{\beta}_i^T}{\sigma^2} + \widetilde{\boldsymbol{\Sigma}}_{si}^{-1} \bigg) \Bigg| =  |\widetilde{\boldsymbol{\Sigma}}_{si}^{-1}| \bigg( 1 + \frac{\boldsymbol{\beta}_i^T \widetilde{\boldsymbol{\Sigma}}_{si} \boldsymbol{\beta}_i}{\sigma^2} \bigg)
\end{align*}
and therefore we have that
\begin{align*}
    \left\{ \prod_{i=1}^n \frac{|\widetilde{\boldsymbol{\Sigma}}_{si}^{-1}|^{1/2}}{\Bigg|\bigg(\frac{\boldsymbol{\beta}_i \boldsymbol{\beta}_i^T}{\sigma^2} + \widetilde{\boldsymbol{\Sigma}}_{si}^{-1} \bigg) \Bigg|^{1/2}} \right\} &= \left\{ \prod_{i=1}^n \frac{|\widetilde{\boldsymbol{\Sigma}}_{si}^{-1}|^{1/2}}{\Bigg( |\widetilde{\boldsymbol{\Sigma}}_{si}^{-1}| \bigg( 1 + \frac{\boldsymbol{\beta}_i^T \widetilde{\boldsymbol{\Sigma}}_{si} \boldsymbol{\beta}_i}{\sigma^2} \bigg) \Bigg)^{1/2}} \right\} \\
    &= \prod_{i=1}^n \bigg( 1 + \frac{\boldsymbol{\beta}_i^T \widetilde{\boldsymbol{\Sigma}}_{si} \boldsymbol{\beta}_i}{\sigma^2} \bigg)^{-1/2}
\end{align*}
Combining this, with the results from above, we can see that the marginal posterior becomes
\begin{align*}
& P(\rho) \left\{ \prod_{i=1}^n \bigg( 1 + \frac{\boldsymbol{\beta}_i^T \widetilde{\boldsymbol{\Sigma}}_{si} \boldsymbol{\beta}_i}{\sigma^2} \bigg)^{-1/2} \right\} \\
&\times \ \text{exp} \Bigg\{ -\frac{1}{2} \Bigg[ \widetilde{\boldsymbol{\mu}}_{si}^T \widetilde{\boldsymbol{\Sigma}}_{si} \widetilde{\boldsymbol{\mu}}_{si} - \bigg(\frac{\widetilde{Y}_i \boldsymbol{\beta}_i}{\sigma^2} + \widetilde{\boldsymbol{\Sigma}}_{si}^{-1} \widetilde{\boldsymbol{\mu}}_{si} \bigg)^T \bigg(\frac{\boldsymbol{\beta}_i \boldsymbol{\beta}_i^T}{\sigma^2} + \widetilde{\boldsymbol{\Sigma}}_{si}^{-1} \bigg)^{-1} \bigg(\frac{\widetilde{Y}_i \boldsymbol{\beta}_i}{\sigma^2} + \widetilde{\boldsymbol{\Sigma}}_{si}^{-1} \widetilde{\boldsymbol{\mu}}_{si} \bigg) \Bigg] \Bigg\}
\end{align*}

\subsection{Identification of correlation parameters}
The additional mild condition required for identification is given in Assumption \ref{ass:mildConditionForCorrelation}. 
Let $G=\{ \boldsymbol{\beta}^T_i\boldsymbol{\Sigma}_S \boldsymbol{\beta}_i,\boldsymbol{\beta}_i^T\boldsymbol{\Sigma}_{i.}:i=1,...,n\}$, which is a set of functions of $\rho$.
\renewcommand{\theassumption}{B\arabic{assumption}}
\begin{assumption}\label{ass:mildConditionForCorrelation}
For any $\rho_0\neq \rho^*$, there exists $g(\rho)\in G$ such that $g(\rho_0)\neq g(\rho^*)$. 
\end{assumption}
While Assumption \ref{ass:mildConditionForCorrelation} is a slightly more technical condition on the values of $\boldsymbol{\beta}_i$, in practice it effectively amounts to $\beta(t,t')\equiv0$ for $t'\in\mathcal{T}_d$, which occurs when principal ignorability holds. 

Given an improper flat prior, Theorem 6.1 is equivalent to showing that
\begin{equation*}
    \rho^\ast=\argmax_\rho \sum_{i=1}^n E[\log 
    p(Y_i,S_i|T_i,\boldsymbol{X}_i)],
\end{equation*}
under some mild conditions, where $p(\cdot|\cdot)$ denotes the conditional probability density function of $Y_i,S_i\mid T_i,\boldsymbol{X}_i$. First we derive the closed form of the conditional probability density function of $Y_i,S_i\mid T_i,\boldsymbol{X}_i$. Since $Y_i\mid T_i,\boldsymbol{S}_i,S_i,\boldsymbol{X}$ and $(\boldsymbol{S}_i,S_i)\mid T_i,\boldsymbol{X}$ both follow Gaussian distributions, then $(Y_i,\boldsymbol{S}_i,S_i)\mid T_i,\boldsymbol{X}$ jointly follows a multivariate Gaussian distribution. Marginally, $(Y_i,S_i)\mid T_i,\boldsymbol{X}$ also follows a bivariate Gaussian distribution. To derive the pdf of $(Y_i,S_i)\mid T_i,\boldsymbol{X}$, $p(y_i,s_i\mid \rho)$, we only need to obtain the mean vector and covariance matrix. For simplicity, we assume that we always condition on $(T_i,\boldsymbol{X}_i)$, and thus, we omit it in the expressions. Let $\boldsymbol{\mu}_{si} = E\boldsymbol{S}_i$ and $\mu_{si}=ES_i$. For the marginal conditional means, we have
\begin{equation*}
    \begin{split}
        E Y_i &= \lambda(T_i,\boldsymbol{X}_i)+\psi(T_i)\mu_{si}+\boldsymbol{\beta}_i^T\boldsymbol{\mu}_{si}, \\
        E S_i &=\mu_{si}.
    \end{split}
\end{equation*}

For the variance of the outcome, we have the following:
\begin{equation*}
    \begin{split}
        Var(Y_i)&=E[E[(Y_i-E(Y_i))^2\mid S_i,\boldsymbol{S_i}]]\\
        &=\sigma^2 + \psi(T_i)^2\sigma^2_S + 2 \psi(T_i)\boldsymbol{\beta}_i^T\boldsymbol{\Sigma}_{i.} + \boldsymbol{\beta}_i^T\boldsymbol{\Sigma}_S \boldsymbol{\beta}_i.
    \end{split}
\end{equation*}
To find the covariance, we can use the tower rule:
\begin{equation*}
    \begin{split}
        Cov(Y_i,S_i)&=E[(Y_i-EY_i)(S_i-ES_i)]\\
        &=E\left[E[(Y_i-EY_i)(S_i-ES_i)\mid S_i]\right]\\
        &= E[\left(\psi(T_i)(S_i-\mu_{si})+\boldsymbol{\beta_i}^T\boldsymbol{\Sigma}_{i.}(S_i-\mu_{si})/\sigma^2_S\right)(S_i-\mu_{si})]\\
        &=E[\left(\psi(T_i)(S_i-\mu_{si})+\boldsymbol{\beta_i}^T\boldsymbol{\Sigma}_{i.}(S_i-\mu_{si})/\sigma^2_S\right)(S_i-\mu_{si})]\\
        &=\psi(T_i)\sigma^2_S+\boldsymbol{\beta}_i^T\boldsymbol{\Sigma}_{i.}.
    \end{split}
\end{equation*}
Lastly, we have that $Var(S_i)=\sigma^2_S$.

Let $\psi_i=\psi(T_i)$, $\tau_i = \boldsymbol{\beta}_i^T\boldsymbol{\Sigma}_S \boldsymbol{\beta}_i$, and $\xi_i=\boldsymbol{\beta_i}^T\boldsymbol{\Sigma}_{i.}$. For simplicity, we assume $E(Y_i)=E(S_i)=0$ and $\sigma_S^2=1$; otherwise, we replace $(Y_i,S_i)$ with $(Y_i-E(Y_i),(S_i-E(S_i))/\sigma_S)$.
Then, the log-likelihood of $(Y_i,S_i)\mid \rho$ is 
\begin{equation*}
    \begin{split}
        l_i(\rho) &= -\frac{1}{2}\log(2\pi(\sigma^2 + \tau_i-\xi_i^2 ))  - \frac{Y_i^2-2(\psi_i+\xi_i)Y_iS_i+(\sigma^2+\psi_i^2+\psi_i\xi_i+\tau_i)S_i^2}{2(\sigma^2 + \tau_i-\xi_i^2 )}.
    \end{split}
\end{equation*}

Next, 
\begin{equation*}
    \begin{split}
        E(l_i(\rho))&=-\frac{1}{2}\log(2\pi(\sigma^2 + \tau_i-\xi_i^2 ))-\\
        &\qquad\frac{(\sigma^2+\psi_i^2+\psi_i\xi_i+\tau_i)^{\ast}-2(\psi_i+\xi_i)(\psi_i+\xi_i)^\ast+(\sigma^2+\psi_i^2+\psi_i\xi_i+\tau_i)}{2(\sigma^2 + \tau_i-\xi_i^2 )},
    \end{split}
\end{equation*}
where $(\cdot)^\ast$ denotes the true value of the parameter $(\cdot)$. The above expression can be simplified as 
\begin{equation*}
    \begin{split}
        E(l_i(\rho))&=-\frac{1}{2}\log(2\pi(\sigma^2 + \tau_i-\xi_i^2 ))-\\
        &\qquad\frac{1}{2}\left(\frac{(\sigma^2 + \tau_i-\xi_i^2)^{\ast}}{(\sigma^2 + \tau_i-\xi_i^2 )}+
        \frac{((\psi_i+\xi_i)-(\psi_i+\xi_i)^\ast)^2}{(\sigma^2 + \tau_i-\xi_i^2 )}+1\right).
    \end{split}
\end{equation*}
To further simplify the notation, let $\nu_i=\sigma^2 + \tau_i-\xi_i^2$ and $\omega_i=\psi_i+\xi_i$. Then,
\begin{equation*}
    \begin{split}
        E(l_i(\rho))&=-\frac{1}{2}\log(2\pi\nu_i)-\frac{1}{2}(\frac{\nu_i^{\ast}}{\nu_i}+
        \frac{(\omega_i-\omega_i^\ast)^2}{\nu_i}+1).
    \end{split}
\end{equation*}
The first derivative of $E(l_i(\rho))$ with respect to $\rho$ is 
\begin{equation*}
    \begin{split}
        \frac{\partial E(l_i(\rho))}{\partial \rho}&=-\frac{1}{2}\left(\frac{\nu_i'}{\nu_i}-\frac{\nu_i^{\ast}\nu_i'}{\nu_i^2}-\frac{(\omega_i-\omega_i^\ast)^2\nu_i'}{\nu_i^2}+2\frac{(w_i-w_i^\ast)w_i'}{\nu_i}\right),
    \end{split}
\end{equation*}
where $\nu_i'=\frac{\partial\nu_i}{\partial\rho}$ and $\omega_i'=\frac{\partial\omega_i}{\partial\rho}$. One can check check that $(\frac{\partial E(l_i(\rho))}{\partial \rho})\big|_{\rho=\rho^\ast}=0$, and check that the second derivative $(\frac{\partial^2 E(l_i(\rho))}{\partial \rho^2})\big|_{\rho=\rho^\ast}=(-\frac{\nu_i'^2}{2\nu_i^2}-\frac{\omega_i'^2}{\nu_i})\big|_{\rho=\rho^\ast}$ is negative. Thus, $\rho^\ast$ is the local maximum.

By Jensen's inequality, for any $\rho\neq\rho^\ast$, we have
\begin{equation*}
    E\left[ \log\left(\frac{p(Y_i,S_i|\rho)}{p(Y_i,S_i|\rho^\ast)}\right)\right] \leq \log \left[E\left( \frac{p(Y_i,S_i|\rho)}{p(Y_i,S_i|\rho^\ast)}\right)\right]=0.
\end{equation*}
This implies that $\rho^\ast=\argmax E\log(p(Y_i,S_i|\rho))$ and
\begin{equation*}
    \rho^\ast=\argmax_\rho \sum_{i=1}^n E[\log 
    p(Y_i,S_i|T_i,\boldsymbol{X}_i)].
\end{equation*}

Note that $G=\{\tau_i,\ \xi_i:i=1,\dots,n\}$. If Assumption \ref{ass:mildConditionForCorrelation} holds, then Jensen's inequality holds with strict inequality for some $i$, leading to $\rho^\ast$ being the unique global maximizer.

\section{Details of MCMC sampling}
\label{sec:AppendixSampling}

In this section, we describe how to update each of the model parameters parameters separately within a Gibbs sampler. Note that the exact updates for each outcome model parameter will depend on how that model is specified. In this section, we show updates for a model of the form:
$$E[Y \mid T=t, \boldsymbol{S} = \boldsymbol{s}, \boldsymbol{X} = \boldsymbol{x}] = \lambda(t) + \boldsymbol{x}\boldsymbol{\gamma} + \psi(t)s(t)+\sum_{t'\in\mathcal{T}_d} \beta(t,t')s(t').$$
Throughout, we adopt the notation that $\boldsymbol{\beta}_i = (\beta(T_i, t_1), \dots, \beta(T_i, t_M))$.
\begin{itemize}
    \item First, we can update mean function of the Gaussian process using SoftBART (\cite{linero2018bayesian}), which also updates $\sigma_S^2$.

    \item Now, we can update $\boldsymbol{S}_i = (S_i(t^{(1)}), \dots, S_i(t^{(M)}))$, the unobserved values of the intermediate at the locations of $t$ that we are interested in. First, we can define $Y_i^* = Y_i - \psi(T_i)S_i- \lambda(T_i) - \boldsymbol{X}_i \boldsymbol{\gamma}$. Next define the mean and variance of the conditional distribution of the potential intermediate, $\boldsymbol{S}_i$, given the observed intermediate, $S_i$, as
    \begin{align*}
\widetilde{\boldsymbol{\mu}}_{si} &=  m(\boldsymbol{t}, \boldsymbol{X}_i) + \boldsymbol{\Sigma}_{i.} (S_i - m(T_i, \boldsymbol{X}_i)) / \sigma_S^2 \\
\widetilde{\boldsymbol{\Sigma}}_{si} &= \boldsymbol{\Sigma}_S - \boldsymbol{\Sigma}_{i.} \boldsymbol{\Sigma}_{i.}^T / \sigma_S^2
\end{align*}
where here $\boldsymbol{\Sigma}_{i.}$ is a vector of values given by $K(T_i, \boldsymbol{t})$. We can now update $\boldsymbol{S}_i$ from its conditional distribution given by
    $$\boldsymbol{S}_i \mid \cdot \sim \mathcal{N}\Bigg(\bigg( \frac{\boldsymbol{\beta}_i^T \boldsymbol{\beta}_i}{\sigma^2} + \widetilde{\boldsymbol{\Sigma}}_{si}^{-1} \bigg)^{-1} \bigg( \frac{Y_i^* \boldsymbol{\beta}_i^T}{\sigma^2} + {\widetilde{\boldsymbol{\Sigma}}_{si}}^{-1} \widetilde{\boldsymbol{\mu}}_{si} \bigg), \bigg( \frac{\boldsymbol{\beta}_i^T \boldsymbol{\beta}_i}{\sigma^2} + \widetilde{\boldsymbol{\Sigma}}_{si}^{-1} \bigg)^{-1} \Bigg)$$

\item To update the effect of the treatment on the outcome, we assume that $\lambda(t) = \sum_{j=1}^{J_2} m_j(t) \delta_j$. We denote the $n \times J_2$ matrix of these basis functions evaluated at the observed $T_i$ values by $\boldsymbol{M}$. We then update the parameters from
\begin{align*}
    \boldsymbol{\delta} \mid \cdot \sim \mathcal{N} \Bigg( \bigg( \frac{\boldsymbol{M}^T \boldsymbol{M}}{\sigma^2} + \boldsymbol{\Sigma}_{\delta}^{-1} \bigg)^{-1} \bigg( \frac{\boldsymbol{M}^T \boldsymbol{Y}^*}{\sigma^2} + \boldsymbol{\Sigma}_{\delta}^{-1} \boldsymbol{\mu}_{\delta} \bigg), \bigg( \frac{\boldsymbol{M}^T \boldsymbol{M}}{\sigma^2} + \boldsymbol{\Sigma}_{\delta}^{-1} \bigg)^{-1} \Bigg)
\end{align*}
where $\boldsymbol{Y}^*$ is a vector of values of $Y_i - \psi(T_i)S_i- \boldsymbol{X}_i \boldsymbol{\gamma} - \boldsymbol{S}_i \boldsymbol{\beta}_i$
\item In order to update $\boldsymbol{\gamma}$ we first can define $\boldsymbol{Y}^*$ as a vector of values of $Y_i - \psi(T_i)S_i - \boldsymbol{M}_i \boldsymbol{\delta} - \boldsymbol{S}_i \boldsymbol{\beta}_i$. Then we have
\begin{align*}
    \boldsymbol{\gamma} \mid \cdot \sim \mathcal{N} \Bigg( \bigg( \frac{\boldsymbol{X}^T \boldsymbol{X}}{\sigma^2} + \boldsymbol{\Sigma}_{\gamma}^{-1} \bigg)^{-1} \bigg( \frac{\boldsymbol{X}^T \boldsymbol{Y}^*}{\sigma^2} + \boldsymbol{\Sigma}_{\gamma}^{-1} \boldsymbol{\mu}_{\gamma} \bigg), \bigg( \frac{\boldsymbol{X}^T \boldsymbol{X}}{\sigma^2} + \boldsymbol{\Sigma}_{\gamma}^{-1} \bigg)^{-1} \Bigg)
\end{align*}
\item If we let the prior distribution for $\sigma^2$ be an inverse-gamma distribution with parameters $a_0$ and $b_0$, then we can update it from an inverse-gamma distribution with parameters $a^*$ and $b^*$ defined by
\begin{align*}
    a^* &= a_0 + \frac{n}{2} \\
    b^* &= b_0 + \frac{{\boldsymbol{Y}^*}^T \boldsymbol{Y}^*}{2}
\end{align*}
where $\boldsymbol{Y}^*$ is a vector of values of $Y_i - \psi(T_i)S_i - \boldsymbol{M}_i \boldsymbol{\delta} - \boldsymbol{X}_i \boldsymbol{\gamma} - \boldsymbol{S}_i \boldsymbol{\beta}_i$

\item In order to update the parameters that dictate both $\psi(t)$ and $\beta(t, t')$ we first can define $\boldsymbol{Y}^*$ as a vector of values of $Y_i - \boldsymbol{M}_i \boldsymbol{\delta} - \boldsymbol{X}_i \boldsymbol{\gamma}$. We also will be using the specification that $\beta(t,t') = \zeta_1 (t-t') + \zeta_2 (t-t')^2$ and $\psi(t) = \iota_0+\iota_1t+\iota_2t^2$ for this derivation, but other functional forms for $\beta(t, t')$ and $\psi(t)$ would apply analogously. It is helpful to write the following:
\begin{align*}
    Y_i^* &= \psi(T_i)S_i +  \sum_{t'} S_i(t') \beta(T_i, t') + \epsilon_i \\
    &= \iota_0S_i + \iota_1T_iS_i + \iota_2T_i^2S_i + \sum_{t'} S_i(t') \left(\zeta_1 (T_i-t') + \zeta_2 (T_i-t')^2\right) + \epsilon_i \\
    & = \iota_0S_i(T_i) + \iota_1T_iS_i(T_i) + \iota_2T_i^2S_i(T_i) + \zeta_1\sum_{t'}S_i(t')(T_i-t') + \\
    &\qquad \zeta_2\sum_{t'}S_i(t')(T_i-t')^2 + \epsilon_i\\
    &= \boldsymbol{W}_i \boldsymbol{\zeta} + \epsilon_i
\end{align*}
Then, if we let $\boldsymbol{\zeta}=(\iota_0,\iota_1,\iota_2,\zeta_1,\zeta_2)^T$ and $\boldsymbol{W}$ represent the $n \times 5$ matrix, where each row corresponds to $\boldsymbol{W}_i$ as above, then we can update the parameters from
\begin{align*}
    \boldsymbol{\zeta} \mid \cdot \sim \mathcal{N} \Bigg( \bigg( \frac{\boldsymbol{W}^T \boldsymbol{W}}{\sigma^2} + \boldsymbol{\Sigma}_{\zeta}^{-1} \bigg)^{-1} \bigg( \frac{\boldsymbol{W}^T \boldsymbol{Y}^*}{\sigma^2} + \boldsymbol{\Sigma}_{\zeta}^{-1} \boldsymbol{\mu}_{\zeta} \bigg), \bigg( \frac{\boldsymbol{W}^T \boldsymbol{W}}{\sigma^2} + \boldsymbol{\Sigma}_{\zeta}^{-1} \bigg)^{-1} \Bigg)
\end{align*}
Here, we are using a multivariate normal prior distribution for $\boldsymbol{\zeta}$ with mean $\boldsymbol{\mu}_{\zeta}$ and covariance $\boldsymbol{\Sigma}_{\zeta}$. 

\item To update the smoothness parameter $\rho$ of the kernel, there is no conjugate prior so we utilize a Metropolis Hastings strategy to updating this parameter. We assume a gamma prior with parameters $a_0$ and $b_0$. We propose a new value centered at the previous one plus normally distributed noise. If our previous value was $\rho$, and the proposal value is $\rho'$, then we calculate 
$$r = \min \Bigg\{ 1, \frac{P(\rho' \vert \mathcal{D}) T(\rho, \rho')}{P(\rho \vert \mathcal{D}) T(\rho', \rho)} \Bigg\}$$
where we let $P(\cdot \vert \mathcal{D})$ be the posterior of $\rho$ conditional on the data and all other unknown parameters, except for $\boldsymbol{S}_i$, which has been integrated out. Similarly, we let $T(\rho, \rho')$ be the transition probability of proposing $\rho'$ from $\rho$. The transition probabilities will cancel out in our case due to our proposal being symmetric, unless the algorithm approaches the boundary value of $0$, in which case a truncated distribution needs to be utilized and the transition probabilities adapted accordingly. We then choose $\rho'$ with probability $r$ and $\rho$ with probability $1-r$.

\end{itemize}

\section{Modeling of potential intermediate incorporating monotonicity and clustering}

Here we specify an alterative approach to modeling of $m(t, \boldsymbol{x})$ that incorporates additional features one may be interested in for certain applications. For this model, we first simplify the mean function as
$$m(t, \boldsymbol{x}) = \phi(t) + \boldsymbol{x \alpha}.$$
Note that nonlinear effects of the covariates in $\boldsymbol{x}$ could be easily incorporated here as well, but our focus will be on the $\phi(\cdot)$ component of the model that shows how the potential intermediate variable is related to the treatment level $t$. Additionally, this is the part of the model where we can impose a monotonicity constraint, as well as clustering to identify different units in terms of how their intermediate variable is affected by treatment.

There is an extensive literature on Bayesian isotonic regression, which can be used to ensure that $\phi(t)$ is increasing in $t$. Similar to \cite{neelon2004bayesian} we let this function be piecewise linear, and we additionally incorporate nonparametric Bayesian prior distributions on the coefficients to induce clustering. This is somewhat related to the nonparametric prior distributions used in \cite{wilson2020bayesian}, however, clustering there was done at the coefficient level, while our prior will induce clustering at the unit level. Specifically, we represent this function using the basis function representation $\phi(t) = \sum_{j=1}^J d_j(t) \theta_j = \boldsymbol{d}(t) \boldsymbol{\theta}.$ If we have knot locations $\kappa_1, \dots, \kappa_{J-1}$, then we can define basis functions as $d_1(t) = t, d_2(t) = (t-\kappa_1)_+, \dots, d_J(t) = (t-\kappa_{J-1})_+$ to obtain a piecewise linear function. To ensure monotonicity of this function, we need $\theta_1 \geq 0, \theta_1 + \theta_2 \geq 0, \dots, \theta_1 + \dots +\theta_J \geq 0$. Alternatively, we need $\boldsymbol{A \theta} \geq \boldsymbol{0}$, where
\begin{equation*}
\boldsymbol{A} = 
\begin{pmatrix}
1 & 0 & 0 &\cdots & 0 \\
1 & 1 & 0 &\cdots & 0 \\
\vdots  & \vdots & \cdots & \ddots & \vdots  \\
1 & 1 & 1 &\cdots & 1 
\end{pmatrix}
.
\end{equation*}
We can re-write our basis expansion as $\phi(t) = \boldsymbol{d}(t) \boldsymbol{A}^{-1} \boldsymbol{A} \boldsymbol{\theta} = \boldsymbol{b}(t) \boldsymbol{\eta}$, where $\boldsymbol{\eta} = \boldsymbol{A} \boldsymbol{\theta}$. To ensure monotonicity, we simply must restrict that $\eta_j \geq 0$ for all $j$. We can enforce this through our prior distribution, by specifying a truncated multivariate normal distribution for $\boldsymbol{\eta}$ that truncates all values to be on the positive real line. Additionally, we can incorporate an intercept as the first basis function, but we do not care whether the intercept is positive or negative, and therefore it will not be truncated below at zero. 

To achieve clustering and model flexibility, we allow the parameters dictating this association, given by $\boldsymbol{\eta}$, to vary by individual and have the following hierarchical formulation:
\begin{align*}
    \boldsymbol{\eta}_i &\sim G, \quad
    G = \sum_{c=1}^{\infty} \pi_c \delta_{\boldsymbol{\xi}_c}, \quad
    \boldsymbol{\xi}_c \sim G_0 \\
    \pi_1 &= V_1, \quad
    \pi_c = V_c \prod_{j < c} (1 - V_j), \quad V_j \sim Beta(1, \kappa)
\end{align*}
This is equivalent to say that $G \sim DP(\kappa, G_0)$, or that $G$ follows a Dirichlet process with concentration parameter $\kappa$ and base distribution $G_0$ \citep{ferguson1973bayesian, sethuraman1994constructive}. We use a truncated multivariate normal distribution for $G_0$ to guarantee monotonicity as described above. In practice, we can set an upper value $C$ on the number of mixture components to approximate the infinite mixture model. A well-known consequence of this Dirichlet process prior is that it induces clustering as some observations will have the same value of $\boldsymbol{\eta}_i$. The size and number of clusters is dictated by the concentration parameter $\kappa$, as smaller values of $\kappa$ will lead to fewer clusters with more observations in them. This is important in our setting as we are aiming to find which observations are in each principal strata, and this nonparametric prior distribution allows observations to have differential effects of the treatment on the potential intermediate, which will lead to them being members of different strata. An additional modification that we choose to make to this prior is to set $\boldsymbol{\xi}_1 = \boldsymbol{0}$, which ensures that the first cluster will have a flat $\phi(t)$ function. This cluster represents observations for which the treatment does not affect the potential intermediate variable, which represents an important principal strata of interest. Lastly, we place a gamma prior distribution on $\kappa$ to let the data inform the degree of clustering. 

Note that there are inherent trade-offs with using this approach compared to the fully nonparametric modeling approach described in the manuscript. If monotonicity is expected to hold, then this approach could improve efficiency in the estimation of the model for the intermediate variable. This model is much simpler, however, and will not perform as well if the true relationships between $t$ and $\boldsymbol{x}$ on the intermediate $S$ are more complex, with nonlinear functions of the covariates, or interactions between $t$ and $\boldsymbol{x}$ not being captured. Combinations of the two that allow for more flexible associations, while still enforcing monotonicity, are certainly possible and would be of interest, though this is beyond the scope of this manuscript. 

\subsection{Update Steps for MCMC Sampling}
In here, we describe how to update each of the model parameters parameters separately within a Gibbs sampler for the model described above that enforces monotonicity. Many of the parameter updates, particularly for the outcome model, are the same as for the Gibbs sampler described in Appendix C and therefore we omit them. Instead, we only include updates unique to this particular modeling strategy. Throughout, we adopt the notation that $\phi_i(t) = \boldsymbol{b}(t) \boldsymbol{\eta}_i$, where $\boldsymbol{\eta}_i = \sum_{c} \boldsymbol{\xi}_c 1(z_i = c)$ and $z_i$ is a latent membership indicator for the stick breaking prior used for the intermediate model. 
\begin{itemize}
    \item We first update the parameters of the mean function of the Gaussian process. First, we can update $\boldsymbol{\alpha}$. Define $\tilde{\boldsymbol{X}}_i$ be the $(M+1) \times (p+1)$ matrix where each row is $\boldsymbol{X}_i$, where here we include the intercept into $\boldsymbol{X}_i$. Now, we can let $\boldsymbol{\tilde{S}}_i = [\boldsymbol{S}_i, S_i]$ and their corresponding treatment values be given by $\boldsymbol{\tilde{t}}_i = [\boldsymbol{t}, T_i]$ Further, define $\boldsymbol{S}_i^* = \boldsymbol{\tilde{S}}_i - \phi_i(\boldsymbol{\tilde{t}}_i)$. We can then update $\boldsymbol{\alpha}$ from the following conditional distribution:
    \begin{align*}
        \boldsymbol{\alpha} \mid \cdot \sim \mathcal{N} \Bigg(\bigg( \sum_{i=1}^n  \tilde{\boldsymbol{X}}_i^T \boldsymbol{\Sigma}_{S,i} \tilde{\boldsymbol{X}}_i \bigg)^{-1} \bigg( \sum_{i=1}^n  \tilde{\boldsymbol{X}}_i^T \boldsymbol{\Sigma}_{S,i} \tilde{\boldsymbol{S}}_i^* \bigg) , \bigg( \sum_{i=1}^n  \tilde{\boldsymbol{X}}_i^T \boldsymbol{\Sigma}_{S,i} \tilde{\boldsymbol{X}}_i \bigg)^{-1} \Bigg)
    \end{align*}
    where $\boldsymbol{\Sigma}_{S,i}$ is the $(M+1) \times (M+1)$ matrix of values of the kernel function evaluated at each combination of $\boldsymbol{\tilde{t}}_i$.
\item We assume an inverse-gamma prior distribution for $\sigma_{S}^2$ with parameters $a_0$ and $b_0$. The full conditional is also an inverse-gamma distribution with parameters
\begin{align*}
    a^* &= a_0 + \frac{n \times (\text{dim}(\boldsymbol{\beta}) + 1)}{2} \\
    b^* &= b_0 + \frac{ \sum_{i=1}^n (\boldsymbol{\tilde{S}}_i - \tilde{\boldsymbol{X}}_i \boldsymbol{\alpha} - \phi_i(\boldsymbol{\tilde{t}}_i))^T \boldsymbol{\Sigma}_{K,i}^{-1} (\boldsymbol{\tilde{S}}_i - \tilde{\boldsymbol{X}}_i \boldsymbol{\alpha} - \phi_i(\boldsymbol{\tilde{t}}_i))}{2},
\end{align*}
where $\boldsymbol{\Sigma}_{K,i} = \frac{1}{\sigma_S^2} \boldsymbol{\Sigma}_{S,i}$. 
\item Now, we must sample the parameters of the stick breaking prior for the regression coefficients, $\boldsymbol{\eta}_i$. First, we need to update the weights $V_j$, which are updated as
$$V_j | \cdot \sim Beta(1 + \sum_{i} 1(z_i = j), \kappa + \sum_{i} 1(z_i > j))$$
Additionally, we need to sample the $z_i$ indicators of group membership. To do so, we sample from the conditional distribution of $z_i$ after $\boldsymbol{S}_i$ has been integrated out to improve computation. 
\begin{align*}
    & P(z_i = c \mid \cdot) \propto \pi_c \ \text{exp} \Bigg\{ -\frac{1}{2} \Bigg[ 
\frac{(S_i - m^{(c)}(T_i, \boldsymbol{X}_i))^2}{\sigma_S^2} +  \Bigg] \Bigg\} \\
& \times \text{exp} \Bigg\{ -\frac{1}{2} \Bigg[ \widetilde{\boldsymbol{\mu}}_{si}^{T(c)} \widetilde{\boldsymbol{\Sigma}}_{si} \widetilde{\boldsymbol{\mu}}_{si}^{(c)} - \bigg(\frac{\widetilde{Y}_i \boldsymbol{\beta}_i}{\sigma^2} + \widetilde{\boldsymbol{\Sigma}}_{si}^{-1} \widetilde{\boldsymbol{\mu}}_{si}^{(c)} \bigg)^T \bigg(\frac{\boldsymbol{\beta}_i \boldsymbol{\beta}_i^T}{\sigma^2} + \widetilde{\boldsymbol{\Sigma}}_{si}^{-1} \bigg)^{-1} \bigg(\frac{\widetilde{Y}_i \boldsymbol{\beta}_i}{\sigma^2} + \widetilde{\boldsymbol{\Sigma}}_{si}^{-1} \widetilde{\boldsymbol{\mu}}_{si}^{(c)} \bigg) \Bigg] \Bigg\}
\end{align*}
Where the $(c)$ superscript denotes that the mean vectors use the parameters dictating the relationship between treatment and intermediate that correspond to the $c^{th}$ group
\item Now we need to update the $\boldsymbol{\eta}_i$ parameters, which is analogous to updating the $\boldsymbol{\xi}_c$ parameters for $c=1, \dots, C$, since $\boldsymbol{\eta}_i = \sum_{c} \boldsymbol{\xi}_c 1(z_i = c)$. As in the manuscript, we define our basis functions as $\boldsymbol{b}(t) = \boldsymbol{d}(t) \boldsymbol{A}^{-1}$. Our model is therefore given by $\phi_i(t) = \boldsymbol{b}(t) \boldsymbol{\eta}_i$ and we require that $\eta_j \geq 0$ for all $j$ to ensure monotonicity of the function. Our prior distribution for $\boldsymbol{\xi}_c$ is assumed to be a truncated normal with mean $\boldsymbol{\mu}_{\xi}$ and covariance $\boldsymbol{\Sigma}_{\xi}$, that is truncated so that all parameters are greater than or equal to zero. We can define  $\boldsymbol{S}_i^* = \boldsymbol{\tilde{S}}_i - \tilde{\boldsymbol{X}}_i \boldsymbol{\alpha}$. Further, we can create the $(M+1) \times J$ matrix $\boldsymbol{B}_i$, where the $k^{th}$ row corresponds to $[b_1(\tilde{t}_{ik}), \dots, b_J(\tilde{t}_{ik})]$. We can now update the parameters from a truncated normal distribution given by
\begin{align*}
    \boldsymbol{\xi}_c \mid \cdot \sim  \mathcal{N}_+ \Bigg( &\bigg( \boldsymbol{\Sigma}_{\xi}^{-1}  + \sum_{i: z_i = c} \boldsymbol{B}_i \boldsymbol{\Sigma}_{S,i}^{-1} \boldsymbol{B}_i \bigg)^{-1} \bigg( \boldsymbol{\Sigma}_{\xi}^{-1} \boldsymbol{\mu}_{\xi} + \sum_{i: z_i = c} \boldsymbol{B}_i \boldsymbol{\Sigma}_{S,i}^{-1} \boldsymbol{S}_i^* \bigg) ,  \\
    & \bigg( \boldsymbol{\Sigma}_{\xi}^{-1}  + \sum_{i: z_i = c} \boldsymbol{B}_i \boldsymbol{\Sigma}_{S,i}^{-1} \boldsymbol{B}_i \bigg)^{-1} \Bigg)
\end{align*}
\item We assign a Gamma prior with parameters $a_{\kappa}$ and $b_{\kappa}$ for $\kappa$, the concentration parameter of the Dirichlet process. Updating $\kappa$ utilizes the latent variable approach described in Section 6 of \cite{escobar1995bayesian}.
\end{itemize}

\section{Positivity assessment and trimming estimators}

The only one of our identifying assumptions that can be explored empirically is the positivity assumption, which states that the conditional density of the treatment given covariates is positive for all possible covariate values at the treatment levels of interest. By treatment values of interest, we refer to the values of $t$ under which we aim to estimate $E(Y(t))$, which in our applied example is a sequence of values from the 10th to 90th quantiles of the observed treatment values. More specifically, let $f_{t|x}(t \mid \boldsymbol{x})$ represent the density of $t$ given $\boldsymbol{X} = \boldsymbol{x}$. For any value of $t$ on the exposure-response curve, we require that
$$f_{t|x}(t \mid \boldsymbol{X}) > 0 \quad \text{with probability 1}.$$
Positivity violations would occur if there are covariate values for which the conditional density is zero.  For binary treatments, a common way to assess positivity violations is to first estimate the propensity score, and then compare the distribution of propensity scores for both treated and control units to ensure that these two distributions are mostly overlapping. Such approaches are not well-established for continuous treatments, but recent work has explored implementing variations of this strategy by binning the continuous treatment into a finite number of levels. 

Throughout the rest of this section, we adopt the approach seen in \cite{flores2012estimating} to evaluating the positivity assumption through overlap of the resulting generalized propensity score (GPS, \cite{imai2004causal}). First, we categorize the continuous treatment into a finite set of values using quintiles (or deciles) of the empirical distribution of $T$.  Let $Q_i \in \{1, \ldots,  5\}$ be the quintile that observation $i$ belongs to, and let $t_q$ be the median value of the treatment within quintile $q$. For each quintile $q$, we compute the value of the GPS at $t_q$, which is given by 
$$\widehat{R}_i^q = \widehat{f}_{t|x}(t_q \mid \boldsymbol{X}_i).$$
Given these values, one can check for overlap of the generalized propensity scores in a similar manner as to what is done for binary treatments. Within each quintile, one can separately examine histograms of the GPS among observations for which $Q_i = q$, and among observations for which $Q_i \neq q$. This can be done for each quintile and then one can examine the extent to which these histograms are overlapping across the quintiles. Another, related metric that can be examined is referred to as the common support subsample with respect to quintile $q$. Intuitively, this represents the subset of the observations that have GPS values in the region where the two histograms described above are overlapping. Mathematically, this set of observations can be defined as
$$	CS_q = \left\{i:  \widehat{R}_i^q \in 
	\left[\max\{\min_{i': Q_{i'}=q}\widehat{R}_{i'}^q, 
	\min_{i': Q_{i'}\neq q}\widehat{R}_{i'}^q\},
	\min\{\max_{i': Q_{i'}=q}\widehat{R}_{i'}^q, 
	\max_{i': Q_{i'}\neq q}\widehat{R}_{i'}^q\}
	\right] \right\}$$
This common support subsample is unique to quintile $q$, but one can also define an overall subsample as $CS = \bigcap_{q=1}^5 	CS_q$. If a very large portion of the data is not included in this set, that would provide evidence of a lack of overlap. 

Additionally, one can use this common support subsample to construct a variation of trimming-based estimators applied to our setting. Propensity score trimming has been used widely in binary treatment settings \citep{crump2009dealing}, and has been recently extended to continuous treatments as well \citep{branson2023causal}. By propensity score trimming, we refer to approaches that restrict estimation to subsets of the population for which the (generalized) propensity score is sufficiently large, and discard observations for which the (generalized) propensity score is small. This changes the estimand being targeted, but can provide much more stable results in cases with positivity violations. Given that in the continuous treatment setting, there is no well-established metric to suggest positivity violations are, or are not, a concern, we can implement a trimming-based estimator and examine whether results vary significantly from the results using the full sample. If results are very similar, this would provide evidence that positivity was not greatly impacting our original findings. In our setting, we will be trimming by conditioning on observations that fall in the common support subsample. Note that the common support subsample is defined by a subject having a generalized propensity score within a particular range across all quintiles, and this can be viewed as a function of the covariates $\boldsymbol{X}$ since the GPS is itself a function of the covariates. Let us define a function $l(X)$ which returns a 1 if that value of $X$ leads to the GPS being inside the common support region and 0 otherwise. The trimming estimand we are targeting in this case is given by
$$E(Y(t) \mid a < g(\boldsymbol{S}) < b, l(\boldsymbol{X}) = 1).$$
Under the same arguments used in Appendix A, we can see that under our standard assumptions, this quantity can be written as
\begin{align*}
    \int_{\boldsymbol{x}} \int_{\boldsymbol{s}} E(Y \mid T=t, \boldsymbol{X}, \boldsymbol{S}=\boldsymbol{s}) \frac{f(\boldsymbol{s} \mid \boldsymbol{X}=\boldsymbol{x}) 1(a < g(\boldsymbol{s}) < b)}{P(a < g(\boldsymbol{S}) < b \vert \boldsymbol{X}=\boldsymbol{x})} d \boldsymbol{s} \  dF_{\boldsymbol{X} \mid a < g(\boldsymbol{S}) < b, l(\boldsymbol{X}) = 1}(\boldsymbol{x}),
\end{align*}
where the only difference is in the distribution of $\boldsymbol{X}$ that we integrate over, which now restricts to those observations in the common support subsample. By restricting to these observations for which positivity is not an issue, we avoid making predictions from the outcome model for combinations of the treatment and covariates that are not well-supported by the data, and therefore this approach should be more robust to issues from positivity violations. 

\subsection{Implementation on policing analysis}

\begin{figure}[ht]
    \centering
    \includegraphics[width=0.99\linewidth]{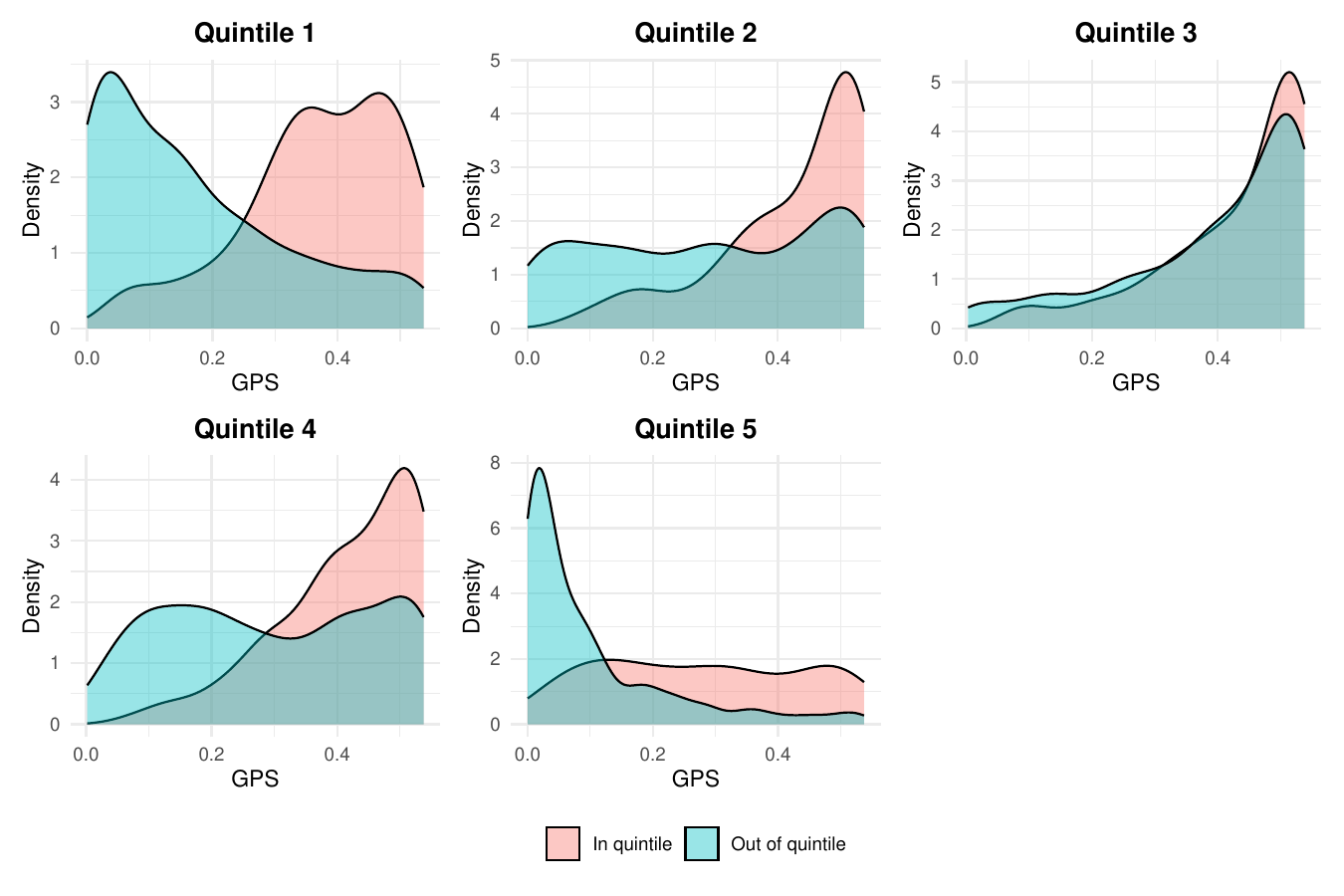}
    \caption{Densities of the estimated GPS under each of five quintiles. The two densities in each panel correspond to the density of the GPS both for observations that fall in that quintile, and for those that do not.}
    \label{fig:AppendixOverlap}
\end{figure}

In this section, we illustrate the aforementioned ideas for positivity assessment on the policing analysis from the manuscript to examine the impact of potential positivity violations on our findings. First, we assess overlap of the estimated generalized propensity score, and then we use these to construct the common support subsample and implement a trimming estimator. To estimate the GPS, we assume that the conditional distribution of the treatment given covariates follows a normal distribution with a linear function of the covariates for the mean, and a constant variance. Figure \ref{fig:AppendixOverlap} shows the density of the estimated GPS for observations that fall within each quintile compared with the density for observations that do not fall within a quintile. As expected, the GPS for observations that fall within a particular quintile are higher than those outside of that quintile, however, there is fairly good overlap of the estimated histograms. The worst overlap occurs in the more extreme values of the treatment, which can be found in the first and last quintiles, however, even at these more extreme values there is a fairly large degree of overlap in the two histograms. This suggests that positivity violations are not a major concern within our analysis. 

\begin{figure}[htbp!]
    \centering
    \includegraphics[width=0.95\linewidth]{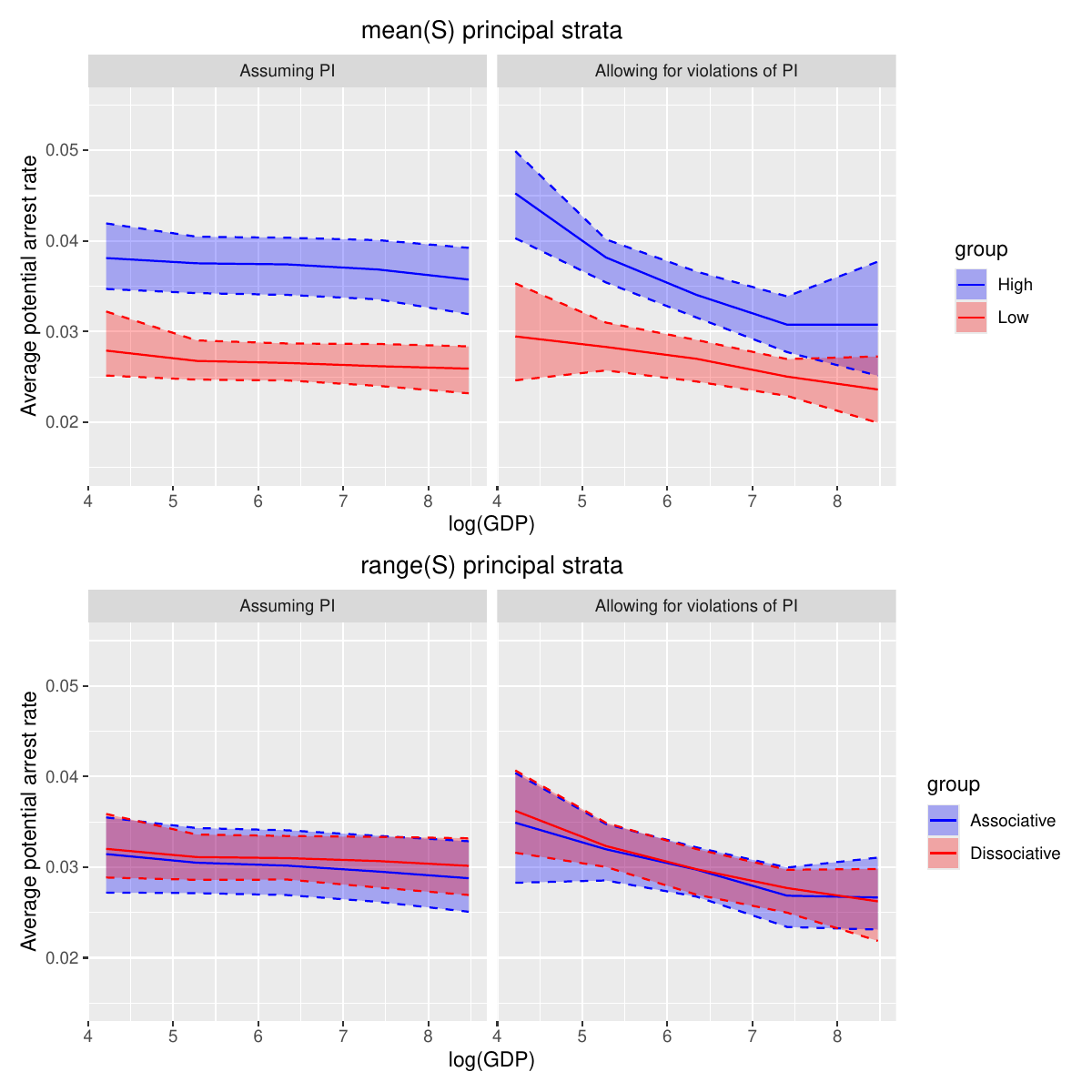}
    \caption{Exposure-response curves across principal strata when implementing trimming during estimation. The left column corresponds to results that assume principal ignorability, while the right column are results when it is allowed to be violated. The first row looks at principal strata defined by mean$(\boldsymbol{S})$, and the second row looks at principal strata defined by range$(\boldsymbol{S})$.}
    \label{fig:AppendixResultsPositivity}
\end{figure}

While positivity violations do not appear to be a large problem in our analysis, there is still a possibility that mild positivity violations could impact our findings. Due to this, we can implement the trimming estimator based on the common support subsample, which focuses on observations with covariate values that are well-supported at all exposure values of interest, and therefore are less affected by positivity issues. In total, 60\% of the observations fall into the common support subsample, and most of those that were discarded were due to the first or last quintile.  The results can be found in Figure \ref{fig:AppendixResultsPositivity}, which show the same exposure-response curves as the manuscript, with the one exception being that trimming is applied. We see that the same substantive conclusions are found in the analysis that implements trimming as the results without trimming. There is a clear separation between the high and low groups for the mean$(\boldsymbol{S})$ principal strata groups and very little difference between the associative and dissociative groups. As with the results in the manuscript, the results also differ between the model assuming principal ignorability and the one that does not. The model assuming principal ignorability has relatively flat exposure-response curves across all principal strata. The model allowing for violations of principal ignorability has a steeper, downward exposure-response curve that is most pronounced in the high mean$(\boldsymbol{S})$ principal strata group. 

\section{Illustrations highlighting failures of principal ignorability}

In this section, we provide additional intuition about when principal ignorability can be violated by highlighting a common example using unmeasured confounding of the relationship between the intermediate and outcome. Note that for simplicity in this discussion, we omit observed covariates $\boldsymbol{X}$, though all ideas would hold conditionally on observed covariates as well. We assume that the unmeasured confounder is normally distributed with variance 1. The scale of the unmeasured variance is arbitrary, so we set it to 1 for simplicity of calculations. Throughout this section we assume the following data generating models for the potential intermediates and potential outcomes:
\begin{align*}
    S(t) &= \alpha_{su} U + \epsilon(t) \\
    Y(t) &= \beta_{ys} S(t) + \beta_{yu}U + \eta(t).
\end{align*}
We assume that the errors $\epsilon(t)$ and $\eta(t)$ both follow mean-zero Gaussian processes and are mutually independent of $U$, $S(t)$, and $Y(t)$. This highlights a situation where only $S(t)$ affects $Y(t)$, but there is an unmeasured confounder of the $S-Y$ relationship. This situation can be depicted by the directed acyclic graph (DAG) in Figure \ref{fig:DAGunmeasured}.

First, we can show how unmeasured confounding in this situation leads to principal ignorability not holding, despite the fact that only $S(t)$ shows up in the generative model for $Y(t)$. To show this, we simply must show that $E(Y(t) \mid S(t) = s(t)) \neq E(Y(t) \mid S(t) = s(t), \boldsymbol{S} = \boldsymbol{s})$. First, we can see that
\begin{align*}
    E(Y(t) \mid S(t) = s(t)) &= E(\beta_{ys} S(t) + \beta_{yu}U + \eta(t) \mid S(t) = s(t)) \\
    &= \beta_{ys} s(t) + \beta_{yu} E(U \mid S(t) = s(t)).
\end{align*}
By similar logic, it holds that
$$E(Y(t) \mid S(t) = s(t), \boldsymbol{S} = \boldsymbol{s}) = \beta_{ys} s(t) + \beta_{yu} E(U \mid S(t) = s(t), \boldsymbol{S} = \boldsymbol{s}).$$
Therefore we can focus on showing what the conditional expectation of $U$ is given the intermediate variables. An important point making this calculation more straightforward is that because the intermediate conditional on $U$ is Gaussian, and $U$ is Gaussian, they jointly follow a Gaussian distribution. We will focus here on a finite set of grid points as in the manuscript and therefore $\boldsymbol{S}$ is finite-dimensional and therefore the joint distribution of $(U, S(t), \boldsymbol{S})$ is a multivariate normal. Using properties of the multivariate normal distribution, we can see that
\begin{align*}
    E(U \mid S(t) = s(t)) &= \frac{\text{Cov(U, S(t))} s(t)}{\text{Var}(S(t))} \\
    &= \frac{\alpha s(t)}{\sigma_S^2 + \alpha^2}
\end{align*}
This shows unsurprisingly that when there is an unmeasured confounder, we obtain biased estimates of the effect of $S(t)$ on $Y$, and the magnitude of bias in the coefficient is given by $(\alpha \beta_{yu}) / (\sigma_S^2 + \alpha^2)$. Now we can perform a similar calculation when conditioning on both $S(t)$ and $\boldsymbol{S}$. To do so, we define $\widetilde{\boldsymbol{S}}  =[S(t), \boldsymbol{S}]$ to be the vector concatenating these two variables.
\begin{align*}
    E(U \mid S(t) = s(t), \boldsymbol{S} = \boldsymbol{s}) = \text{Cov}(U, \widetilde{\boldsymbol{S}}) \text{Var}(\widetilde{\boldsymbol{S}})^{-1} \widetilde{\boldsymbol{s}}.
\end{align*}
While writing down an explicit expression for the coefficients for each element in $\widetilde{\boldsymbol{S}}$ is difficult in general due to the presence of $\text{Var}(\widetilde{\boldsymbol{S}})^{-1}$, it is clear from this expression that the coefficients are nonzero for $\boldsymbol{S}$. This confirms that the presence of the unmeasured confounder leads to violations of principal ignorability in this setting. Interestingly, it also shows that the outcome model will satisfy the linear, additive structure that we have placed in our manuscript for the effect of $\boldsymbol{S}$ on $Y$. This is a commonly assumed model for unmeasured confounding and is much weaker than assuming no unmeasured confounding, and it falls within the parametric framework that we have presented throughout for addressing violations of principal ignorability.

\begin{figure}[ht]
\begin{center}
\begin{tikzpicture}[ node distance=2.5cm and 2.5cm, every node/.style={draw, circle, minimum size=1cm}, >=stealth ] 
\node[draw, circle] (T) at (0,0) {$T$}; 
\node[draw, circle] (S) at (3,0) {$S$}; 
\node[draw, circle] (Y) at (6,0) {$Y$};
\node[draw, circle] (U) at (1.5,-1.5) {$U$};
\draw[->] (T) -- (S); 
\draw[->] (S) -- (Y); 
\draw[->] (T) to[bend left=35] (Y); 
\draw[->] (U) -- (S); 
\draw[->] (U) -- (Y); 
\end{tikzpicture}
\end{center}
\caption{Causal diagram illustrating the relationship between the unmeasured confounder and the observed variables. }
\label{fig:DAGunmeasured}
\end{figure}

\section{Simulations under model misspecification}

In this section we consider the situation where principal ignorability is violated, but where our proposed model is misspecified and the linear-additive relationship between $\boldsymbol{S}$ and the outcome does not hold. We generate covariates and treatment in the same manner as in Section 7.2 of the manuscript. Similarly, we also set $\rho = 3$ and the mean function of the intermediates to be $m(t,\boldsymbol{x})=0.3t+0.4tx_4+0.7tx_2$. Much of the outcome data generating process is the same as we set $\lambda(t,\boldsymbol{x}) = t-0.3x_2-0.2x_3+0.3x_5$, $\psi(t)=0.5t$, and $\beta(t,t')=0.05(t-t')+0.025(t-t')^2$. We also assume a constant residual variance of 1. This simulation differs in that the data are generated with the following mean function:
$$E[Y \mid T=t, \boldsymbol{S} = \boldsymbol{s}, \boldsymbol{X} = \boldsymbol{x}] = \lambda(t,\boldsymbol{x}) + \psi(t)s(t)+\sum_{t'\in\mathcal{T}_d} \beta(t,t')s(t') +\sum_{t'\in\mathcal{T}_d} \beta_2(t,t')s(t')^2,$$
where we now have included a quadratic function of $s(t')$. For this section, we let $\beta_2(t,t')=c(t-t')^2$ and we vary the magnitude of $c$ to increase the degree of model misspecification. Specifically, we vary $c \in \{ 0, 0.009, 0.012 \}$ and refer to these as no misspecification, mild misspecification, and large misspecification. The values of 0.009 and 0.012 are chosen so that the variability of the nonlinear component of the mean function is 25\% and 50\% of the variability of the linear component, respectively. We estimate the effects across 100 data sets and report posterior means across data sets for all metrics. Throughout we use the weighted posterior distribution that avoids issues due to MCMC chains getting stuck in suboptimal models. As in the manuscript, our goal is to estimate  1) $E(Y(t) \mid \text{range}(\boldsymbol{S}) < 1.5)$, and 2) $E(Y(t) \mid \text{range}(\boldsymbol{S}) >2.5)$, and we look at exposure values $t\in\{-1,-0.5,0,0.5,1\}$. 

\begin{figure}[htbp!]
    \centering

    \includegraphics[width=0.95\linewidth]{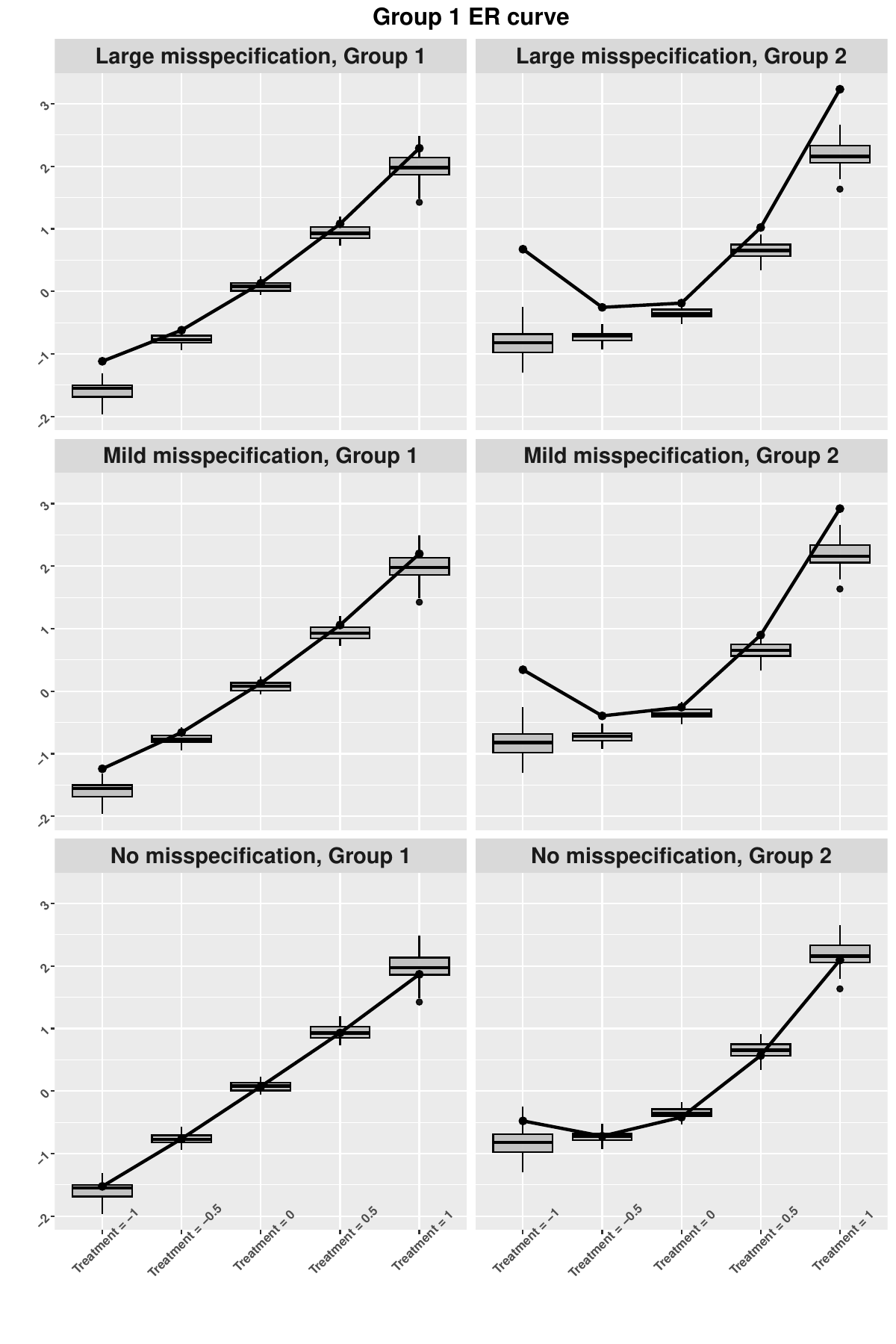}
    \caption{Estimates of the causal effect under differing degrees of model misspecification }
    \label{fig:SimMisspecified}
\end{figure}

The results for estimation of the exposure-response curves in each of the two principal strata considered are provided in Figure \ref{fig:SimMisspecified}. We see that results are relatively robust for the first principal strata considered, but large degrees of misspecification do lead to bias in the second principal strata considered, particularly at lower exposure values. 

\section{Additional data analysis results}

Here we present additional results beyond those seen in Section 8 of the manuscript. Specifically, we examine the relationship between GDP and police capacity from the fit of the SoftBart model for the intermediate, we assess the robustness of our findings to the choice of $\rho$ when assuming principal ignorability holds, we evaluate whether the size of the grid when discretizing $\boldsymbol{S}$ affects inferences, and we provide assessment of MCMC convergence through trace plots of key parameters under both model specifications. 

\subsection{Relationship between GDP and police capacity}

\begin{figure}[htbp!]
    \centering

    \includegraphics[width=0.5\linewidth]{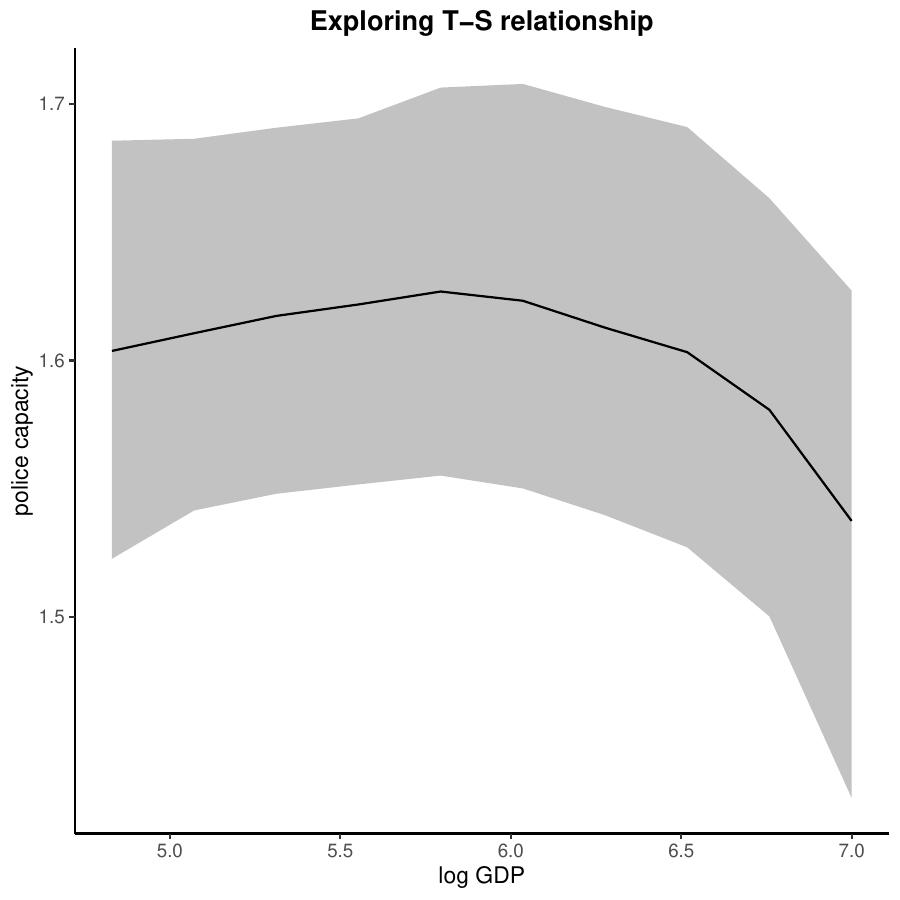}
    \caption{Estimates of the relationship between log GDP and police capacity after averaging over the covariate distribution. }
    \label{fig:TSrelationship}
\end{figure}

While the manuscript mostly focuses on the effect of GDP on police rates and how that is potentially modified by principal strata, here we illustrate how GDP is associated with police capacity. This is important because if police capacity is to mediate the effect of GDP on arrest rates, there must be an effect of GDP on police capacity first. As a reminder, we denote 
$$m(t, \boldsymbol{x}) = E(S(t) \mid T=t, \boldsymbol{X} = \boldsymbol{x}).$$
To examine how treatment affects the mediator, we can integrate over the covariate distribution at a range of potential values of the treatment. Specifically, for any treatment value $t$, we can calculate
$$\overline{m}(t) = \frac{1}{n} \sum_{i=1}^n m(t, X_i).$$
The point estimates and 95\% credible regions for this quantity, evaluated at a grid of values spanning the 25th and 75th quantiles of log GDP, can be seen in Figure \ref{fig:TSrelationship}. We can see that the curve is relatively flat until larger levels of GDP, where the curve trends downwards, though this does not appear significant given the widths of the posterior intervals. This can help explain why the associative and dissociative strata in the manuscript have similar exposure-response curves, as GDP does not strongly impact police capacity. 

\subsection{Sensitivity of results to choice of correlation parameter}

\begin{figure}[htbp!]
    \centering

    \includegraphics[width=0.85\linewidth]{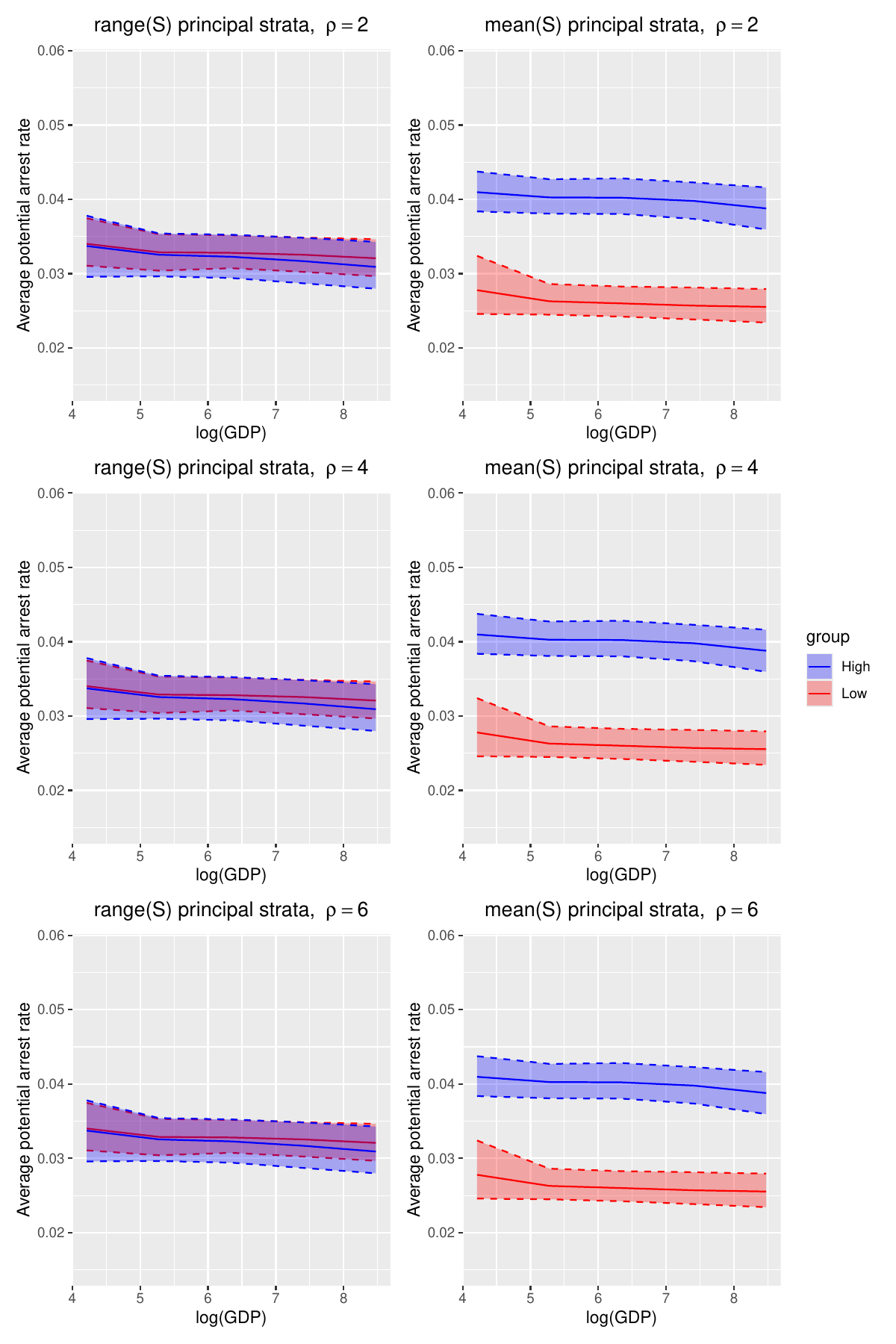}
    \caption{Sensitivity to the choice of $\rho$ when assuming principal ignorability. }
    \label{fig:SensitivityRho}
\end{figure}

When principal ignorability is assumed, we additionally rely on knowledge of the joint distribution of $\boldsymbol{S}$, which includes the correlation parameter dictating the correlation between $S(t)$ and $S(t')$ for any $t,t'$. Throughout the manuscript, we use a choice of $\rho = 4$, though here we evaluate other degrees of dependence by looking at $\rho=2$ and $\rho = 6$ to assess whether results change under reduced or increased levels of dependence. These can be found in Figure \ref{fig:SensitivityRho} where we present results for both the mean and range principal strata. We see that results are effectively indistinguishable between the different correlation specifications highlighting a degree of robustness to this choice within our analysis. 

\subsection{Sensitivity of results to grid size}
\label{App_grid}
\begin{figure}[htbp!]
    \centering

    \includegraphics[width=0.85\linewidth]{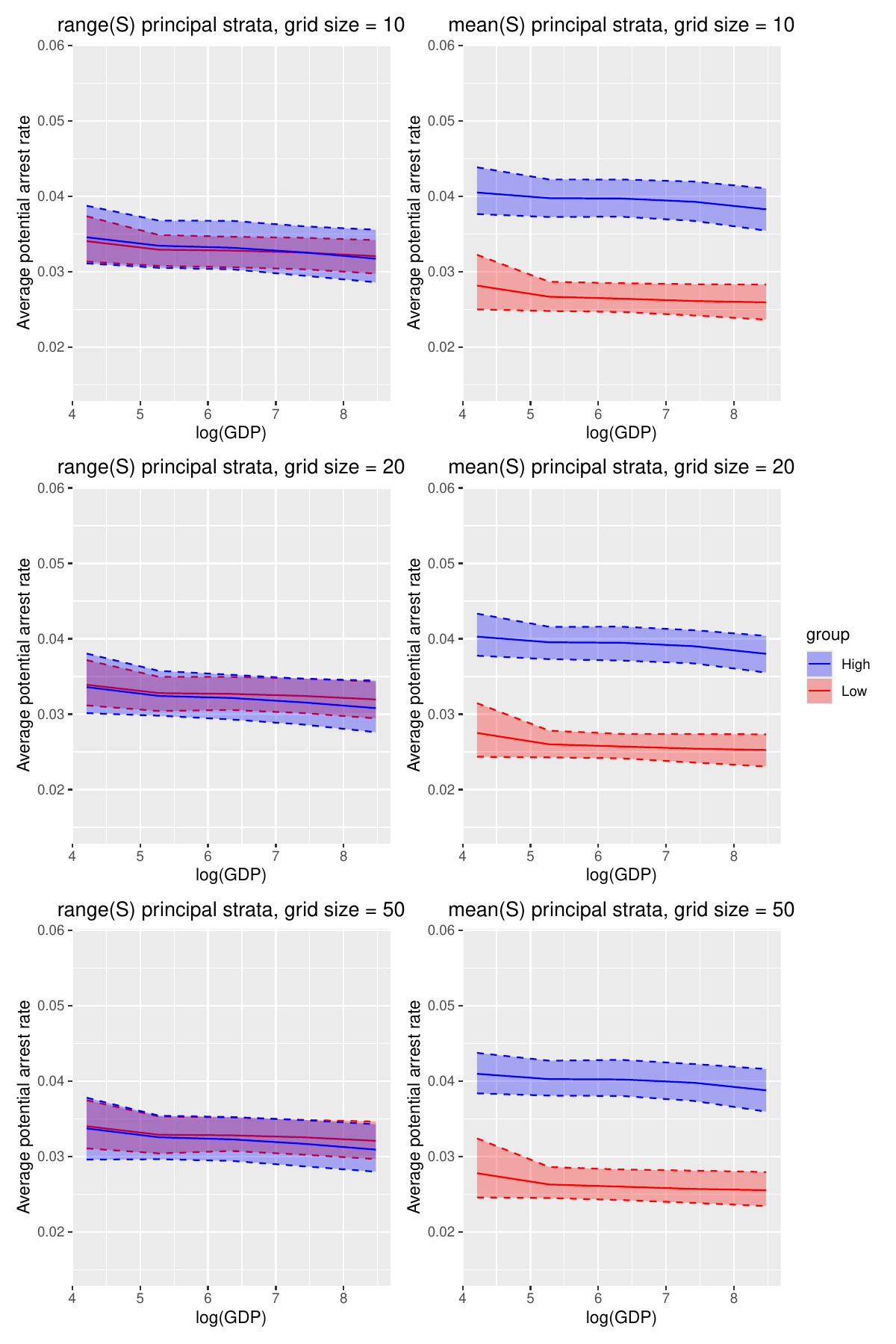}
    \caption{Sensitivity to the size of the discretization grid for models assuming principal ignorability.}
    \label{fig:SensitivityGridPI}
\end{figure}

\begin{figure}[htbp!]
    \centering

    \includegraphics[width=0.85\linewidth]{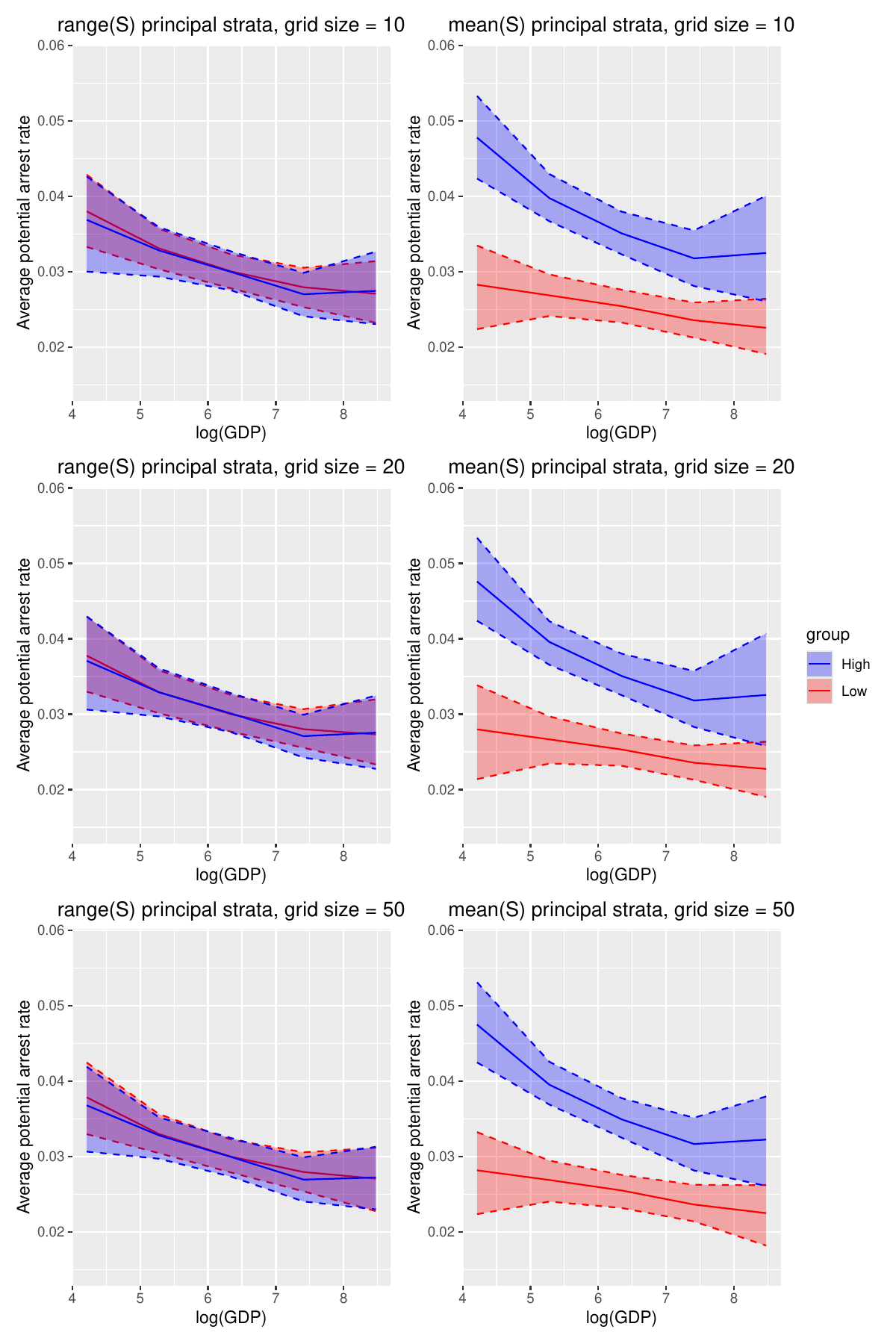}
    \caption{Sensitivity to the size of the discretization grid for models allowing for violations of principal ignorability.}
    \label{fig:SensitivityGridViolations}
\end{figure}

Throughout the manuscript, for all analyses we discretized the treatment-intermediate trajectories given by $\boldsymbol{S}$ into a finite set of locations for practical implementation and to increase computation speed. Here, we evaluate whether results are robust to this choice by examining grid sizes of 10, 20, and 50. The results for analyses assuming principal ignorability can be found in Figure \ref{fig:SensitivityGridPI}, and the results for analyses that allow for violations of principal ignorability are in Figure \ref{fig:SensitivityGridViolations}. The results are similar for both approaches, which is that the size of the grid plays effectively no role in the resulting causal effect estimates. This shows that as long as the grid is sufficiently large to cover the range of possible treatment values, increasing the denseness of the grid does not lead to materially different findings. For this reason, we recommend a choice of $|\mathcal{T}_d| = 10$ as it leads to similar results as larger choices, but is computationally faster.

\subsection{Convergence assessment}

\begin{figure}[htbp!]
    \centering

    \includegraphics[width=0.95\linewidth]{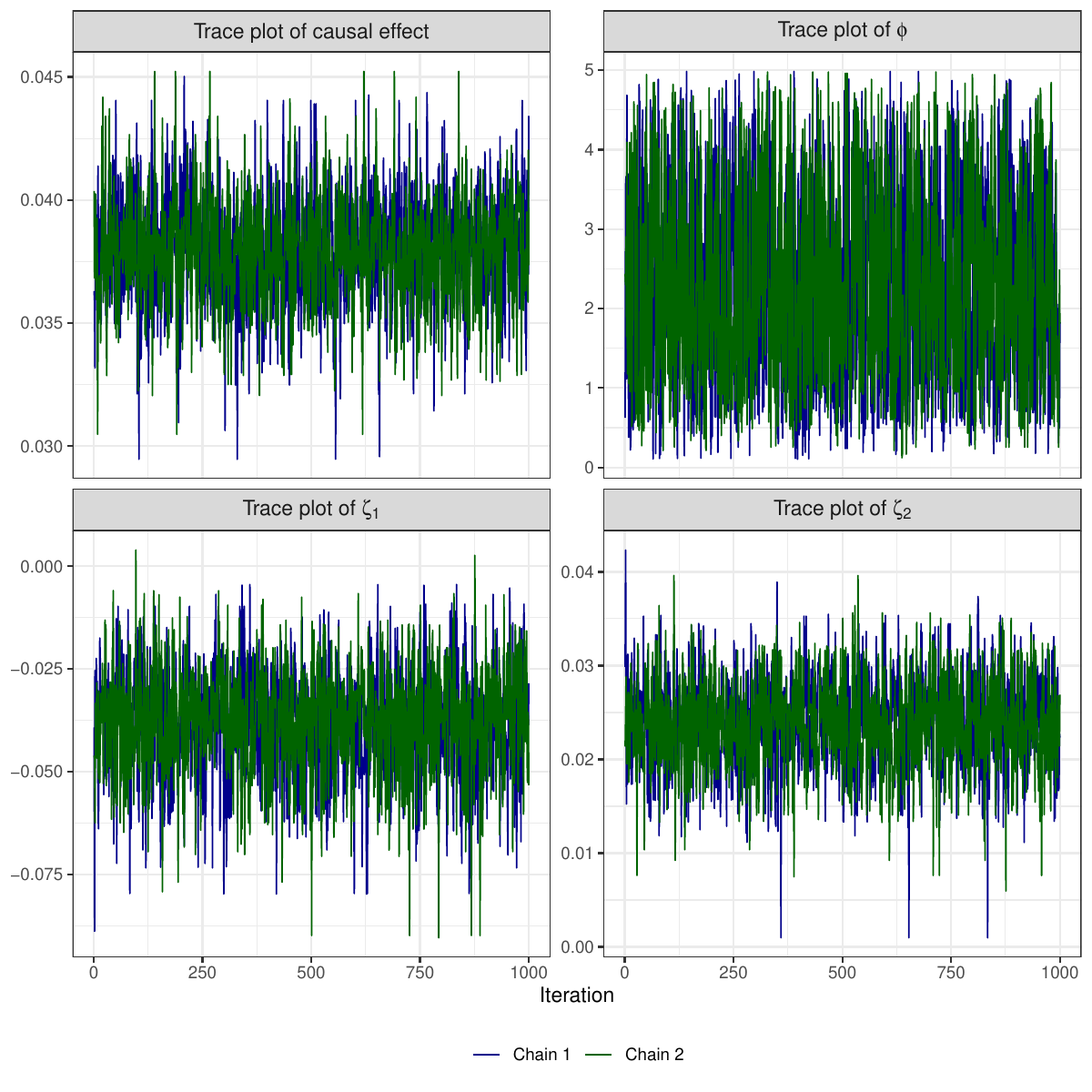}
    \caption{Trace plots for the model allowing for violations of principal ignorability. }
    \label{fig:TraceViolations}
\end{figure}

\begin{figure}[htbp!]
    \centering

    \includegraphics[width=0.95\linewidth]{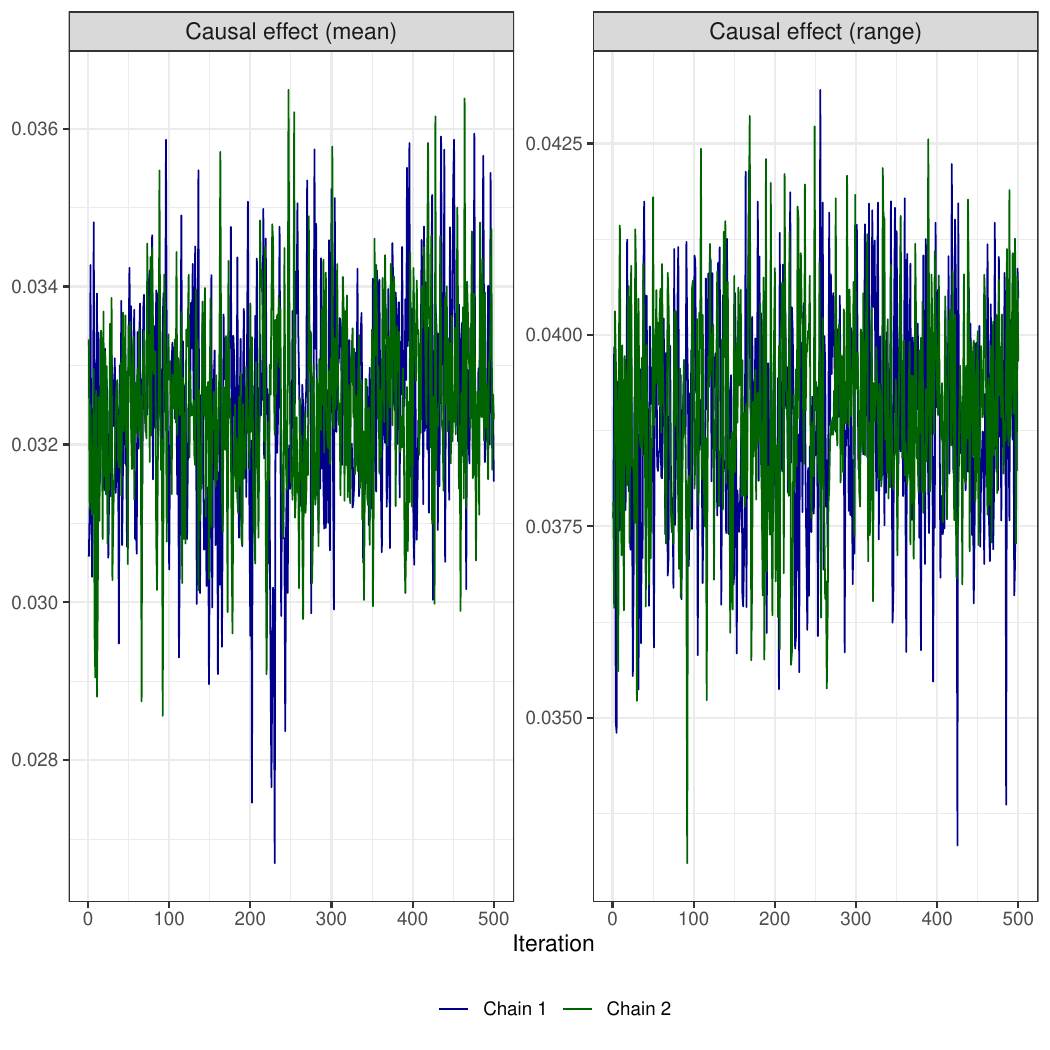}
    \caption{Trace plots for the model assuming principal ignorability holds.}
    \label{fig:TracePI}
\end{figure}

In this section, we provide trace plots for MCMC algorithms from both of the proposed approaches. For the approach allowing for violations of principal ignorability, we show trace plots for the model parameters given by $(\zeta_1, \zeta_2, \rho)$ as well as for $E(Y(t) \mid \text{mean}(\boldsymbol{S}) \leq c_m)$ for a single value of $t$. Similar traceplots were observed at all possible values of $t$ considered, as well as across all principal strata of interest, and therefore are omitted for space considerations. For the model assuming principal ignorability holds, we simply examine $E(Y(t) \mid \text{mean}(\boldsymbol{S}) \leq c_m)$ and $E(Y(t) \mid \text{range}(\boldsymbol{S}) \leq c_s)$ for the same choice of $t$. These results can be found in Figures \ref{fig:TraceViolations} and \ref{fig:TracePI}, where in both cases we see very similar behavior in both chains suggesting that the MCMC algorithm has converged. 

\section{Assessment of computation time}

In this section we present computation times for MCMC algorithms under both of the proposed identification and estimation strategies in order to assess how scalable the algorithms are to large data sets. Throughout, we keep track of the computation time required to run 1000 MCMC iterations. In practice, we have found that the MCMC algorithms have generally converged to the target posterior distribution by this point, though we recommend running as many MCMC samples as is feasible and to assess convergence on a case-by-case basis. The computation times under our proposed approach that allows for violations of principal ignorability, as well as the approach assuming principal ignorability, can be found in Table \ref{tab:computation_times}. One thing to note is that computation times are generally higher for the approach assuming principal ignorability, which may seem counterintuitive at first, but this difference arises because of differences in how the outcome model is estimated. Both approaches use a SoftBart model for the mean of the intermediate given treatment and covariates. For the outcome model, however, the approach assuming principal ignorability again uses a SoftBart model, while the approach allowing for violations uses the parametric model introduced in the manuscript. Fitting the SoftBart outcome model is actually quite fast for the sample sizes proposed. The computational burden comes because the calculation of the target estimand requires Monte Carlo integration that involves repeatedly drawing new values of $\boldsymbol{S}$, and calculating the outcome model at those values. It is very heavy computationally to repeatedly evaluate predictions from a SoftBart model at new values of the intermediate. This problem can be avoided using other approaches or flexible parametric models, but we use SoftBart due to its flexibility and robustness to model misspecification. The more parametric outcome model is much faster to produce predictions at new values of the inputs, and does not suffer from this issue. The main computational burden of the approach allowing for violations of PI is that at every MCMC iteration, we must impute the missing $\boldsymbol{S}_i$ values, which is a rather large number of calculations to perform per MCMC iteration. This is why the approach under violations of PI is more sensitive computationally to the size of the grid, $|\mathcal{T}_d|$, since it affects the size of the parameters being imputed. The model assuming PI is less affected by $|\mathcal{T}_d|$ as it does not require such difficult imputation in the outcome model MCMC algorithm. One last thing to note is that computation time appears to scale linearly with the sample size for both approaches. This shows that inference is possible for larger sizes than those seen here as computation time does not grow quickly with the sample size. Finally, these results, coupled with our results in Section \ref{App_grid} showing that results are not sensitive to the grid dimension $|\mathcal{T}_d|$, indicate that we should use a smaller value of $|\mathcal{T}_d|$, such as 10, when using the model that allows for violations of PI. Increasing the grid size further does not materially change the results, though does significantly increase computation time. 

\begin{table}[ht] 
\centering 
\begin{tabular}{ccccc | ccc} 
\hline
& & \multicolumn{3}{c}{Violations of PI} & \multicolumn{3}{c}{PI assumed} \\
\cline{3-5} \cline{6-8}  & & $|\mathcal{T}_d| =10$ & $|\mathcal{T}_d| =20$ & $|\mathcal{T}_d| =50$ & $|\mathcal{T}_d| =10$ & $|\mathcal{T}_d| =20$ & $|\mathcal{T}_d| =50$ \\
\hline $n = 200$ & & 4.48 & 11.99 & 31.65 & 21.76 & 34.13 & 33.33 \\
$n = 500$ & & 11.13 & 29.76 & 77.64 & 55.67 & 86.66 & 88.17 \\
$n = 1000$ & & 22.12 & 60.83 & 159.75 & 122.52 & 192.87 & 194.14 \\
$n = 2000$ & & 44.31 & 124.32 & 311.94 & 244.49 & 389.02 & 407.98 \\
\hline
\end{tabular} 
\caption{Computation times (in minutes) for estimation when PI is violated and when PI is assumed.} 
\label{tab:computation_times} 
\end{table}

\bibliographystyle{apalike}
\bibliography{PScTcM}

\end{document}